\documentclass[useAMS,usenatbib]{mnras}

\usepackage[english]{babel}
\usepackage{amsmath}
\usepackage{amssymb}
\usepackage{txfonts}
\usepackage{graphicx}
\usepackage{url}
\usepackage{color}

\newcommand{\ud}{\ensuremath{\mathrm{d}}}

\title{A Simple Approach to the Supernova Progenitor-Explosion Connection}

\author[B.~M\"uller et al.]{
  Bernhard~M\"uller$^{1,2,4}$\thanks{E-mail: b.mueller@qub.ac.uk},
  Alexander~Heger$^{2,3,4,5}$,
David~Liptai$^{2}$,
Joshua~B.~Cameron$^{2,4}$
\\
$^1${Astrophysics Research Centre, School
  of Mathematics and Physics, Queen's University
  Belfast, Belfast, BT7~1NN, United Kingdom} \\
$^2${Monash Centre for Astrophysics, School of
  Physics and Astronomy, Monash University, Victoria
  3800, Australia} \\
$^3${School of Physics \& Astronomy,
  University of Minnesota, Minneapolis, MN 55455, U.S.A.} \\
$^4${Joint Institute for Nuclear Astrophysics, 225
  Nieuwland Science Hall Department of Physics, University of Notre
  Dame, Notre Dame, IN 46556, U.S.A.} \\
$^5${Center for Nuclear Astrophysics, Department of
  Physics and Astronomy, Shanghai Jiao-Tong University, Shanghai
  200240, P. R. China.} \\
}

\begin{document}

\label{firstpage}
\pagerange{\pageref{firstpage}--\pageref{lastpage}}

\maketitle

\begin{abstract}
 We present a new approach to understand the landscape of supernova
 explosion energies, ejected nickel masses, and neutron star birth
 masses.  In contrast to other recent parametric approaches, our model
 predicts the properties of neutrino-driven explosions based on the
 pre-collapse stellar structure without the need for hydrodynamic
 simulations.  The model is based on physically motivated scaling laws
 and simple differential equations describing the shock propagation,
 the contraction of the neutron star, the neutrino emission, the
 heating conditions, and the explosion energetics.  Using model
 parameters compatible with multi-D simulations and a fine grid of
 thousands of supernova progenitors, we obtain a variegated landscape
 of neutron star and black hole formation similar to other
 parameterised approaches and find good agreement with semi-empirical
 measures for the ``explodability'' of massive stars.  Our predicted
 explosion properties largely conform to observed correlations between
 the nickel mass and explosion energy.  Accounting for the coexistence
 of outflows and downflows during the explosion phase, we naturally
 obtain a positive correlation between explosion energy and ejecta
 mass.  These correlations are relatively robust against parameter
 variations, but our results suggest that there is considerable leeway
 in parametric models to widen or narrow the mass ranges for black
 hole and neutron star formation and to scale explosion energies up or
 down.  Our model is currently limited to an all-or-nothing treatment
 of fallback and there remain some minor discrepancies between model
 predictions and observational constraints.
\end{abstract}


\section{Introduction}
\label{sec:intro}
The connection between the properties of progenitors of core-collapse
supernovae (SNe) and the properties of the resulting explosion and the
compact remnant is one of the outstanding problems in stellar
astrophysics.  Systematically understanding this connection from first
principles is difficult because the problem of the core-collapse
supernova explosion mechanism has not yet been finally solved (see,
e.g.,~\citealt{janka_12,burrows_12} for reviews).  Supernova theory,
however, now becomes testable due to observational findings and
indirect constraints on the supernova explosion mechanism on three
completely different fronts.

Over the recent years, direct observations of core-collapse supernova
explosions based on large surveys and the combination of observations
with archival data have yielded important insights about properties
such as the minimum and maximum progenitor mass of Type II-P
supernovae \citep{smartt_09a,smartt_09b,smartt_15}, the demography of
progenitors of different supernova types in the HR diagram (see
\citealp{smartt_09b} for a review), and possible correlations, e.g.,
between explosion energy and nickel mass \citep{hamuy_03} and between
progenitor mass and explosion energy
\citep{poznanski_13,chugai_14,pejcha_15b}.

The distribution of neutron star and black holes masses
\citep{kiziltan_13,oezel_12,oezel_10}, remnant kicks, and spins of
young neutron stars \citep{faucher_06,ng_07,repetto_12} provides
additional constraints on the inner workings of the supernova engine
and the progenitor--remnant connection
\citep[e.g.,][]{schwab_10,fryer_12,pejcha_12,janka_13,kochanek_15,clausen_15}.

The progenitor--explosion connection and the explosion mechanism are
intimately linked with the nucleosynthetic contribution of
core-collapse supernovae to the chemical evolution of galaxies.
Supernova theory
needs to account not only for the population-integrated yields of all
massive stars (in the vein of \citealt{rauscher_02}); it must also
explain the non-uniformity of heavy-element nucleosynthesis channels emerging
from stellar abundance studies
\citep[e.g.,][]{travaglio_04,ting_12,hansen_12,hansen_14}, which is
thought to be related to the existence of core-collapse supernova
sub-populations.  More indirect constraints on the fate of massive
stars come from, for example, the comparison of the observed star
formation and supernova rates \citep{horiuchi_11,botticella_12} and
the limit for the diffuse supernova neutrino background
\citep{beacom_10}.

To interpret these observational findings and constraints and their
implications for the core-collapse supernova explosion mechanism in a
systematic and statistical way, we are still largely relegated to
simplified analytic or parameterised numerical models, and this is
also the approach we follow here.  It is exceedingly difficult to
connect first-principle simulations of core-collapse supernova
explosions to the observable explosion properties and the remnant mass
distribution for several reasons: Despite recent successes in 3D
explosion modelling
\citep{melson_15a,melson_15b,lentz_15,mueller_15b}, obtaining
explosions has proved more difficult in 3D multi-group neutrino
transport models than in 2D \citep{hanke_13,tamborra_14a,takiwaki_14}.
Even in 2D, extending successful multi-group models to
sufficiently late times to obtain saturated values for the explosion
energy and remnant mass remains difficult
\citep{mueller_12a,mueller_12b,janka_12b,bruenn_13,suwa_10,summa_16,oconnor_16}, although
the models of \citet{bruenn_15} and \citet{mueller_15b} come close to
this point.  Even ignoring these obstacles, scanning the full
parameter space of progenitor models in zero-age main sequence
(ZAMS) mass, metallicity, and rotation rate with 3D simulations will
remain impractical in the near future.

For this reason, approximate analytic models or parameterised
simulations presently remain indispensable for understanding the
connection between progenitor and explosion properties.  Indeed, they
are arguably becoming more useful as fully-fledged simulations provide
an impetus and corrective for their improvement.  Earlier studies
\citep{fryer_99,fryer_01,heger_03} relied on a simple comparison of a
parameterised explosion energy (obtained from a fit to -- now outdated
-- 2D SPH simulations) to the binding energy of the envelope to
predict the ultimate fate of the remnant (neutron star vs.\ black
hole).  Recent studies have taken some steps to improve this simple
approach to various degrees in order to obtain a more consistent
estimate of the time of shock revival, the ``initial'' explosion
energy pumped into the ejecta during the first few seconds by the
supernova engine, and the resulting fallback and residual accretion
onto the compact remnant.  \citet{fryer_12} and \citet{belczysnki_12}
used an analytic estimate of the internal energy in the gain region at
the time of shock revival as a proxy for the explosion energy and then
calculated the fallback numerically to obtain the remnant mass
distribution, but their choice of the time of shock revival remains
very much \emph{ad hoc}.  \citet{pejcha_15a} used analytic scaling
laws for the critical neutrino luminosity required for shock revival
and various contributions to the explosion energy (recombination of
neutrino-heated material, explosive burning, and the neutrino-driven
wind) to predict the time of shock revival and the explosion
parameters.  Neutrino luminosities and mean energies from spherically
symmetric (1D) simulations were required as input.  Whereas the
approach of \citet{pejcha_15a} still leaves considerable freedom in
the choice of parameters, they account for this by an extended
statistical analysis of the remnant and explosion properties and their
dependence on the free parameters of their model.

Several works have relied on parameterised 1D simulations to
investigate the progenitor-explosion connection
\citep{oconnor_11,ugliano_12,perego_15,ertl_15,sukhbold_15}.
\citet{oconnor_11} used a simple trapping scheme and artificially
increased neutrino heating to determine the demarcation line between
neutron star and black hole formation for several sets of progenitor
models with 1D simulations of the first few seconds after collapse,
and formulated an approximate criterion $\xi_{2.5}\gtrsim 0.45$ for
the explodability in terms of a ``compactness parameter'' $\xi_{2.5}$.
\citet{ugliano_12} performed 1D simulations of 101 progenitors with
grey transport and an excised neutron star core using a cooling model
for the core and a prescribed contraction law to supply the necessary
inner boundary conditions.  The cooling model was calibrated to match
the explosion properties of SN~1987A.  Different from
\citet{oconnor_11}, they extended their simulations well beyond shock
breakout, thus filtering out ``failed explosions'' in which shock
revival occurs, but the energy input by the supernova engine is
insufficient to unbind the envelope.  Their long-time simulations
allowed them to predict the nature of the remnant (neutron star/black
hole), the explosion energies, nickel masses, the amount of fallback,
and the remnant mass function.  Using the same modelling approach
(with a few improvements), \citet{ertl_15} derived a more reliable and
physically motivated explosion criterion based on the mass coordinate
$M_4$ of the shell with entropy $s=4k_b/\mathrm{nucleon}$ and another
parameter $\mu_4$ related to the density and radius at that mass
coordinate, and the follow-up project of \citet{sukhbold_15}
  studied the nucleosynthesis, light curves, and explosion properties
  for a wide range of progenitors using their improved 1D approach.
Recently, \citet{perego_15} used a combination of the isotropic
diffusion source approximation \citep{liebendoerfer_09} with a
trapping scheme for heavy flavor neutrinos and a rather \emph{ad hoc}
enhancement of the neutrino heating to study the variation of
explosion energies and nucleosynthesis conditions for progenitors in
the limited range between $18\,M_\odot$ and $21\,M_\odot$.  Similar to
\citet{oconnor_11}, their simulations were limited to the first few
seconds after collapse.

These parameterised 1D simulations undoubtedly represent a step
forward in terms of consistency and rigour.  Replacing the simple
analytic arguments of
\citet{fryer_99,fryer_01,heger_03,fryer_12,belczysnki_12,pejcha_15a}
with numerical calculations has an obvious downside, however, since
this approach abandons the attractive, though very optimistic, idea of
a direct prediction of explosion properties based on the progenitor
structure alone.  It does not provide a fast way to assess the impact
of variations in stellar evolution models (wind mass loss, rotation,
magnetic fields, binary interaction, metallicity, mixing, etc.)
on the supernova explosion properties, unless the results can be
boiled down to readily computable criteria like the progenitor
compactness introduced by \citet{oconnor_11}.  Stellar evolution
modellers may also want to bypass 1D simulations of the collapse and
the initial explosion phase altogether and instead use a simpler model
for the explosion and remnant properties as input for nucleosynthesis
studies \citep{woosley_02} and population synthesis.  For these
purposes, parameterised 1D simulations are not a viable option even if
they are only used to provide time-dependent input data for an
analytic model as in \citet{pejcha_15a}.  Furthermore, simulation-based
approaches often make it difficult to disentangle how changes of the
input physics improve or degrade the heating conditions and affect the
explosion conditions.  Breaking the operation of the supernova engine
down to an overseeable number of simple equations is potentially helpful
for this purpose.

On a different note, \emph{all} the (semi-)analytic and numerical
approaches to the progenitor--explosion connection ignore the role of
multi-dimensional (multi-D) effects in the supernova explosion
mechanism.  Multi-D effects are responsible for improving the heating
conditions sufficiently to allow an explosive runaway, and it is by no
means clear that an artificial enhancement of neutrino heating in 1D
simulations will result in shock revival for similar progenitors and
at similar times as would a full multi-D simulation.  The situation is
even more serious after shock revival, where accretion funnels and
neutrino-driven outflows can persist for hundreds of milliseconds.
Since the bulk of the explosion energy is pumped into the ejecta
precisely in the phase during which downflows and outflows coexist
\citep{bruenn_15}, predictions of supernova explosion energies based
on 1D simulations (or analytic considerations relying essentially on a
spherically symmetric picture of the engine) remain problematic.

For these reasons, we present a somewhat different approach to the
progenitor-explosion connection in this paper.  In contrast to the
recent studies of \citet{oconnor_11,ugliano_12,pejcha_15a,perego_15},
our model is based on analytic predictions for the heating conditions
in the pre-explosion phase and simple ordinary differential equations
(ODEs) for the final explosion and remnant properties.  Moreover, we
improve the prediction of the initial explosion energy as a pivotal
quantity for the progenitor--explosion connection by taking the
co-existence of accretion downflows and neutrino-driven outflows
during the first $\mathord{\sim}1\,\mathrm{s}$ after shock revival
into account relying on guidance from recent multi-D simulations.

Our model is able to provide a quick estimate for the explosion
properties using only the stellar structure at the onset of collapse
as input.  This allows us to study the landscape of supernova
progenitors in unprecedented detail using a set of 2120
solar-metallicity stellar model with ZAMS masses ranging from
$10\,M_\odot$ to $32.5\,M_\odot$ computed with the stellar evolution
code \textsc{Kepler} \citep{weaver_78,heger_10}.  The extremely fine
grid of initial ZAMS masses with a spacing of $0.01\,M_\odot$ allows us
to assess the robustness of general trends in the explosion properties
with ZAMS mass and the magnitude of stochastic variations
\citep{clausen_15} more reliably than hitherto possible (although we
do not explore variations in other stellar parameters like rotation
and metallicity yet).  Moreover, with an analytic model, we can more
easily assess the sensitivity to any of the physical assumptions
inherent the model, such as the neutron star contraction law, and to
dimensionless efficiency parameters, e.g., for the conversion of
accretion power into neutrino luminosity and for the conversion of
neutrino heating into an outflow rate.  Indeed, we do not attempt to
predict ``the'' progenitor-explosion connection, which is arguably
impossible for \emph{any} model at this stage in the light of all the
uncertainties inherent both in the models and in the observational
constraints that can be used for calibrating them.  With a model like
ours, one can realistically hope to predict \emph{trends and
  tendencies}; and if these are roughly in line with observations,
this provides some corroboration for the underlying theory.

Our paper is structured as follows: In Section~\ref{sec:model}, we
introduce the analytic model used to estimate the heating conditions
in the pre-explosion phase, the onset of shock revival, and the ODE
model for the explosion properties for progenitors for which we predict
shock revival.  In Section~\ref{sec:results}, we discuss how our
model aligns with previous theoretical models for the landscape of
supernova explosion and remnant properties and observational findings
about core-collapse supernova explosion energies and the neutron star
mass function.  We then explore the sensitivity of our analytic/ODE
model to the most important adjustable parameters to assess the
robustness of our findings.  Finally, we summarise our results and
discuss their wider implications for supernova physics and stellar
evolution in Section~\ref{sec:summary}. In the Appendix,
we provide supplementary
information on the dependence of explosion properties on
helium and carbon/oxygen core mass.

\section{Model for Shock Revival and Explosion Properties}
\label{sec:model}

Before we formulate our analytic/ODE model for the evolution of
core-collapse supernovae from collapse through shock revival to
the point when neutrino energy input effectively ceases, it is
advisable to briefly review the different phases of this process,
highlighting the relevant physics that our model needs to capture.

After the collapse of the iron core to a neutron star, the core
bounce, and the formation of a shock wave, the shock quickly stalls
due the to the photodisintegration of heavy nuclei and neutrino
losses. Even once the shock has stalled, it continues to expand for a
few tens of milliseconds, however, as matter is piled onto the
proto-neutron star.  After a phase of $\gtrsim 50\, \mathrm{ms}$
during which cooling dominates in the entire post-shock region,
a region of net neutrino heating (gain region) behind the shock emerges.

Somewhat later the shock radius reaches a maximum and then recedes
again. Shock revival by the neutrino-driven mechanism is expected no
earlier than this juncture.  We therefore need a model of the subsequent
phase only, which can essentially be treated as a stationary accretion
problem with a time-varying mass accretion rate $\dot{M}$ and neutron
star mass $M$, and an appropriate inner boundary condition at the
neutron star surface. The neutrino heating conditions can then be
evaluated for the solution of this accretion problem.  For the sake
  of simplicity, we shall refer to this period of quasi-stationary
  accretion as \emph{pre-explosion phase} in the remainder of this
  paper; the first $\mathord{\sim} 50 \ldots 100 \, \mathrm{ms}$
  after bounce are not considered in this work since one does not
  expect neutrino-driven explosions to develop that early.

Shock revival occurs roughly once the accreted material spends
sufficient time in the gain region to receive enough energy from
neutrinos to negate its binding energy
\citep{janka_01,murphy_08b,fernandez_12}. This point marks the beginning
  of the \emph{explosion phase}, which we sub-divide further as follows:
Once the explosion sets in,
neutrino-driven outflows and accretion downflows coexist for a
considerable time (phase~I).  Due to continued accretion onto the proto-neutron
star, high neutrino luminosities comparable to the pre-explosion phase
can be maintained, which power outflows at a rate proportional to the
volume-integrated neutrino heating rate \citep{mueller_15b}.  Phase~I
continues roughly until the newly shocked matter is accelerated to a
sufficiently high velocity (roughly the escape velocity) to avoid
falling back onto the proto-neutron star.  Once accretion subsides
and the shock sweeps up the remaining shells of the envelope
without significant further input of energy from neutrino heating (phase~II),
the
explosion energy is expected to level out, or decline slowly to its
final value if the pre-shock matter still has a considerable binding
energy.  In the following, we shall present a quantitative model for
these different phases.

\subsection{Pre-Explosion Phase}

\label{sec:preexplosion}
During the pre-explosion phase, we largely follow \citet{janka_01} and model the gain region as an
adiabatically stratified layer dominated by radiation pressure
($P\propto T^4$) so that the pressure $P$, density $\rho$, and
temperature $T$ approximately follow power laws,
\begin{equation}
P \propto r^{-4}, \quad
\rho \propto r^{-3},\quad
T \propto r^{-1}.
\end{equation}
The Rankine-Hugoniot jump conditions at the shock and the balance of
heating and cooling at the gain radius (which needs to be specified by
a model for the contraction of the neutron star) provide an outer and
inner boundary condition. Once the neutrino luminosity $L_\nu$ and
mean energy $E_\nu$, the gain radius $r_\mathrm{g}$, the
proto-neutron star mass $M$, and the mass accretion rate $\dot{M}$
are known, we can first solve for the shock radius $r_\mathrm{sh}$ and
then finally evaluate the neutrino heating conditions.  To this end, we
compute the advection time-scale $\tau_\mathrm{adv}$ (the time-scale
over which the accreted matter is exposed to neutrino heating in the
gain region) and the heating time-scale $\tau_\mathrm{heat}$ (the time
required to inject enough energy into the gain region to make it
unbound).  Once the condition $\tau_\mathrm{adv}/\tau_\mathrm{heat} >1$
is met, we assume that runaway shock expansion takes place
 \citep{thompson_00,janka_01,thompson_05,murphy_08b,fernandez_12} and then estimate the
explosion energy and the residual accretion onto the proto-neutron
star in detail in Section~\ref{sec:explosion_phase}.  In order to
account for multi-dimensional effects, we correct the shock radius as
well as the accretion and heating time-scale by approximately
accounting for the effect of turbulent stresses in the post-shock
region \citep{mueller_15a}.

\subsubsection{Infall and Accretion Rate}
During the pre-explosion phase, we assume that matter reaches the
neutron star within a constant multiple of the free-fall time-scale
for a given mass shell, The infall time $t$ is thus related to the
mass coordinate $M$ of the infalling shell as,
\begin{equation}
  \label{eq:tff}
t=C \tau_\mathrm{ff}(M)
=\sqrt{\frac{\pi}{4 G \bar{\rho}}},
\end{equation}
where $\bar{\rho}$ is the average density inside a given mass shell
located at an initial radius $r$ (i.e., $\bar{\rho}=4/3\pi M r^{-3}$).  The
resulting mass accretion rate $\dot{M}$ is given by
\citep{woosley_15},
\begin{equation}
\label{eq:macc}
\dot{M}=\frac{2 M }{t} \frac{\rho}{\bar{\rho} - \rho},
\end{equation}
where $\rho$ is the initial density of a given mass shell prior to
collapse.  We note that the non-dimensional coefficient in our
definition of the free-fall time-scale has been chosen such that our
analytic estimate for the accretion rate fits the results form
numerical simulations in the late accretion phase.  At early times
($\lesssim 100\,\mathrm{ms}$ after bounce), there are significant
quantitative deviations from Equation~(\ref{eq:macc}), but both
simulations as well as tight constraints on the amount of ejected
material that has undergone explosive silicon burning
\citep{woosley_73,arnett_96} indicate that shock revival should not
occur at such an early stage yet anyway.

\subsubsection{Jump Conditions at the Shock}
In the pre-explosion phase, we can assume the shock to be
quasi-stationary, i.e., the shock velocity is negligible
(although the shock radius slowly changes due to
secular changes in the parameters of the accretion problem).
Using the strong-shock approximation and
neglecting the thermal pressure in the pre-shock region, the Rankine-Hugoniot
conditions for the post-shock density $\rho_\mathrm{sh}$ and pressure
$P_\mathrm{sh}$ can be written as
\begin{equation}
\rho_\mathrm{sh}=\beta \rho_\mathrm{pre},
\end{equation}
\begin{equation}
\label{eq:p_shock}
P_\mathrm{sh}=\frac{\beta-1}{\beta} \rho_\mathrm{pre} v_\mathrm{pre}^2,
\end{equation}
in terms of the pre-shock velocity $v_\mathrm{pre}$ and density
$\rho_\mathrm{pre}$, and the compression ratio $\beta$ at the shock.
Simulations indicate that $v_\mathrm{pre}$ reaches a large
fraction of the free-fall velocity, and we thus use
\begin{equation}
\label{eq:v_pre}
v_\mathrm{pre}
=
\sqrt{\frac{2 G M}{r_\mathrm{sh}}},
\end{equation}
for further calculations. $\rho_\mathrm{pre}$ can then obtained from $\dot{M}$ as
$\rho_\mathrm{pre}= \dot{M}/(4\pi r^2 v_\mathrm{pre})$.

\subsubsection{The Inner Boundary Condition}
\label{sec:ibc_gain}
For formulating the inner boundary condition for the gain region, we
require a model for the evolution of the gain radius $r_\mathrm{g}$
and the neutrino luminosity $L_\nu$ and mean energy $E_\nu$ of
electron neutrinos and antineutrinos as a function of time,
proto-neutron star mass and accretion rate. It is convenient to start
with the neutrino mean energy (or, specifically,
the electron antineutrino mean energy), for which one
finds a very simple relationship from first-principle neutrino
hydrodynamics simulations \citep{mueller_14},
\begin{equation}
\label{eq:e_nu}
E_\nu \propto M.
\end{equation}
At the level of this work, we do not distinguish between electron neutrinos
and antineutrinos and use this as a proxy for the mean energy of either
species.
Since the cooling layer is roughly isothermal, the same proportionality
also holds  for the temperature at the gain radius, $T_\mathrm{g} \propto M$.

\begin{figure}
\includegraphics[width=\linewidth]{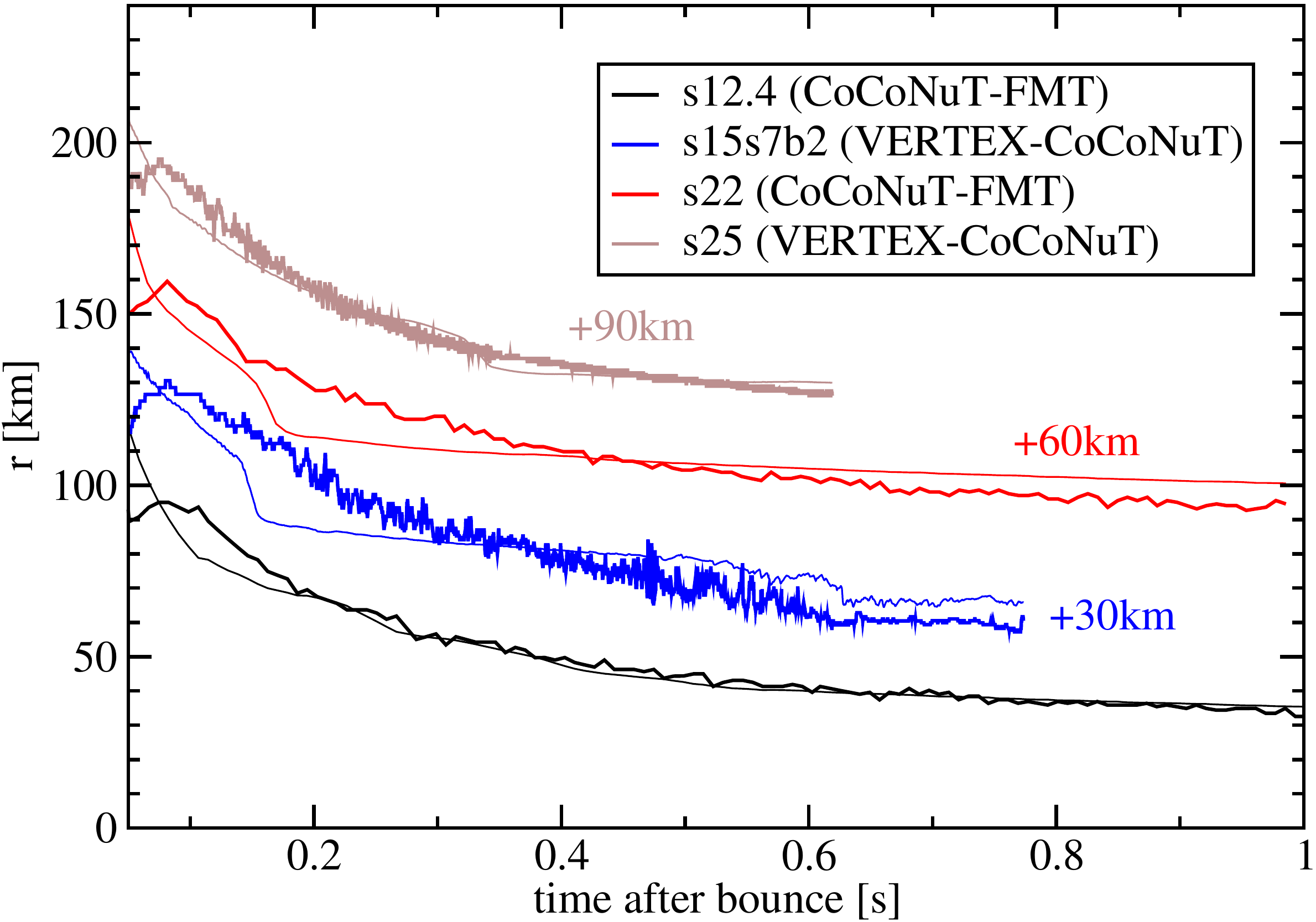}
\caption{ Comparison of the gain radius in models computed with the
  \textsc{Vertex-CoCoNuT} and \textsc{CoCoNuT-FMT} codes (thick lines)
  and the analytic contraction law~(\ref{eq:pns_radius}) (thin lines)
  for progenitor models from \citet{woosley_02} (black:
  \texttt{s12.4}, red: \texttt{s22}, light brown: \texttt{s25}) and
  \citet{woosley_95} (blue: \texttt{s15s7b2}).  All models have been
  computed using the nuclear equation of state of \citet{lattimer_91}
  with a nuclear incompressibility modulus $K=220\,\mathrm{MeV}$,
  except for \texttt{s15s7b2} ($K=180\,\mathrm{MeV}$).  The
  \textsc{Vertex} models have been discussed previously in
  \citet{mueller_12a,mueller_13,mueller_14}.  Note that the curves for
  \texttt{s15s7b2}, \texttt{s22}, and \texttt{s25} are offset by
  $30\,\mathrm{km}$, $60\,\mathrm{km}$, and $90\,\mathrm{km}$,
  respectively.  Overall, the analytic contraction law reproduces the
  contraction of the gain radius reasonably well. It somewhat
  overestimates the recession after the infall of composition
  interfaces with strong density jumps, which is more gradual in
  realistic models.
\label{fig:contraction}
}
\end{figure}

The gain radius $r_\mathrm{g}$ can then be determined by noting
that the accreted matter loses roughly half of its
gravitational potential energy as accretion luminosity
$G M \dot{M}/(2r_\mathrm{g})$ close to the gain radius, and by
equating this luminosity contribution to the
grey-body luminosity at the gain radius \citep[cp.][]{janka_12} we find
\begin{equation}
E_\nu^4 r_\mathrm{g}^2 \propto T_\mathrm{g}^4 r_\mathrm{g}^2 \propto M^4 r_\mathrm{g}^2
\propto \frac{G M \dot{M}}{2r_\mathrm{g}}.
\end{equation}
This leads to $r_\mathrm{g} \propto \dot{M}^{1/3} M^{-1}$. Obviously,
this approximation breaks down for small $\dot{M}$, and we
therefore interpolate smoothly between this
solution and the radius of a cold neutron star $r_0$ as a floor
value,
\begin{equation}
\label{eq:pns_radius}
r_\mathrm{g}=\sqrt[3]{r_1^3\left(\frac{\dot{M}}{M_\odot \, \mathrm{s}^{-1}}\right)  \left(\frac{M}{M_\odot}\right)^{-3}+r_0^3}.
\end{equation}
In this work, we use $r_0=12\,\mathrm{km}$ and $r_1 =
120\,\mathrm{km}$. Figure~\ref{fig:contraction} shows that
Equation~(\ref{eq:pns_radius}) provides a reasonably good fit to the
contraction of the proto-neutron star except for brief phases when the
accretion rate drops abruptly after the infall of a shell interface.

The neutrino luminosity $L_\nu$ (of electron neutrinos and
antineutrinos) is modelled as consisting of an accretion component
$L_\mathrm{acc}$,
\begin{equation}
\label{eq:lum_acc}
L_\mathrm{acc}=\zeta \frac{G M \dot{M}}{r_\mathrm{g}},
\end{equation}
where the conversion of accretion energy into luminosity is specified
by an adjustable efficiency parameter $\zeta$, and a diffusive
component $L_\mathrm{diff}$ originating from deeper layers of the
proto-neutron star (compare \citealt{fischer_09,mueller_14}). Based on
the results of \citet{mueller_14}, we typically use
$\zeta=0.7$.\footnote{ \citet{mueller_14} define $\zeta$ by comparing
  the accretion luminosity to the gravitational potential at a density
  of $10^{11}\,\mathrm{g}\,\mathrm{cm}^{-3}$, and therefore obtain
  slightly smaller values of $\zeta \approx 0.5$. If $\zeta$ is
  defined using the potential at the gain radius, a larger value is
  needed.}  We estimate $L_\mathrm{diff}$ by assuming that the binding
energy of a cold neutron star \citep{lattimer_89,lattimer_01},
\begin{equation}
\label{eq:e_bind}
E_\mathrm{bind} \approx 0.084\, M_\odot c^2 (M_\mathrm{NS}/M_\odot)^2
\end{equation}
is radiated away as diffusion luminosity over a time-scale
$\tau_\mathrm{cool}$. Here $M_\mathrm{NS}$ is the
  gravitational mass of the neutron star. An additional
  power-law dependence on the baryonic neutron star mass is introduced
for $\tau_\mathrm{cool}$ to account (somewhat
\emph{ad hoc}) for the fact that the higher densities, temperatures,
and (for electron neutrinos) chemical potentials in high-mass neutron
stars increase the diffusion time-scale,
\begin{equation}
\tau_\mathrm{cool}=\tau_\mathrm{1.5}\,\mathrm{s} \times
\left(\frac{M}{1.5\,M_\odot}\right)^{5/3},
\end{equation}
\begin{equation}
  \label{eq:ldiff}
L_\nu=-0.3\dot{E}_\mathrm{bind}
\approx 0.3\times  \frac{E_\mathrm{bind}(M)}{\tau_\mathrm{cool}(M)}.
\end{equation}
The pre-factor $0.3$ accounts for the fact that only roughly one third
of the binding energy is emitted in the form of electron neutrinos and
antineutrinos that contribute to neutrino heating in the gain
layer. Moreover, the material accreted onto the proto-neutron star has
already lost part of its binding energy as accretion luminosity in the
cooling region. The value of the proportionality constant
$\tau_\mathrm{1.5}$ (cooling time-scale for a $1.5\,M_\odot$ mass
neutron star) has to be determined from simulations; the recent
results of \citet{huedepohl_phd} suggest that $\tau_{1.5} \approx
1.2\,\mathrm{s}$.\footnote{Note that this is the $e$-folding
  time-scale for the luminosity, whereas the cooling time-scale often
  refers to the time it takes for the proto-neutron star to cool down
  sufficiently to become transparent to neutrinos (which is
  considerably longer).} Our choice of parameters results in diffusion
luminosities of a few $10^{52}\,\mathrm{erg}\,\mathrm{s}^{-1}$ and a
tendency towards slightly higher diffusion luminosities for higher
neutron star masses, which is in agreement with systematic studies of
the progenitor dependence of the heavy flavour neutrino emission
\citep{oconnor_13}.

Neglecting secular changes in $M$ and $\tau_\mathrm{cool}$, we
simply use the exponential solution for the diffusion luminosity for
constant $M$ to estimate the instantaneous value of $L_\mathrm{diff}$:
\begin{equation}
\label{eq:lum_diff}
L_\mathrm{diff}=E_\mathrm{bind}(M) e^{-t/\tau_\mathrm{cool}(M)}.
\end{equation}
We note that $E_\mathrm{bind}$ (equation~\ref{eq:e_bind}) can be expressed
explicitly in terms of the current baryonic neutron star mass $M$ instead of
the gravitational neutron star mass $M_\mathrm{NS}$,
\begin{equation}
E_\mathrm{bind}=
\left(M-\frac{\left(-1+\sqrt{1+0.336\, M/M_\odot}\right)M_\odot}{0.168}
\right) c^2.
\end{equation}

For the total electron (anti-)neutrino luminosity, we also
include a factor accounting for general relativistic redshift
of neutrinos originating from close to the proto-neutron star radius:
\begin{equation}
  \label{eq:ltot}
L_\mathrm{\nu}=\sqrt{1-\frac{2GM}{r_\mathrm{PNS}}}(L_\mathrm{acc}+L_\mathrm{diff}),
\end{equation}
where we use $r_\mathrm{PNS} \approx 5/7 r_\mathrm{g}$. The redshift
factor also needs to be applied to the neutrino mean energies.

Once the neutrino luminosity and mean energy and
the gain radius are determined, we can
formulate the second (inner) boundary condition for
the pressure stratification in the gain region.
If the neutrino heating and cooling rate per
unit mass are to balance each other
at the gain radius, we must have
\begin{equation}
T_\mathrm{g}^6 \propto \frac{L_\nu E_\nu^2}{r_\mathrm{g}^2},
\end{equation}
for the temperature $T_\mathrm{g}$ at the gain radius
since the cooling and heating rates per baryon
scale as $T^6\propto P^{3/2}$ and $L_\nu E_\nu^2/r_\mathrm{g}^2$,
respectively. With $P_\mathrm{g} \propto T_\mathrm{g}^{4}$, the
pressure at the gain radius $P_\mathrm{g}$ scales as,
\begin{equation}
\label{eq:p_gain}
P_\mathrm{g}^{3/2} \propto \frac{L_\nu E_\nu^2}{r_\mathrm{g}^2},
\end{equation}
which is our second (inner) boundary condition
for the pressure stratification in the gain region.

\subsubsection{Solution for the Shock Radius}
Solving Equations~(\ref{eq:p_shock},\ref{eq:p_gain})
using $P \propto r^{-4}$, we obtain a scaling
relation for the shock radius \citep{janka_12},
\begin{equation}
\label{eq:shock_proportionality}
r_\mathrm{sh} \propto
\frac{(L_\nu  E_\nu^2)^{4/9} r_\mathrm{g}^{16/9}}
{\dot{M}^{2/3} M^{1/3}}
\propto
\frac{L_\nu  ^{4/9} M^{5/9} r_\mathrm{g}^{16/9}}
{\dot{M}^{2/3}},
\end{equation}
where we have used $E_\nu \propto M$ to obtain the second form.

Multi-dimensional effects are not yet taken into account in this
formula for the shock radius.   \citet{mueller_15a} pointed out
  that the different multi-D effects that have been proposed as
  beneficial for shock revival, such as shock expansion due to
  turbulent stresses \citep{burrows_95,murphy_12}, the increased
  advection time-scale \citep{buras_06b,murphy_08b,marek_09}, and the
  increased heating efficiency compared to 1D are inseparably related
  and coupled to each other by feedback processes. They suggested that
  they can effectively be captured in a 1D model by modifying the
  equations for the hydrostatic structure and the jump conditions at
  the shock. To this end, they proposed to account for turbulent
  stresses in a rather simple fashion by a correction factor
containing the root-mean-square averaged turbulent Mach number
$\langle \mathrm{Ma}^2\rangle$ in the gain region,
\begin{equation}
r_\mathrm{sh} \rightarrow r_\mathrm{sh}
\left(1+\frac{4 \langle\mathrm{Ma}^2 \rangle }{3}\right)^{2/3},
\end{equation}
which then also implies an increase in $M_\mathrm{g}$ and hence in the
heating efficiency (Equation 16 in \citealt{mueller_15a}) and the
advection time-scale (see Equation~\ref{eq:tadv} below). Using a large
number of axisymmetric supernova simulations of different progenitors,
\citet{summa_16} recently showed that the effect of turbulent stresses
on the critical neutrino luminosity required for shock revival can be
captured remarkably well by such a simple modification.

Since we merely use the shock radius to solve for the
point in time where the critical explosion condition
$\tau_\mathrm{adv}/\tau_\mathrm{heat}=1$ is met, we may
as well replace the turbulent Mach number
with its critical value $\langle \mathrm{Ma}^2 \rangle\approx 0.4649$
\citep{mueller_15a}, which implies that
the shock radius obtained from
Equation~(\ref{eq:shock_proportionality}) can be
consistently multiplied with a constant factor
$\alpha_\mathrm{turb}$,
\begin{equation}
r_\mathrm{sh} \rightarrow \alpha_\mathrm{turb} r_\mathrm{sh}.
\end{equation}
\citet{mueller_15a} derived a value of $\alpha_\mathrm{turb} \approx
1.38$ in the absence of strong seed perturbations in the progenitor,
which they found to be in good agreement with 2D simulations.
  There is obviously justification for varying $\alpha_\mathrm{turb}$
  within reasonable bounds on several grounds: While the
  underlying scaling law for the turbulent Mach number likely holds in
  3D as well, the relevant dimensionless efficiency parameters
  (e.g.\ for turbulent dissipation) and hence $\alpha_\mathrm{turb}$
  are bound to be slightly different, although the difference in
  $\alpha_\mathrm{turb}$ between 2D and 3D cannot be excessive given
  that the critical luminosity is very similar in both cases
  \citep{hanke_12,dolence_13,couch_12b,handy_14}. Moreover, since the
record of 3D supernova simulations in obtaining explosions is mixed so
far, and some crucial ingredients that boost the turbulent motions
behind the shock may still be missing (such as strong seed
perturbations from convective burning in the progenitor;
\citealt{couch_13,couch_15b,mueller_15a}), Finally, since our fits
for the shock radius, and the advection and heating time-scales are
already based on 2D and 3D simulations, and since the fits never
perfectly reproduce the heating conditions in self-consistent models,
$\alpha_\mathrm{turb}$ needs to be renormalised, and we will use
values in the range $\alpha_\mathrm{turb}=1.08\ldots 1.28$ with a
  standard value of $\alpha_\mathrm{turb}=1.18$. Because of this
  renormalisation, $\alpha_\mathrm{turb}=1$ no longer has any special
  significance, and cannot be interpreted as the limit where multi-D
  effects are ``switched off''.  Using the theoretically inferred
  value of $\alpha_\mathrm{turb} \approx 1.38$ at shock revival in
  multi-D, the ``1D'' limit would more likely correspond to
  $\alpha_\mathrm{turb} \approx 0.86$, in which case we only obtain
  four explosions at the lower mass end, which is not implausible and
  in line with the fact that 1D simulations do not show explosions
  except at the low-mass end
  \citep{kitaura_06,janka_08,fischer_10,melson_15a}. This finding
  should not be interpreted as more than a rough consistency check for
  our model, since the role of multi-D effects is a lot more subtle in
  reality.

The proportionality constants for the final scaling law for
$r_\mathrm{sh}$ are once again chosen to obtain a good fit to
simulation results,
\begin{eqnarray}
\nonumber
r_\mathrm{sh}
&=&  \alpha_\mathrm{turb} \times 55\,\mathrm{km}
\times
\left(\frac{L_\nu}{10^{52}\,\mathrm{erg}\,\mathrm{s}^{-1}}\right)^{4/9}
\times
\left(\frac{M}{M_\odot}\right)^{5/9} \\
\label{eq:rsh}
&&\times
\left(\frac{r_\mathrm{g}}{10\,\mathrm{km}}\right)^{16/9}
\times
\left(\frac{\dot{M}}{M_\odot\,\mathrm{s}^{-1}}\right)^{-2/3}.
\end{eqnarray}

\subsubsection{Heating Conditions}
From the shock radius, we immediately find a scaling law for
the advection time-scale,
\begin{eqnarray}
\label{eq:tadv}
\tau_\mathrm{adv}&=&
\frac{M_\mathrm{g}}{\dot{M}}
=\frac{\int_{r_\mathrm{g}}^{r_\mathrm{sh}} 4 \pi r^2 \beta \rho_\mathrm{pre} (r_\mathrm{sh}/r)^3 \, \ud r}{\dot{M}}
\\
\nonumber
&\approx&
18\,\mathrm{ms}
\left(\frac{r_\mathrm{sh}}{100\,\mathrm{km}}\right)^{3/2}
\left(\frac{M}{M_\odot}\right)^{-1/2}
\ln \frac{r_\mathrm{sh}}{r_\mathrm{g}},
\end{eqnarray}
where the proportionality constant has been chosen
to fit the results of first-principle simulations
\citep{mueller_15a}.

The heating time-scale $\tau_\mathrm{heat}$ can be expressed in terms
of the average mass-specific neutrino heating rate $\dot{q}_\nu$ and
the average net binding energy (i.e., thermal, kinetic, and potential
energy) $e_\mathrm{g}$ of matter in the gain region.
It is relatively easy to obtain a robust scaling law for $\dot{q}_\nu$
\citep{janka_01,janka_12,mueller_15a},
\begin{equation}
\label{eq:qheat}
\dot{q}_\nu
\propto
\frac{L_\nu E_\nu^2}
{r_\mathrm{g}^2}.
\end{equation}
The average binding energy is a slightly more complicated
case. Neither the assumption of a constant, time-independent binding
energy \citep{mueller_15a}, nor the assumption that $e_\mathrm{g}$
scales with the gravitational potential energy at the shock
\citep{janka_12} provides an optimal fit to simulation results.  A
better estimate for $e_\mathrm{g}$ can be obtained by invoking
Bernoulli's theorem for a stationary compressible flow in spherical
symmetry \citep{mueller_15b}: Since the sum of the total enthalpy $h$
(including rest-mass contributions), the kinetic energy density, and
the gravitational potential are conserved during the infall, it is
roughly equal to its (negligibly small) value at the initial position
of a given mass shell,
\begin{equation}
h+v^2/2+\Phi=0,
\end{equation}
Neglecting the kinetic energy in the post-shock region, we therefore
find
\begin{equation}
\epsilon_\mathrm{therm}
+\epsilon_\mathrm{diss}
+\frac{P_\mathrm{sh}}{\rho_\mathrm{sh}}
-\frac{GM}{r_\mathrm{sh}}
\approx 0,
\end{equation}
for the thermal energy per unit mass $\epsilon_\mathrm{therm}$ just
behind the shock. Note that rest-mass contributions are excluded from
$\epsilon_\mathrm{therm}$ and lumped into the dissociation energy
$\epsilon_\mathrm{diss}$. With radiation pressure dominating in the
post-shock region, we have
$P_\mathrm{sh}/\rho_\mathrm{sh}=\epsilon_\mathrm{therm}/3$, and hence
obtain
\begin{equation}
\frac{4}{3}\epsilon_\mathrm{therm}
+\epsilon_\mathrm{diss}
=\frac{GM}{r_\mathrm{sh}},
\end{equation}
\begin{equation}
\epsilon_\mathrm{therm}
=
\frac{3}{4}
\left(\frac{GM}{r_\mathrm{sh}}
-\epsilon_\mathrm{diss}\right),
\end{equation}
which leads to
\begin{equation}
\label{eq:ebind}
|e_\mathrm{g}|=
\left |
\epsilon_\mathrm{therm}-\frac{GM}{r_\mathrm{sh}}
\right |
=\frac{3}{4}\epsilon_\mathrm{diss}
+\frac{GM}{4r_\mathrm{sh}},
\end{equation}
for the post-shock binding energy \emph{without} rest-mass
contributions. Assuming complete dissociation
of the infalling heavy nuclei into nucleons, we have
$\epsilon_\mathrm{diss} \approx 8.8\,\mathrm{MeV}$.
Note that the value of the total energy per unit mass
immediately behind the shock is used as a proxy for
the \emph{entire} gain region.

After combining Equations~(\ref{eq:qheat}) and (\ref{eq:ebind})
and choosing the appropriate value for the proportionality
constant, we obtain our final expression for the heating
time-scale,
\begin{eqnarray}
\label{eq:theat}
\tau_\mathrm{heat}
&=&
\nonumber
150\,\mathrm{ms}
\times \left(\frac{|e_\mathrm{g}|}{10^{19}\,\mathrm{erg}\,\mathrm{g}^{-1}}\right)
\times \left(\frac{r_\mathrm{g}} {100\,\mathrm{km}}\right)
\\
&&
\nonumber
\times\left(\frac{L_\nu}{10^{52}\,\mathrm{erg}\,\mathrm{s}^{-1}}\right)^{-1}
\times
\left(\frac{M}{M_\odot}\right)^{-2} \\
\end{eqnarray}

We note that Equations~(\ref{eq:rsh},\ref{eq:tadv},\ref{eq:theat}) also implicitly
determine the mass in the gain region $M_\mathrm{g}$, the average neutrino heating
rate per unit mass $\dot{q}_\nu$ and the volume-integrated
neutrino heating rate $\dot{Q}_\nu= \dot{q}_\nu M_\mathrm{g}$. For
our treatment of the explosion phase, it will be convenient
to define an efficiency parameter $\eta_\mathrm{acc}$ relating the
mass accretion rate $\dot{M}$ onto the proto-neutron star
to the volume-integrated heating rate $\dot{Q}_\nu$,
\begin{equation}
\eta_\mathrm{acc}=
\frac{\dot{Q}_\nu}{\dot{M}}.
\end{equation}

\subsection{Explosion Phase}
\label{sec:explosion_phase}
The analytic model for the heating conditions during the pre-explosion
phase presented in Section~\ref{sec:preexplosion} allows us to compute
the critical time-scale ratio $\tau_\mathrm{adv}/\tau_\mathrm{heat}$
as a function of the mass coordinate of the infalling shells. If we
find $\tau_\mathrm{adv}/\tau_\mathrm{heat}<1$
throughout the star or at least for all $M$ smaller than the (unknown)
maximum baryonic neutron star mass $M_\mathrm{max}$, we assume
that a stellar model forms a black hole without ever undergoing
shock revival. In this work, we use a maximum gravitational
neutron star mass of $2.05\,M_\odot$, which is compatible with
the best current lower limits for $M_\mathrm{max}$ \citep{antoniadis_13,demorest_10}.

Otherwise, we take the minimum $M$ for which
$\tau_\mathrm{adv}/\tau_\mathrm{heat}=1$ as an ``initial mass cut''
$M_\mathrm{ini}$ and then proceed to estimate the residual accretion
onto the proto-neutron star and the explosion parameters.  We achieve
this by relating the amount of accretion after shock revival, the
shock propagation, and the energetics of the incipient explosion
(quantified by the ``diagnostic explosion energy'', viz.\ the total
energy of the material that is nominally unbound at a given stage
after shock revival) to each other.

\subsubsection{Accretion after Shock Revival}
Except for the least
massive supernova progenitors \citep{kitaura_06,mueller_12b},
successful first-principle simulations of supernova explosions
\citep{buras_06b,marek_09,mueller_12a,mueller_12b,mueller_13,mueller_14,janka_12b,bruenn_13,bruenn_15,suwa_10,nakamura_15,takiwaki_12,takiwaki_14,summa_16,oconnor_16}
consistently show the persistence of accretion downflows for many
hundreds of milliseconds after shock revival --- and in many cases to the
very end of the simulations so that the final explosion parameters of
the models cannot be determined yet. A more quantitative analysis of
the mass fluxes $\dot{M}_\mathrm{out}$ and  $\dot{M}_\mathrm{acc}$ in neutrino-driven outflows and
cold accretion downflows in
the long-time simulations of \citet{mueller_15b} revealed that
the accretion through downflows
completely outweighs the outflow rate for a long time,
\begin{equation}
\label{eq:accretion}
\dot{M}_\mathrm{acc} \gg \dot{M}_\mathrm{out}.
\end{equation}
While the long persistence of accretion is a major technical
obstacles for simulations, it simplifies the treatment of the
post-explosion phase in the sense that it allows us to use the same
estimate for the accretion rate onto the proto-neutron star (and hence
for the neutron star contraction, the neutrino luminosity, and the
neutrino heating rate) as in the pre-explosion phase. During this
initial phase of the explosion (henceforth called \emph{phase~I of the
  explosion}), the primary contribution to the explosion energy comes
from neutrino-heated outflows that are driven by a relatively high
accretion luminosity.

Eventually, the residual accretion will cease and
Equation~(\ref{eq:accretion}) breaks down. In the subsequent phase
(\emph{phase~II of the explosion}), the proto-neutron star will still
radiate neutrinos as it cools over a time-scale of several seconds,
and the neutrino-driven wind will still contribute somewhat to the
explosion energy.

One can estimate that the transition from phase~I to phase~II occurs
roughly when the newly shocked material is accelerated to the local
escape velocity \citep{marek_09} because this precludes accretion onto the neutron star
on a short time-scale (although the interaction with the rest of the
ejecta may still lead to late-time fallback).  This can be translated
into a condition for the shock velocity: At the transition point, the
shock will already have propagated to several thousands or even tens
of thousands of kilometres and the immediate post-shock velocity will
be high compared to the small pre-shock infall velocity. For a
negligible pre-shock velocity, the post-shock velocity
$v_\mathrm{post}$ of the newly shocked material is given in terms
of the shock velocity $v_\mathrm{sh}$ and
the compression ratio $\beta_\mathrm{expl}$ as
\begin{equation}
v_\mathrm{post}=\frac{\beta_\mathrm{expl}-1}{\beta_\mathrm{expl}}v_\mathrm{sh}.
\end{equation}
Accretion will thus subside roughly once the criterion
\begin{equation}
\label{eq:freezeout}
\frac{\beta_\mathrm{expl}-1}{\beta_\mathrm{expl}}v_\mathrm{sh}
=\sqrt{\frac{2GM}{r}}
\end{equation}
is met. The radius $r$ in this equation
is the \emph{initial} radius of the mass shell $M$, which cannot have
moved very far from its initial position when it is hit by the shock.
Furthermore, we note that the compression ratio $\beta_\mathrm{expl}$
in the explosion phase can be different from the pre-explosion phase
and will generally be \emph{smaller} than the compression ratio
$\beta=(\gamma+1)/(\gamma-1)$ for an ideal gas with a $\gamma$-law
equation of state because of nuclear burning in the shock.
Values around $\beta_\mathrm{expl}=4$ are typical for the early explosion phase
\citep{mueller_15b}.

\subsubsection{Shock Velocity}
The propagation of the shock depends on the energetics of the
explosion. \citet{mueller_15b} showed that despite the enormously
complicated multi-dimensional flow structure after shock revival, it
turns out that the average shock velocity (defined as the time
derivative of the average shock radius $r_\mathrm{sh}$) closely
follows the analytic formula derived by \citet{matzner_99} for shock
propagation in spherical symmetry,\footnote{In this paper, we use
the original formula of \citet{matzner_99} although \citet{mueller_15b}
found slightly larger values (by $\approx 30 \%$) for the average shock velocity.
This does not fundamentally change the results, and would merely require
a slight adjustment of the standard set of parameters that we shall introduce
later to produce more or less the same results.}
\begin{equation}
\label{eq:vshock}
v_\mathrm{sh}
=
0.794
\left(\frac{E_\mathrm{diag}}{M-M_\mathrm{ini}}\right)^{1/2}
\left(\frac{M-M_\mathrm{ini}}{\rho r^3}\right)^{0.19}.
\end{equation}
Here, $E_\mathrm{diag}$ is the diagnostic explosion energy,
and the density $\rho$ and radius $r$ again refer to the
initial progenitor model.

\subsubsection{Evolution of Explosion and Remnant
Parameters-- Phase I}
Combined with a model for the evolution of the diagnostic explosion
energy in phase~I and phase~II, we can use
Equations~(\ref{eq:freezeout}) and (\ref{eq:vshock}) to determine both
the amount of residual accretion (and hence the final neutron star
mass) as well as the final explosion energy.

During phase~I, both strong neutrino heating powered by the accretion
downflows as well as nuclear burning in the shock contribute to the
explosion energy. Simulation results suggest that the contribution
from neutrino heating can be estimated as follows:
As the outflowing material just barely reaches positive
total energy, the outflow rate is roughly given by the ratio
of the volume-integrated neutrino-heating rate $\dot{Q}_\nu$
and the initial binding energy at the gain radius
$|e_\mathrm{g}|$,
\begin{equation}
\label{eq:mout}
\dot{M}_\mathrm{out}
=
\frac{\eta_\mathrm{out} \dot{Q}_\nu}{|e_\mathrm{g}|}
=
\frac{\eta_\mathrm{out} \eta_\mathrm{acc} \dot{M}_\mathrm{acc}}{|e_\mathrm{g}|}.
\end{equation}
Here, $\eta_\mathrm{out}$ is a dimensionless efficiency parameter for
the conversion of neutrino heating into an outflow rate. The recent 3D
simulation of \citet{mueller_15b} suggests $\eta_\mathrm{out}=1$, and
we adopt this value throughout our work. $\eta_\mathrm{out}$ needs to
be carefully distinguished from the surface fraction
$\alpha_\mathrm{out}$ occupied by neutrino-driven outflows far away from
the gain radius, which we will need later.
Note that we use the heating rate computed
in Section~\ref{sec:preexplosion} during the pre-explosion phase because the accretion
onto the proto-neutron star and hence the neutrino heating
are hardly affected by the outflows initially.

The energy input by neutrino heating into the outflow is essentially
used up completely to unbind the material, and the net contribution
from the explosion energy comes from the recombination of nucleons
(compare \citealt{scheck_06,mueller_12a}).  Therefore, the evolution
of the diagnostic explosion energy $E_\mathrm{diag}$ is given
by
\begin{equation}
\label{eq:dot_eexpl}
\dot{E}_\mathrm{diag}=
\epsilon_\mathrm{rec}
\dot{M}_\mathrm{out},
\end{equation}
where $\epsilon_\mathrm{rec}$ is the recombination energy.  For
recombination into iron group nuclei, we would have
$\epsilon_\mathrm{rec}\approx 8.8\,\mathrm{MeV}$, but for the high
entropies in neutrino-driven outflows, recombination will mostly go
into $\alpha$-particles and not to iron group nuclei, and some of the
energy gained from recombination is lost due to turbulent energy
exchange between the outflows and downflows \citep{mueller_15b}. In
this work, we therefore use the value of $\epsilon_\mathrm{rec}\approx
5\,\mathrm{MeV}$ found by \citet{mueller_15b} for the asymptotic total
energy of the neutrino-heated ejecta.

It is convenient to rewrite Equation~(\ref{eq:dot_eexpl}) as an
equation for the time derivative $\ud E_\mathrm{expl}/\ud M_\mathrm{sh}$
instead, where $M_\mathrm{sh}$ is the mass coordinate reached by the
shock at a given time. Assuming that a fraction $1-\alpha_\mathrm{out}$
(where $\alpha_\mathrm{out}$ is the surface fraction occupied by
neutrino-driven outflows) of the shocked material is eventually
accreted, the diagnostic energy should grow as
\begin{equation}
\frac{\ud E_\mathrm{diag}}{\ud M_\mathrm{sh}}
\label{eq:de_dm0}
=
\epsilon_\mathrm{rec} \frac{\ud {M}_\mathrm{out}}{\ud {M}_\mathrm{acc}}
\frac{\ud M_\mathrm{acc}}{\ud M_\mathrm{sh}}
=
\frac{(1-\alpha_\mathrm{out}) \epsilon_\mathrm{rec} \eta_\mathrm{acc}}{|e_\mathrm{g}|}.
\end{equation}
While the \emph{eventual} contribution to the explosion energy
from the accretion of a given mass shell can be computed according to
Equation~(\ref{eq:de_dm0}), the diagnostic explosion energy will still
be lower at the time when the mass shell is shocked, and this lower
value is needed to determine (via the post-shock velocity) when
accretion subsides.

To calculate the diagnostic energy $E_\mathrm{imm}$ at the time
when the shock reaches a given mass shell, we assume that the
accretion rate at this point is still given by Equation~(\ref{eq:accretion})
as for non-exploding models.
Since the shock sweeps
up matter at a rate of $\ud M_\mathrm{sh}/\ud t=
4\pi r^2 v_\mathrm{sh} \rho$, we obtain
\begin{eqnarray}
\frac{\ud E_\mathrm{imm}}{\ud M_\mathrm{sh}}
&=&
\frac{\ud E_\mathrm{imm}}{\ud t}
\frac{\ud t}{\ud M_\mathrm{sh}}
=
\frac{1}{4 \pi r^2 v_\mathrm{sh} \rho}\frac{\ud E_\mathrm{diag}}{\ud t}
\\
\label{eq:de_dm1}
&=&
\frac{1}{4 \pi r^2 v_\mathrm{sh} \rho}
\frac{\epsilon_\mathrm{rec}
 \dot{Q}_\nu}{|e_\mathrm{g}|}
=
\frac{\epsilon_\mathrm{rec} \eta_\mathrm{acc} \dot{M}}{4 \pi r^2 v_\mathrm{sh} \rho
|e_\mathrm{g}|},
\end{eqnarray}
in the regime where the shock velocity is considerably larger than the pre-shock
infall velocity.\footnote{Strictly speaking, one would need to compute $\dot{M}$
according to Equation~(\ref{eq:macc}) for the shell $M'$ that falls onto
the proto-neutron star at the
time when the shock hits the mass shell $M$. In practice, this makes little
difference because one typically finds only a slow variation of the accretion rate
outside the Si/O interface (where shock revival typically occurs), so that
we are justified in approximating $M'=M$.
}
 Immediately after shock revival, this is not the case, and we can instead
assume that the shocked matter is immediately accreted onto the proto-neutron star.
To accommodate both regimes, we solve the following equation for $E_\mathrm{imm}$,
\begin{equation}
\label{eq:de_dm_nu}
\frac{\ud E_\mathrm{imm}}{\ud M_\mathrm{sh}}
=
\frac{ \epsilon_\mathrm{rec} \eta_\mathrm{acc}}{|e_\mathrm{g}|}
\min \left(1,\frac{\dot{M}}{4 \pi r^2 v_\mathrm{sh} \rho}\right).
\end{equation}
$E_\mathrm{imm}$ is then used to compute the shock velocity according to
Equation~(\ref{eq:vshock}) and to determine the
amount of explosive burning (see below).

Aside from energy input by neutrino heating, we
also need to take into account that the
shocked material is initially bound and that
nuclear burning in the shock contributes to
the explosion energy provided that the post-shock
temperatures are high enough. It is straightforward
to take this into account by including additional source
terms $\epsilon_\mathrm{bind}$ for the binding energy per unit
mass
of the unshocked material and $\epsilon_\mathrm{burn}$
for nuclear burning,
\begin{equation}
\label{eq:de_dm}
\frac{\ud E_\mathrm{diag}}{\ud M_\mathrm{sh}}
=
\frac{(1-\alpha_\mathrm{out})\epsilon_\mathrm{rec} \eta_\mathrm{acc}}{|e_\mathrm{g}|}
+\alpha_\mathrm{out} \left(\epsilon_\mathrm{bind}
+\epsilon_\mathrm{burn}\right).
\end{equation}
Unlike neutrino heating powered by the accretion of shocked
material,
$\epsilon_\mathrm{bind}$ and
$\epsilon_\mathrm{burn}$ contribute to the
explosion energy without any delay, so that
the  equation for $E_\mathrm{imm}$ becomes:
\begin{eqnarray}
\nonumber
\frac{\ud E_\mathrm{imm}}{\ud M_\mathrm{sh}}
&=&
\frac{\epsilon_\mathrm{rec} \eta_\mathrm{acc}}{|e_\mathrm{g}|}
\min \left(1,\frac{\dot{M}}{4 \pi r^2 v_\mathrm{sh} \rho}\right) \\
&&+\alpha_\mathrm{out} \left(\epsilon_\mathrm{bind}
+\epsilon_\mathrm{burn}\right).
\end{eqnarray}
Note that $\epsilon_\mathrm{bind}$
and $\epsilon_\mathrm{burn}$
are multiplied with the surface fraction occupied
by outflows, $\alpha_\mathrm{out}$, to account for
the fact that some of the shocked material is channelled
into downflows and not swept up by the ejecta.

$\epsilon_\mathrm{burn}$ is given in terms of the initial and final
mass fractions $X_i$ and $X'_i$ prior to and after explosive burning
and the rest-mass contributions $\epsilon_\mathrm{rm}$ per unit mass for
nucleus $i$,
\begin{equation}
\epsilon_\mathrm{burn}
=
\sum_i (X_i-X'_i) \epsilon_{\mathrm{rm},i}.
\end{equation}
To obtain $X_i$, we apply the ``flashing'' method of \citet{rampp_02},
i.e., we assume that the different burning processes (C-, O-,
Si-burning) occur instantaneously at certain ignition temperatures.
To this end, we compute the post-shock temperature $T_\mathrm{sh}$ by
assuming that radiation pressure dominates behind the shock and that
the infall velocity is negligible compared to the shock velocity. With
the post-shock pressure $P_\mathrm{sh}$ determined by the jump
conditions, we then obtain
\begin{equation}
P_\mathrm{sh}= \frac{a T_\mathrm{sh}^4}{3}=
\frac{\beta_\mathrm{expl}-1}{\beta_\mathrm{expl}} \rho v_\mathrm{sh}^2,
\end{equation}
or,
\begin{equation}
T_\mathrm{sh}=
\sqrt[4]{\frac{3 \beta_\mathrm{expl}-1 }{a \beta_\mathrm{expl}} \rho v_\mathrm{sh}^2},
\end{equation}
where $a$ is the radiation constant.

Depending on the post-shock temperature $T_\mathrm{sh}$,
the initial composition is then changed as follows:
\begin{enumerate}
\item For $2.5 \times 10^9\,\mathrm{K} \leq T_\mathrm{sh}
< 3.5 \times 10^9\,\mathrm{K}$, we burn elements
lighter than O to ${}^{16} \mathrm{O}$.
\item For $3.5 \times 10^9\,\mathrm{K} \leq T_\mathrm{sh}
< 5 \times 10^9\,\mathrm{K}$, we burn elements
lighter than Si to ${}^{28} \mathrm{Si}$.
\item For $5 \times 10^9\,\mathrm{K} \leq T_\mathrm{sh} < T_\alpha$,
  we burn everything to ${}^{56} \mathrm{Ni}$.  Note that we follow
  \citet{iliadis} in choosing a different temperature threshold
  for complete Si than \citet{rampp_02}.
\end{enumerate}
Here $T_\alpha$ denotes the density-dependent temperature for which
the mass fraction of $\alpha$-particles reaches $0.5$ in nuclear
statistical equilibrium. $T_\alpha$ is implicitly given by
\citep{shapiro_83,rampp_02},
\begin{equation}
\log_{10} \rho=11.62+1.5 \log_{10} \left(\frac{T_\alpha}{10^9\,\mathrm{K}}\right)
-39.17 \left(\frac{T_\alpha}{10^9\,\mathrm{K}}\right)^{-1}.
\end{equation}

The proto-neutron star also grows due to continued accretion
during phase~I. The fraction of the shocked material
that ends up in the proto-neutron star roughly corresponds
to the surface fraction of the downflows. Moreover,
a fraction $\eta_\mathrm{acc}/|e_\mathrm{g}|$ of the
accreted material is re-ejected by neutrino heating,
so that we obtain the following differential equation
for the (baryonic) neutron star mass $M_\mathrm{by}$
as a function of $M_\mathrm{sh}$,
\begin{equation}
\label{eq:dm_dm}
\frac{dM_\mathrm{by}}{dM_\mathrm{sh}}
=(1-\alpha_\mathrm{out})(1-\eta_\mathrm{acc}/|e_\mathrm{g}|)
\end{equation}.

Using Equations~(\ref{eq:freezeout},\ref{eq:vshock},\ref{eq:de_dm},\ref{eq:dm_dm}),
we can follow the evolution of the explosion energy and determine the
mass $M_\mathrm{by}$ of the proto-neutron star at the end of phase~I.

\subsubsection{Evolution of Explosion and Remnant
Parameters-- Phase II}
During phase~II, the explosion energy can still change
due to explosive burning in the shock, the accumulation
of bound material by the shock, and the energy input
from the neutrino-driven wind (which also reduces
the proto-neutron star mass).

In recent self-consistent simulations of the wind phase in electron
capture supernova explosion \citep{janka_08,groote_phd}, the wind
contributes only $\mathord{\sim} 10^{48}\,\mathrm{erg}$ to the explosion energy
 and the integrated mass loss is $\Delta
M_\mathrm{wind}\lesssim 10^{-4}\,M_\odot$. Even for more massive
progenitors that leave behind more massive neutron stars with hotter
neutrinospheres, the integrated mass loss in the wind remains well below
$10^{-3}\,M_\odot$ \citep{huedepohl_phd}, implying a contribution to the
  explosion energy of $\ll 10^{50}\,\mathrm{erg}$.

We therefore feel justified in neglecting the effect
of the neutrino-driven wind on the final explosion and
remnant properties in this work, and
consider only the two remaining contributions. Aside
from the fact that \emph{all} of the matter swept
up by the shock now contributes to the energy
budget of the ejecta (and not just a fraction
$\alpha_\mathrm{out}$), these can be treated
exactly as in phase~I, and the equation for
the explosion energy becomes,
\begin{equation}
\label{eq:de_dm_phase2}
\frac{\ud E_\mathrm{diag}}{\ud M_\mathrm{sh}}
=
\epsilon_\mathrm{bind}
+\epsilon_\mathrm{burn}.
\end{equation}
The baryonic remnant mass $M$ is left unchanged
during this phase.

\subsubsection{Final Explosion Properties and Neutron Star Mass}
Integrating Equation~(\ref{eq:de_dm_phase2}) out to the
stellar surface yields the final explosion energy $E_\mathrm{expl}$.
If $E_\mathrm{expl}$ is positive, we compute the final gravitational
mass $M_\mathrm{NS}$ of the neutron star using the approximate formula \citep{lattimer_89,lattimer_01}
\begin{equation}
  \label{eq:mgrav}
M_\mathrm{NS}=M_\mathrm{by}- 0.084\,M_\odot (M_\mathrm{NS}/M_\odot)^2.
\end{equation}
If $E_\mathrm{diag}$ becomes negative at any $M_\mathrm{sh}$, if
the remnant mass $M_\mathrm{NS}$ exceeds the maximum neutron star mass $M_\mathrm{max}$,
or (as discussed earlier) if the condition $\tau_\mathrm{adv}/\tau_\mathrm{heat}=1$
 was never met, we assume that the entire star collapses to a black hole and
 set $E_\mathrm{expl}=0$.
 In that case,
   the gravitational remnant mass $M_\mathrm{BH}$ is
   set to the pre-collapse mass of the star. This is only a very crude estimate,
   and in the presentation of our results, we include $M_\mathrm{BH}$ primarily
   to indicate non-exploding models without attaching too much significance
   to the actual values. Even without shock revival, the actual black hole
   mass could be lower because the reduction of the gravitational
   mass of the interior shells by neutrino losses could lead
   to the (partial) ejection of the hydrogen envelope
   \citep{nadyoshin_80,lovegrove_13}, so that the helium core
   mass may be the more appropriate estimator for the
   black hole mass \citep{sukhbold_14}.
 Moreover, the possibility of fallback is  considered only
as an all-or-nothing event -- it will involve the entire star if the
diagnostic energy becomes negative, and no fallback is assumed to happen
for successful explosions. The reality is thus obviously more complicated
  than our model, and
the systematics of fallback will need to be studied in greater detail
in a future continuation of this work.

During phase~I and phase~II, we also integrate the mass of iron
group elements $M_\mathrm{IG}$ produced by explosive nuclear burning
(taking into account that only a fraction $\alpha_\mathrm{out}$ out
these will be ejected during phase~I).
$M_\mathrm{IG}$ can be taken as a rough proxy for the nickel mass,
but needs to be interpreted with caution:  ${}^{56} \mathrm{Ni}$
is not the only iron group element produced by explosive burning at sufficiently
high temperatures, and the very crude ``flashing'' treatment
based on an estimate of the post-shock temperature cannot be
expected to yield quantitatively reliable results. For these
reasons, $M_\mathrm{IG}$ can \emph{at best} be expected to agree
with the actual nickel mass within a factor
of $\mathord{\sim} 2$.

\begin{table*}
  \begin{center}
  %\begin{minipage}{\textwidth}
  %\centering
  \caption{Adjustable model parameters.
    \label{tab:parameters}}
  \begin{tabular}{cccc}
    \hline\hline
    parameter & explanation & standard value & typical range  \\
    \hline
    $\alpha_\mathrm{out}$ & volume fraction of outflows & $0.5$ & $0.3 \ldots 0.7$ \\
    $\alpha_\mathrm{turb}$ & shock expansion due to turbulent stresses & $1.18$ & $1\ldots 1.4$ \\
    $\beta_\mathrm{expl}$ & shock compression ratio during explosion phase & $4$  &$3 \ldots 7$ \\
    $\zeta$ & efficiency factor for conversion of accretion energy into $\nu$ luminosity & $0.8$ &$0.5 \ldots 1 $ \\
    $\tau_{1.5}$ & cooling time-scale for $1.5\,M_\odot$ neutron star
    & $1.2\,\mathrm{s}$ & $0.6\,\mathrm{s} \ldots 3\,\mathrm{s}$ \\
\hline
  \end{tabular}
\medskip
%  \end{minipage}
\end{center}
\end{table*}

\begin{figure}
  \includegraphics[width=\linewidth]{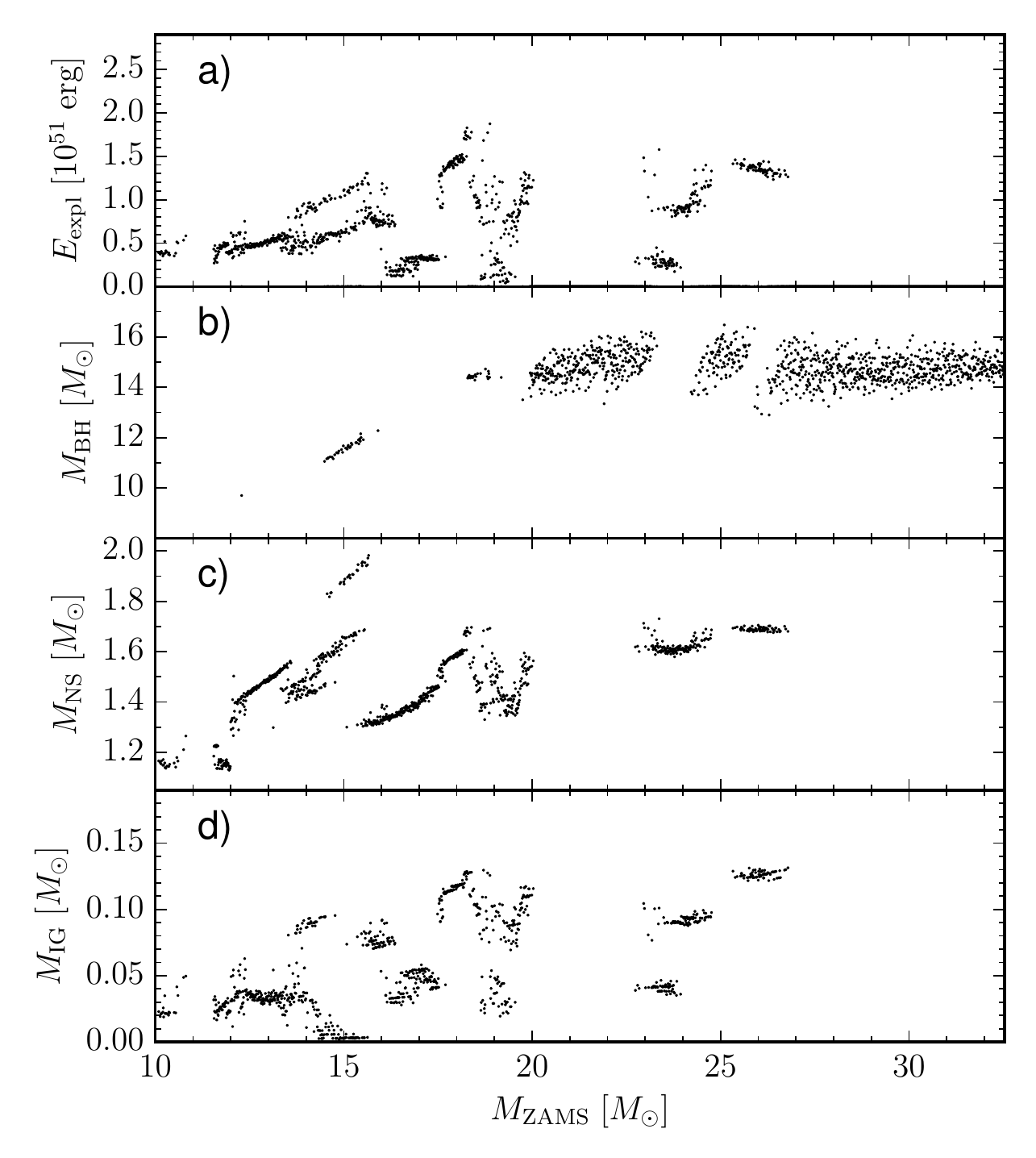}
\caption{Explosion energy ($E_\mathrm{expl}$, Panel a),
  gravitational remnant masses for black holes ($M_\mathrm{BH}$,
    Panel b), neutron stars ($M_\mathrm{NS}$, Panel c), and the
  iron-group mass ($M_\mathrm{IG}$, Panel d) as a function of ZAMS
  mass for the standard case.  Note that there is a gap in our set of
  progenitors around $11\,M_\odot$; missing data points in this region
  are \emph{not} indicative of black hole formation.
\label{fig:standard}}
\end{figure}

\begin{figure}
  \includegraphics[width=\linewidth]{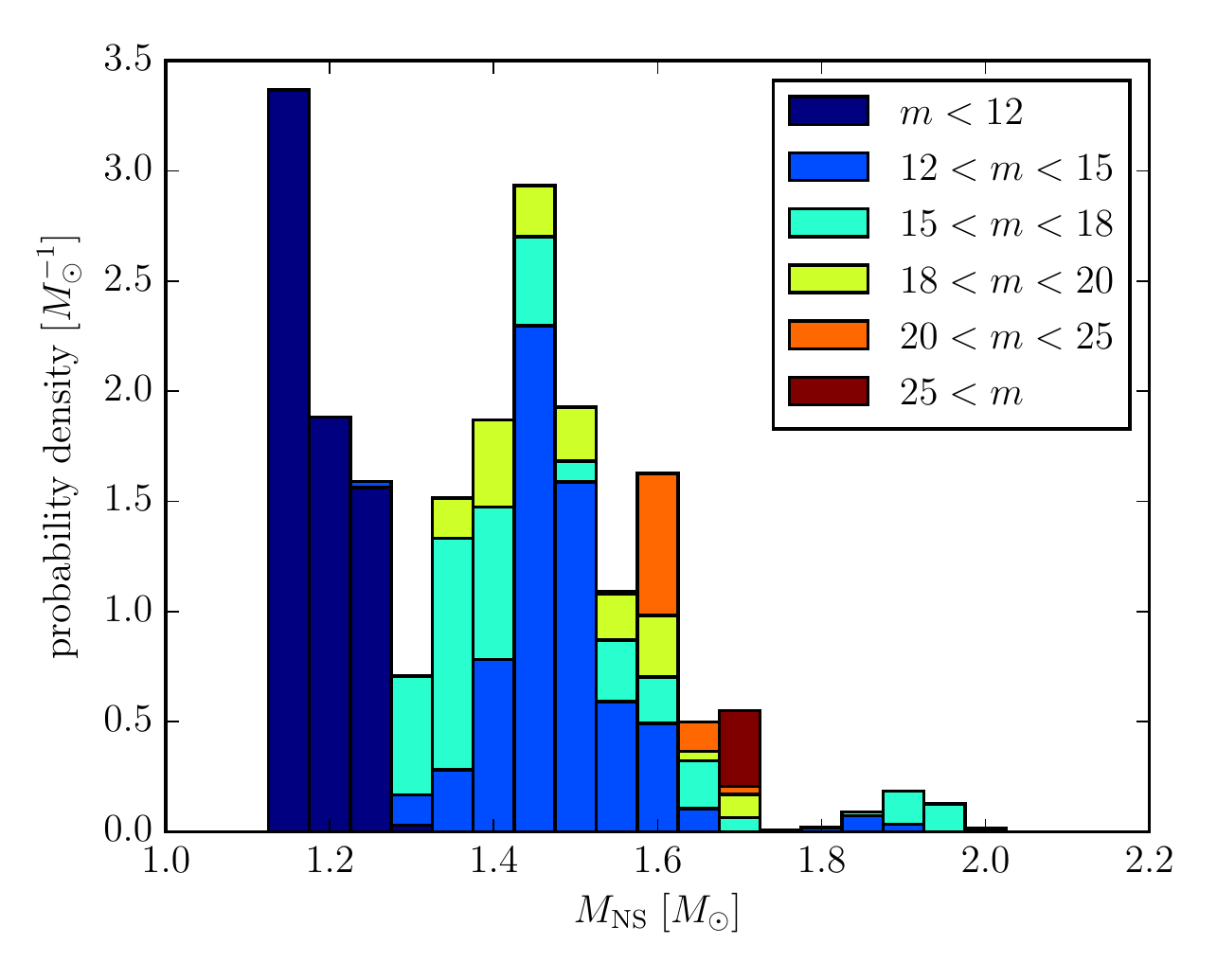}
  \caption{Histogram of the distribution of gravitational neutron star
    masses for the standard case. The stacked bars in different
    colours give the contribution of progenitors from different ranges
    of the ZAMS mass $m$ (measured in solar masses) to the average
    probability density in a given bin.
    \label{fig:hist}}
\end{figure}

\begin{figure}
  \includegraphics[width=\linewidth]{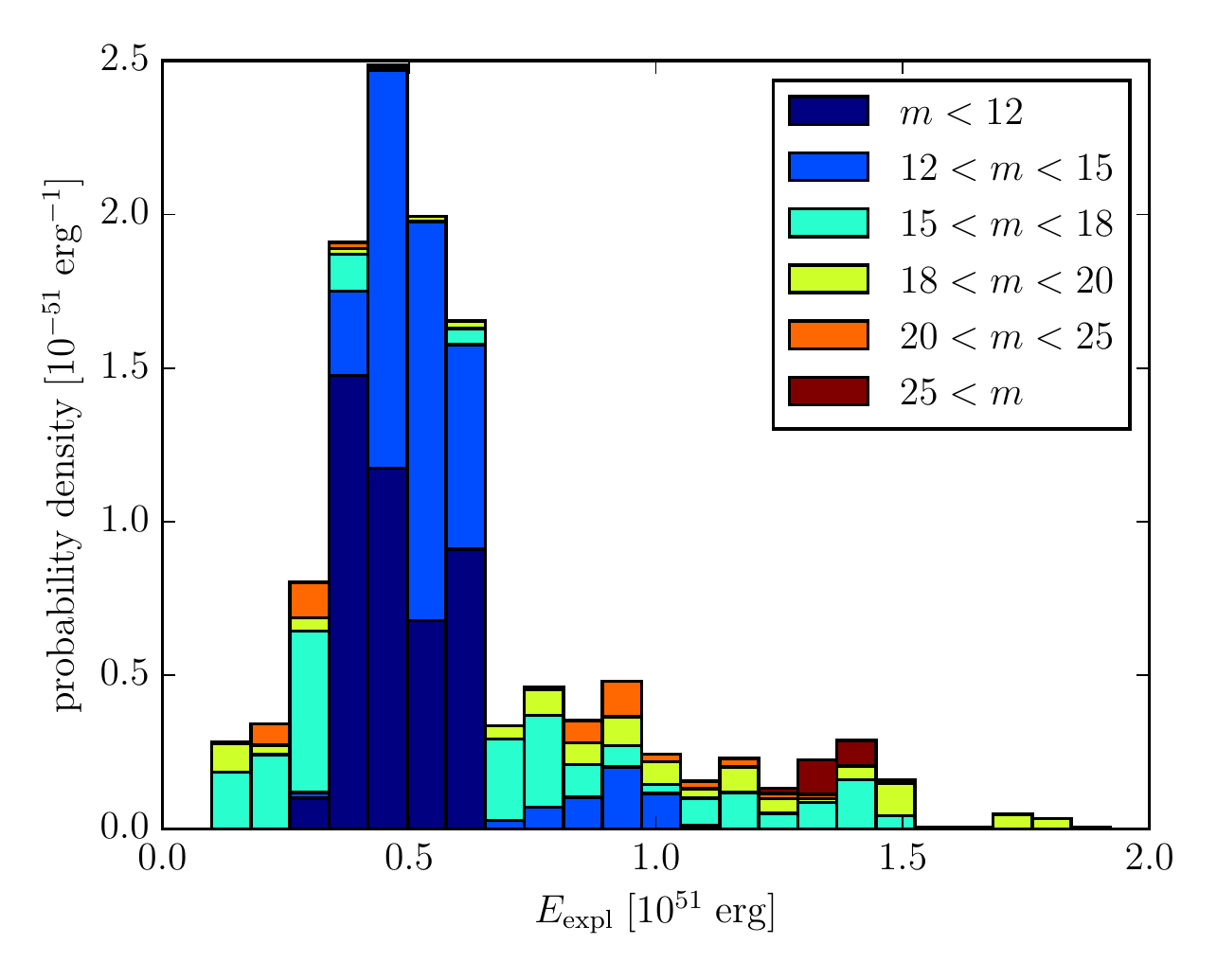}
  \caption{Histogram of the distribution of explosion energies for the
    standard case. The stacked bars in different colours give the
    contribution of progenitors from different ranges of the ZAMS mass
    $m$ (measured in solar masses) to the average probability
    density in a given bin.
    \label{fig:hist_energy}}
\end{figure}

\begin{figure}
  \includegraphics[width=\linewidth]{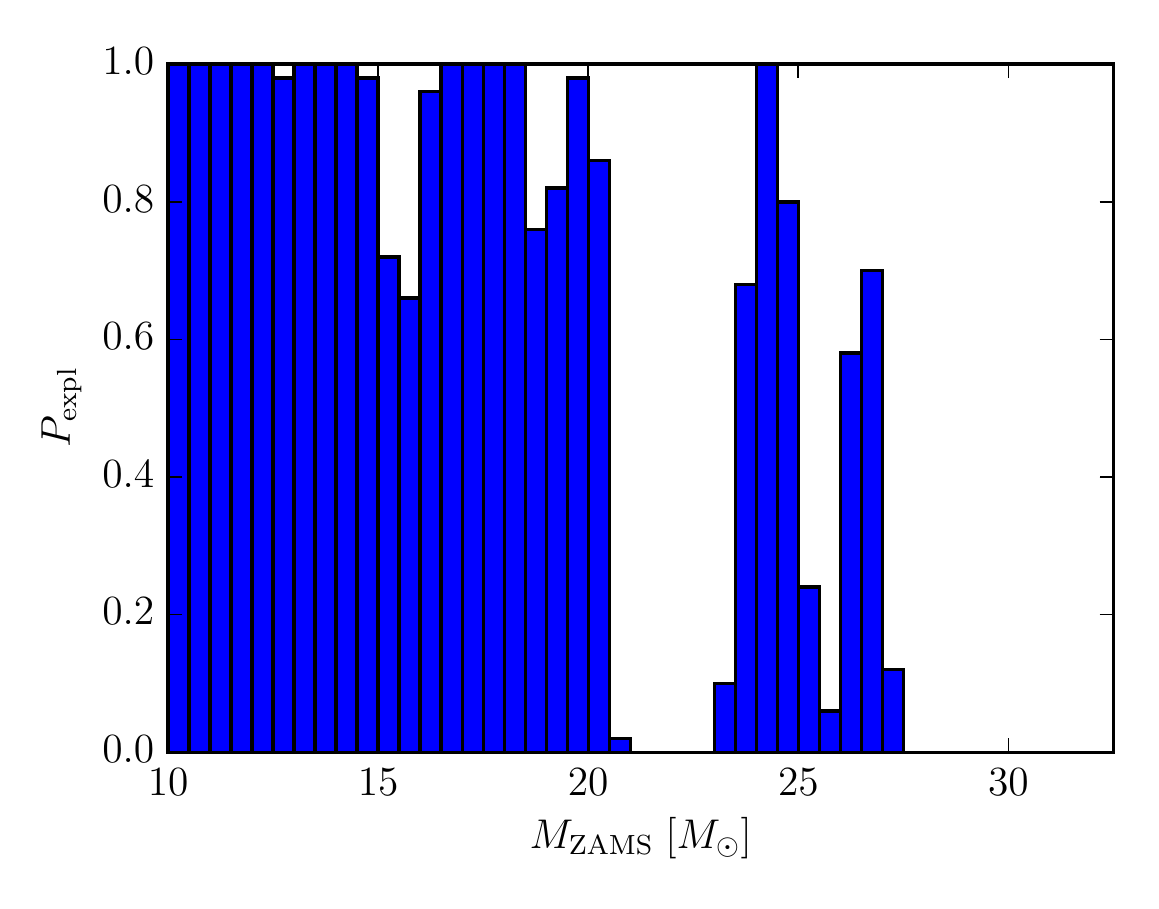}
\caption{Fraction $P_\mathrm{expl}$ of successful explosions within
  bins of $0.5\,M_\odot$ for the standard case.
  There are no models in the bin between $11\,M_\odot$ and $11.5\,M_\odot$,
  but we assume that $P_\mathrm{expl}$ can be interpolated in this region,
i.e., that there is no black hole formation.
\label{fig:expl_frac}}
\end{figure}

\section{Results}
\label{sec:results}
We apply our model to a set of 2120 solar-metallicity progenitor
models computed with an up-to-date version of the stellar evolution
code \textsc{Kepler} \citep{weaver_78,heger_10}. The models cover a
range from $10\,M_\odot$ and $32\,M_\odot$ in ZAMS mass with a
typical spacing of $0.01\,M_\odot$
except for the mass range between
$11\,M_\odot$ and $11.5\,M_\odot$, where not all of the models
could be run up to collapse because
of time constraints (see \citealt{woosley_07,woosley_15b} for a more detailed
study of the lowest-mass supernova progenitors at solar metallicity).
The input physics is very similar to
the models of \citet{sukhbold_14} and \citet{sukhbold_15}, except
for updates in the neutrino loss rates \citep{itoh_96} and the initial solar
composition \citep{asplund_09}. The
overall effect of these updates is a downward shift of the
transition of structural features in the pre-SN evolution (like the transition
between convective and radiative central carbon burning)
by $\mathord{\approx} 1.5\,M_\odot$ in ZAMS mass.

The analytic/ODE model has been implemented in \textsc{Python~3}.
Once the \textsc{Kepler} model files are loaded, all progenitors can
be processed within $35\,\mathrm{s}$ on a modern laptop computer.  As
our standard set of parameters, we adopt values of $\beta_\mathrm{expl}=4$,
$\zeta=0.7$, $\alpha_\mathrm{out}=0.5$, $\alpha_\mathrm{turb}=1.18$,
and $\tau_{1.5}=1.2\,\mathrm{s}$ for the five
adjustable parameters of the model.
Different from some coefficients and parameters that
have implicitly been fixed in the preceding section,
these parameters are beset with larger uncertainties, and we therefore explore variations itof each of
these within a reasonable and justifiable range. Limits for
the different parameters are listed in Table~\ref{tab:parameters}
along with our preferred values. These limits represent extremes that
could be justified under certain physical assumptions (e.g., a strong
reduction of neutrino opacities); if we require agreement with
observational constraints the limits are in fact much tighter. We
shall first discuss salient features of the explosion and remnant
properties for our standard case before exploring the sensitivity to
such parameter variations in Section~\ref{sec:variations}.

\subsection{Landscape of Neutron Star and Black Hole Formation -- Standard Case}
\label{sec:standard}

Figure~\ref{fig:standard} shows the explosion energy
$E_\mathrm{expl}$, the gravitational mass of the remnant
($M_\mathrm{NS}$ for neutron stars and $M_\mathrm{BH}$ for black
  holes), and the estimated mass $M_\mathrm{IG}$ of iron group
elements in the ejecta as a function of ZAMS mass, and the
distribution of the explosion and remnant properties is further
illustrated by IMF-weighted histograms in Figure~\ref{fig:hist} for
$M_\mathrm{NS}$ and Figure~\ref{fig:hist_energy} for
$E_\mathrm{expl}$.  We find a range of explosion energies from a few
$10^{49}\,\mathrm{erg}$ to above $2 \times 10^{51}\,\mathrm{erg}$,
neutron star masses between $1.15\,M_\odot$ and $2\,M_\odot$, and iron
group masses up to $0.15\,M_\odot$ similar to the parameterised 1D
studies of \citet{ugliano_12,ertl_15} and \citet{sukhbold_15}.
Different from these works, we do not include blue supergiant
progenitors for the well-studied case of SN~1987A as a benchmark
case. Given the uncertainty in the provenance of SN~1987A, whose
progenitor may have originated from a merger event
\citep{podsiadlowski_89,podsiadlowski_90}, and the range of stellar
evolution models available for SN~1987A (see, e.g.,
\citealp{sukhbold_15}), the only firm constraints that can be derived
from this event is that \emph{some} progenitor in the mass range
between $15\,M_\odot$ and $20\,M_\odot$ with a helium core mass of
$\sim 6\,M_\odot$ should explode with an energy of $(1 \ldots
1.5)\times 10^{51}\,\mathrm{erg}$ and produce $\sim 0.07\,M_\odot$ of
nickel
\citep{shigeyama_90,utrobin_93,blinnikov_00,utrobin_05,tanaka_09}.
Given the large diversity of progenitor models in our samples, it is
not surprising that a very rough fit to SN~1987A can be found even
though we did not specifically construct one to match its surface
properties and its metallicity; for example the $19.7\,M_\odot$
progenitor explodes with $1.24 \times 10^{51}\,\mathrm{erg}$ and
produces $0.11 M_\odot$ of iron group elements (see
also~Appendix~\ref{sec:appendix} for plots of the explosion properties
as a function of helium core mass).

The similarities to recent numerical and analytic studies
\citep{ugliano_12,pejcha_15a,ertl_15,sukhbold_15} also extend to the
prediction of a variegated landscape of regions of black hole
formation interspersed with ``islands of explodability'' at masses
above $\gtrsim 15\,M_\odot$. In Figure~\ref{fig:expl_frac}, we further
illustrate this landscape by showing the fraction $P_\mathrm{expl}$ of
exploding progenitors within bins of $0.5\,M_\odot$.  Although some of
the ``islands of explodability'' have cores with $P_\mathrm{expl}=1$,
Figure~\ref{fig:expl_frac} shows that they are smeared out
considerably with a gradual transition between them, which supports
the case for a probabilistic description of black hole and neutron
star formation \citep{clausen_15}.

We note that the islands of explodability are slightly shifted
compared to previous works, and the black hole formation probability
around $15\,M_\odot$ is relatively small. Such changes are not
unexpected for a different set of progenitors, and are not indicative
of a fundamental disagreement between our model and other approaches.

Given the uncertainties in the determination of progenitor masses
using HR tracks, our standard case is also appears broadly consistent
with observational evidence for missing explosions above ZAMS masses
of $\mathord{\approx}18\,M_\odot$ \citep{smartt_15} despite a drop of
the explosion probability at a slightly higher mass of $\approx
20\,M_\odot$ in our model, whose robustness will be further discussed
in Section~\ref{sec:variations}.

\begin{figure}
  \includegraphics[width=\linewidth]{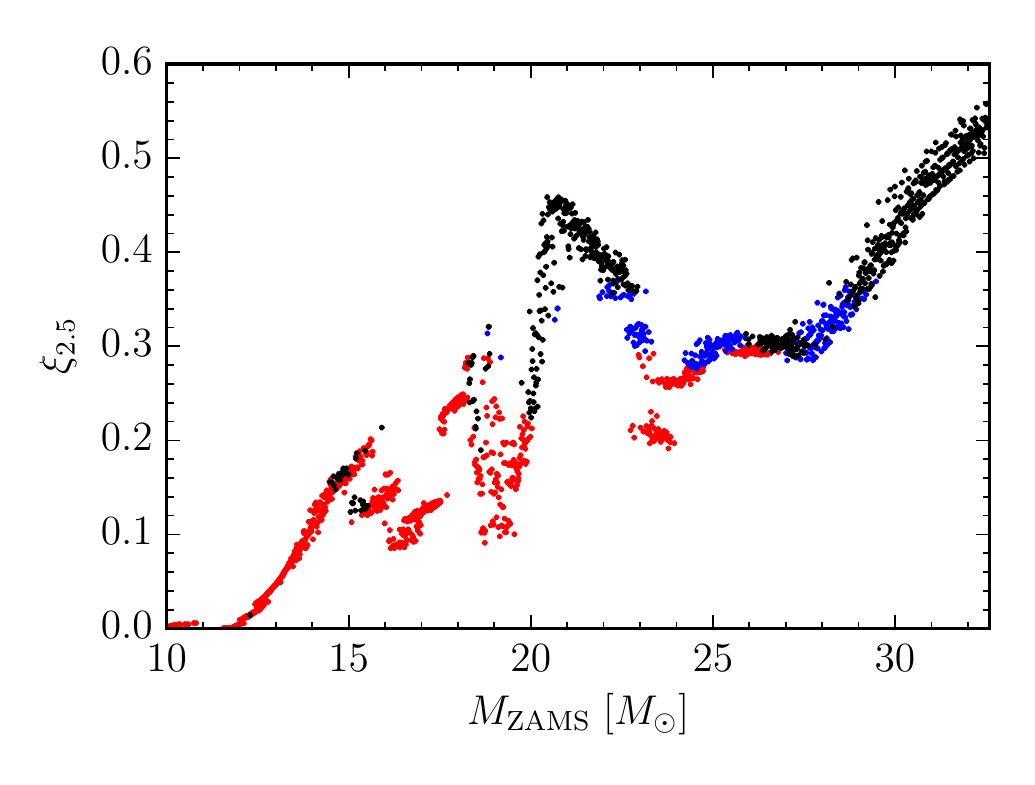}
  \caption{Compactness parameters $\xi_{2.5}$ for exploding (red) and
    non-exploding (black) models as a function of ZAMS mass.  Blue
    dots denote models where shock revival is initiated, but the
    explosion eventually fails because the diagnostic energy becomes
    negative as the shock propagates out or the neutron star mass
    exceeds the maximum neutron star mass due to ongoing accretion in
    the explosion phase.
    \label{fig:xi}}
\end{figure}

\begin{figure}
  \includegraphics[width=\linewidth]{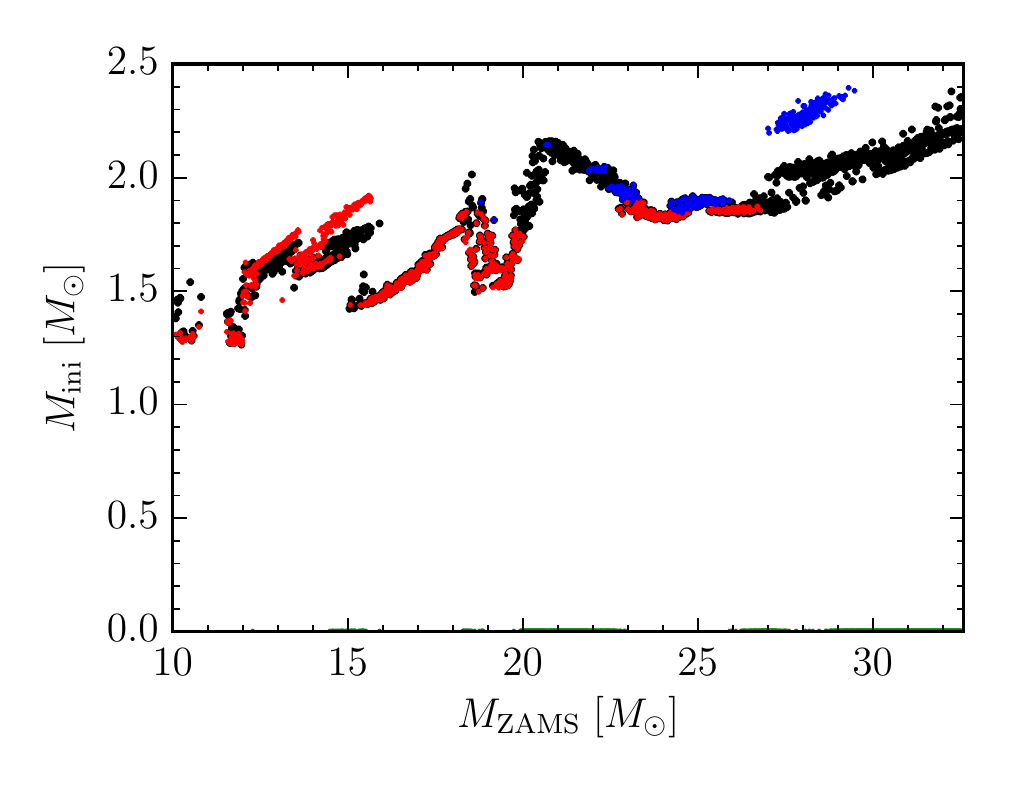}
\caption{Mass coordinate $M_\mathrm{ini}$ of the initial mass cut
as a function of ZAMS mass (red/blue dots). Red
  dots denote models that explode successfully, while blue dots are
  used for models where we predict black hole formation due to
  continued accretion after shock revival. The mass of the silicon
  core is shown in black.
\label{fig:m_init}}
\end{figure}

\begin{figure}
  \includegraphics[width=\linewidth]{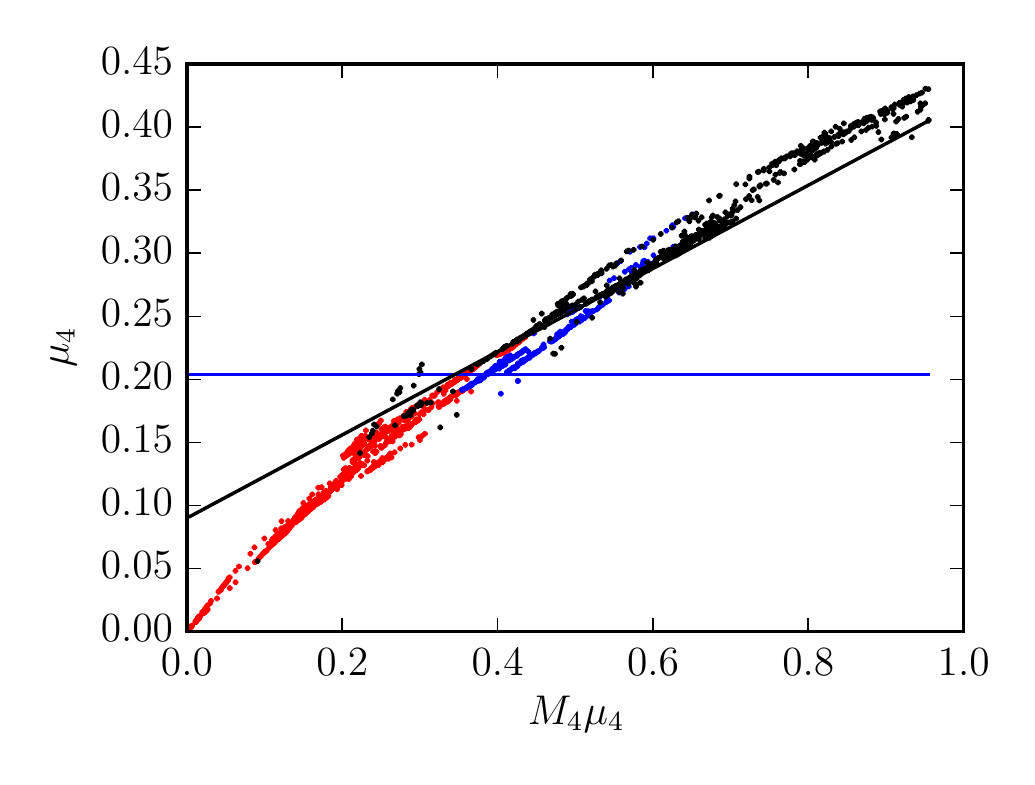}
  \caption{Distribution of the progenitor models
in the $M_4 \mu_4$-$\mu_4$ plane introduced
by \citet{ertl_15}. Cases of neutron star
formation, black hole formation without shock revival
and black hole formation after shock revival
are shown as red, black, and blue dots, respectively.
Exploding and non-exploding can be reasonably
well separated by $\mu_4=0.203$
(black line). A better discrimination between
explosions and failures using an Ertl-type
line $\mu_4=k_1 \mu_4 M_4+k_2$ with positive
slope is precluded
by the location of models that form black holes
after shock revival due to continued accretion.
Models with and without shock revival are nicely
distinguished by a modified Ertl criterion
$\mu_4 =0.33 \mu_4 M_4+0.09$ with only 133 false
identifications.
    \label{fig:ertl}}
\end{figure}

\subsection{Comparison to Proposed Explosion Criteria}

The qualitative similarity of the regions of neutron star and black
hole formation with approaches that rely on 1D hydrodynamics simulations
\citep{oconnor_11,ugliano_12,ertl_15,sukhbold_15} is reassuring as these models
arguably treat the phase up to shock revival more accurately than our
analytic model in Section~\ref{sec:preexplosion}. The fundamental
agreement about the conditions for shock revival (as opposed
to the explosion and remnant properties in case of successful
explosion that we discuss in Section~\ref{sec:expl_prop}) is borne
out by an analysis of several phenomenological explosion criteria
that have been proposed on the basis of 1D models.

\subsubsection{Compactness Parameter}
\citet{oconnor_11} introduced
the compactness parameter $\xi_{M}$, which is defined
for a given mass coordinate $M$ as
\begin{equation}
\xi_M
=
\frac{M/M_\odot}{r(M)/1000\,\mathrm{km}},
\end{equation}
where $r(M)$ is the radial coordinate of this mass shell at the time
of core bounce.  They suggested $\xi_{2.5} \lesssim 0.45$ as a rough
condition for successful explosions Subsequently, the parameterised 1D
study of \citet{ugliano_12} revealed a broad transition region between
neutron star and black hole formation in the range $\xi_{2.5}\approx
0.15 \ldots 0.35$.  The distribution of the compactness parameters for
our exploding and non-exploding models is shown in
Figure~\ref{fig:xi}. In line with the weaker tendency for black hole
formation around $15\,M_\odot$, the transition region between neutron
star and black hole formation is located at somewhat higher values
than in \citet{ugliano_12}, i.e., $\xi_{2.5}=0.2 \ldots 0.4$ with some
outliers of black hole formation at even lower $\xi_{2.5}$. A choice
of $\xi_{2.5,\mathrm{crit}}=0.278$ for the critical value best
discriminates between explosions and non-explosions (with 158 false
identification).

\subsubsection{Ertl Criterion}
While $\xi_{2.5}$ has been justified empirically as a measure of
``explodability'', it evidently provides no sharp dividing line
between explosion and failure, and aside from a vague connection with
the maximum neutron star mass it lacks an intuitive theoretical
basis. \citet{ertl_15} therefore proposed a different criterion with
higher discriminating power which is based on the structure of the
progenitor near the outer edge of the Si core, whose infall typically
results in a considerable improvement in heating conditions and is
often closely associated with the transition to explosion in 1D
\citep{ugliano_12,ertl_15} and self-consistent multi-D simulations
\citep{buras_06b,marek_09,mueller_12a,suwa_16}. They considered the
two parameters $M_4$, the mass coordinate corresponding to an entropy
of $s=4 \ k_\mathrm{b}/ \mathrm{nucleon}$ (which typically defines the
interface between the Si core and the O shell), and $\mu_4$,
\begin{equation}
\mu_4=\frac{0.3}{[r(M_4+0.3\,M_\odot)-r(s=4)]/1000\,\mathrm{km}}
\propto \left.  \frac{\ud M}{\ud r} \right |_{s=4},
\end{equation}
which can be related to the accretion rate $\dot{M} \propto
\mu_4$ and the
accretion luminosity $L_\mathrm{acc} \propto \mu_4 M_4$  shortly after the infall of the Si/O interface. \citet{ertl_15} further argue that
a calibrated linear inequality
\begin{equation}
\mu_4 < k_1 \mu_4 M_4+k_2,
\end{equation}
can then be used to decide whether the critical neutrino luminosity
for explosion \citep{burrows_93}
is reached or exceeded around the infall of the shell interface
so that it can be used as a predictor for shock revival (provided
that the heating conditions do not improve significantly later on).
Since $\mu_4$ and $\mu_4 M_4$ are loosely correlated with the
accretion rate and the accretion luminosity around the infall
of the Si/O interface, one expects the coefficient $k_1$ to be
positive to reflect the monotonic increase of the critical
luminosity with $\dot{M}$.

While it has more of a physical justification than the
compactness parameter, the Ertl parameter rests on two important
assumptions: It presupposes that successful shock revival generally
also leads to a successful explosions, which is by no means to be
taken for granted considering that some long-time multi-D supernova
models show continued accretion over seconds \citep{mueller_15b},
which implies that many progenitor could undergo delayed black hole
formation even after successful shock revival. Furthermore, in some
multi-D simulations \citep{marek_09,mueller_12a,melson_15b,lentz_15},
shock revival is delayed considerably beyond the infall of the Si/O
interface and is instead triggered by a continuing improvement of the
heating conditions due to the increase of the mean energy with neutron
star mass \citep[cp.][]{mueller_15a}.

Our model allows for both of these scenarios, and they are in fact
realised in the standard case as demonstrated by
Figure~\ref{fig:m_init}, which compares the mass coordinate
$M_\mathrm{ini}$ for which we predict shock revival with the mass of
the iron and Si core.  Although shock revival generally occurs at
or shortly outside the Si/O interface, there are progenitors with
considerable delays between $27\,M_\odot$ and
$30\,M_\odot$. Most of these, as well
as some cases at slightly smaller ZAMS masses undergo delayed
black hole formation after shock revival.

This turns out to be somewhat problematic for formulating an optimal
Ertl criterion because the cases of late black hole formation tend to
lie at higher $\mu_4 M_4$ for a given $M_4$, i.e., one would expect
the ratio of accretion luminosity to critical luminosity to be higher
for these after the infall of the Si/O interface.  We illustrate this
in Figure~\ref{fig:ertl}, which shows the distribution of our
progenitors in the $\mu_4 M_4$-$\mu_4$ plane.  For an optimal
discrimination between exploding and non-exploding cases, we are
forced to resort to an extreme choice $k_1=0$ for the
slope, so that the Ertl criterion again becomes a one-parameter
criterion,
\begin{equation}
\mu_4 \lesssim 0.204.
\end{equation}
This still furnishes a relatively good dividing line between
exploding and non-exploding progenitors, but counterintuitively with a
few more false predictions (188) than the
compactness parameter. As a predictor for shock revival
alone, the Ertl criterion fares considerably better, with just 133
($6.3 \%$) false predictions with the modified criterion
\begin{equation}
  \mu_4 =0.33 \mu_4 M_4+0.09.
\end{equation}

Our results can of course not be taken as a test or a comparison of
these criteria, since they are based on a very simplified model
themselves. The numbers of false identifications mostly provide
a consistency check between different approaches, and at best
help to bolster these phenomenological criteria under different
physical assumptions for the energetics and dynamics of the
explosion phase: Despite the complications introduced by accretion after
shock revival, both the compactness parameter and the Ertl criterion
can still be relied upon for rule-of-thumb estimates for
explodability. False positives and false negative never lie far
away from the dividing line, and it is doubtful whether a reliable
calibration of these criteria using multi-D or even only 1D
simulations is possible at the present state of supernova theory. If
different criteria and models agree for $90\%$ of all
progenitor models, this rather points to a high level of
compatibility.

\begin{figure}
  \includegraphics[width=\linewidth]{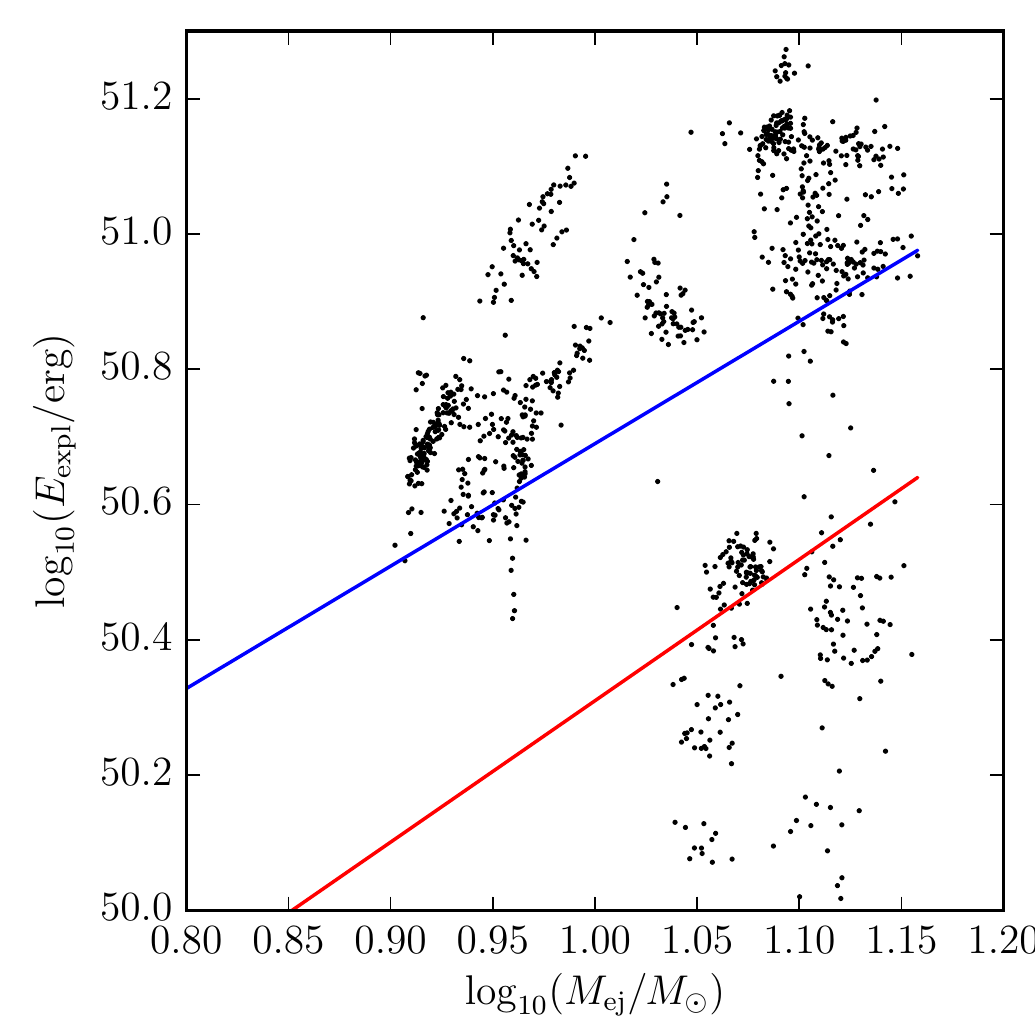}
  \caption{Plot of explosion energy $E_\mathrm{expl}$ versus ejecta
    mass $M_\mathrm{ej}$ for successfully exploding models. Fitted
    power laws for observed supernovae from \citet{pejcha_15b} using
    two different calibrations of their light curve model based on
    \citet{litvinova_85} and \citet{popov_93} are shown in red and
    blue, respectively.  While the bulk of our progenitor models
    conform to the observed correlation between $E_\mathrm{expl}$
    versus ejecta mass $M_\mathrm{ej}$, there is also a sub-population
    of underenergetic supernovae from high-mass progenitors. This
    sub-population would likely exhibit considerable fallback, which
    could bring it back in line with the general trend.
    \label{fig:e_mej}}
\end{figure}

\begin{figure}
  \includegraphics[width=\linewidth]{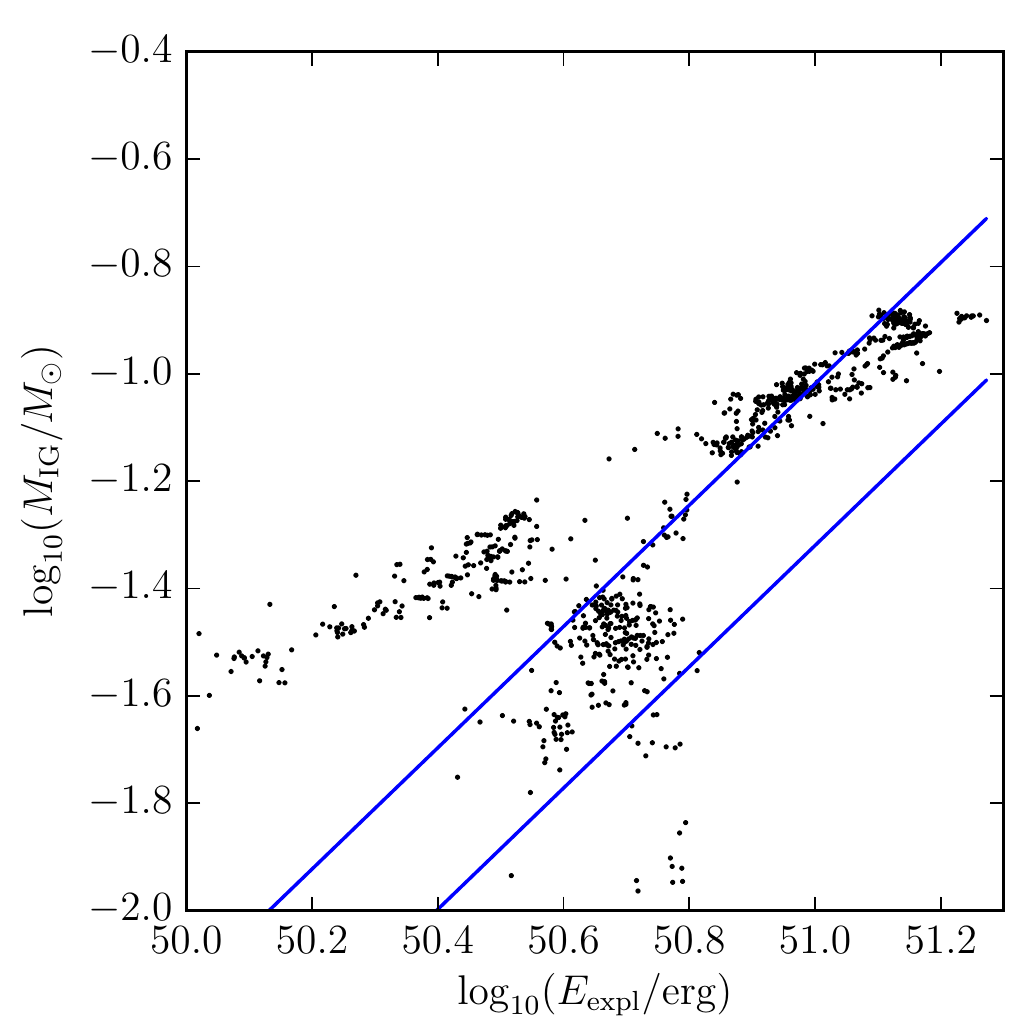}
  \caption{Explosion energy $E_\mathrm{expl}$ versus iron group mass $M_\mathrm{IG}$.
    A fitted power
    law for the dependence of $E_\mathrm{expl}$
    on the nickel mass $M_\mathrm{Ni}$ for observed supernovae from
    \citet{pejcha_15b} using their light curve model
calibrated against
\citet{popov_93} is shown in blue (lower line). The
upper blue line corresponds to twice the value
of the fit of \citet{pejcha_15b} and roughly
defines a band where $M_\mathrm{IG}$ is expected
to lie considering that nickel will only make up
part of the iron group elements produced by explosive burning.
    \label{fig:e_mig}}
\end{figure}

\begin{figure}
  \includegraphics[width=\linewidth]{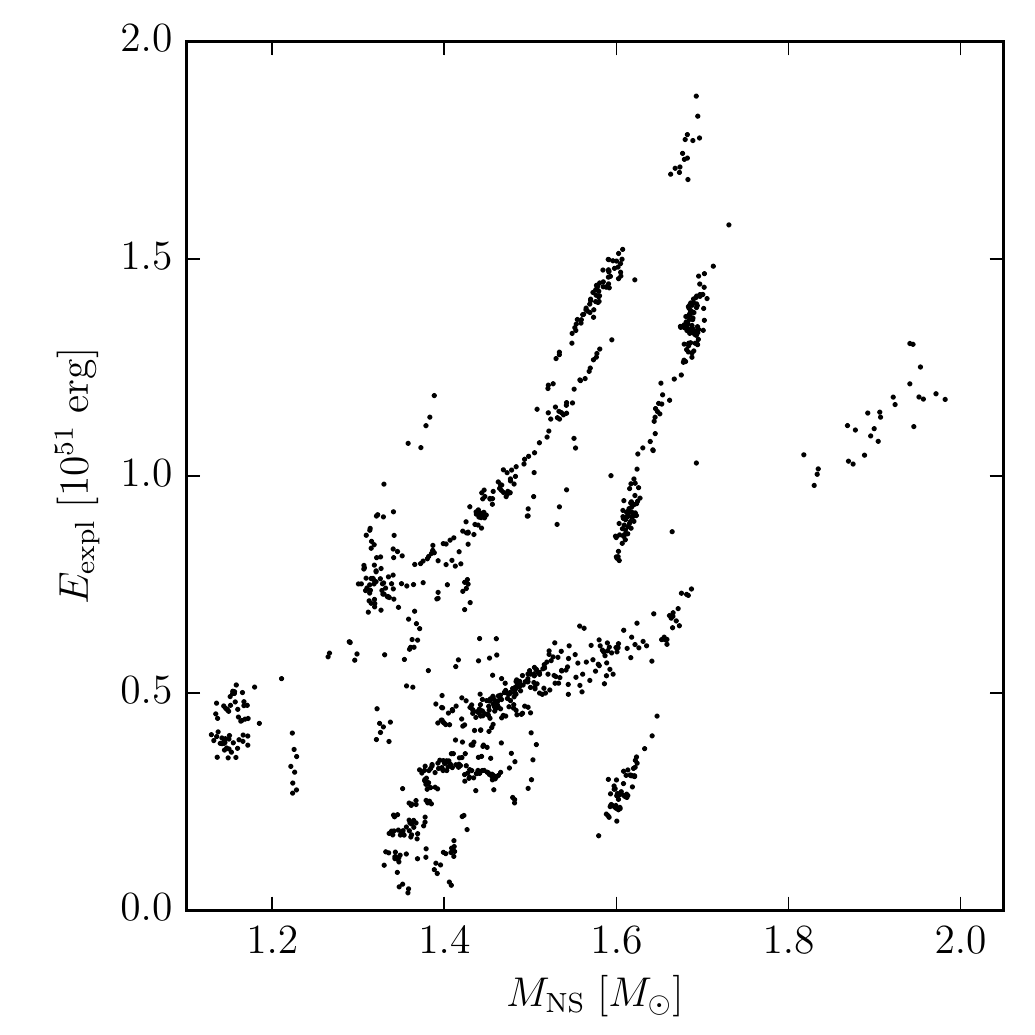}
  \caption{Explosion energy versus gravitational neutron star mass for
    successful explosions. Typically, more energetic
    explosions also tend to produce more massive neutron
    stars, but there is considerable scatter. Moreover,
    our model yields, perhaps spuriously, a clump
    of high-mass neutron stars from supernovae with moderate
explosion energy.
    \label{fig:e_mgrav}}
\end{figure}

\begin{figure}
  \includegraphics[width=\linewidth]{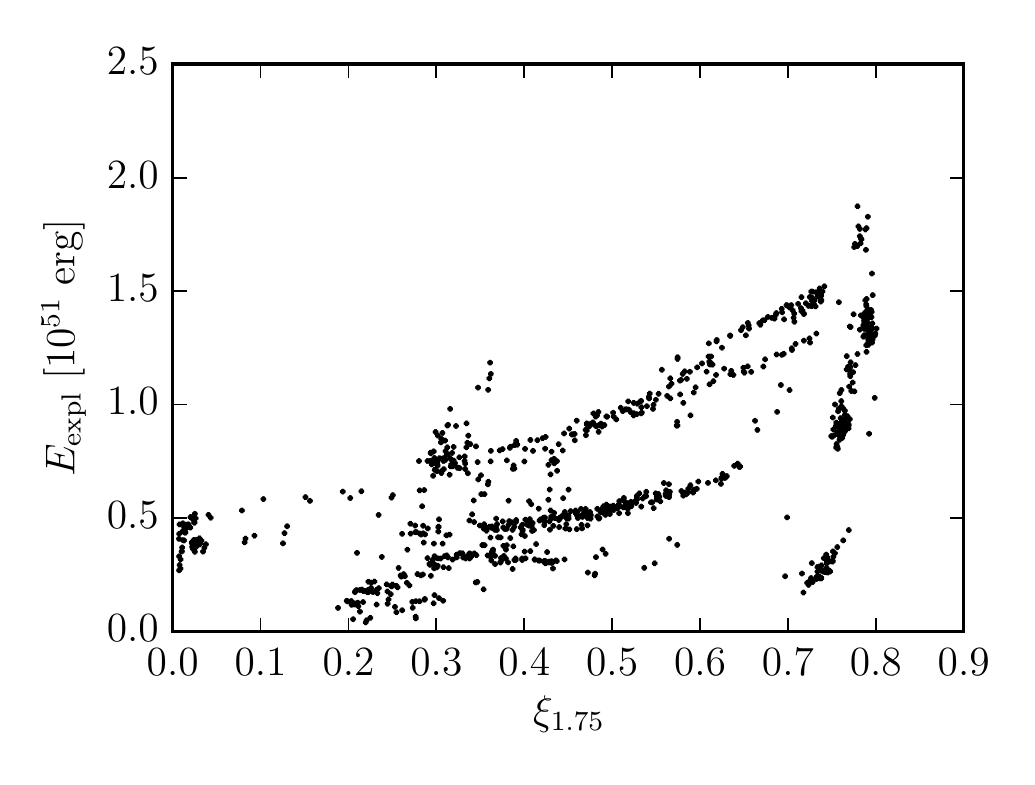}
  \caption{Explosion energy versus $\xi_{1.75}$. Highly
    energetic explosions only come from progenitors
    with large $\xi_{1.75}$, but aside form this there
    is no tight correlation between $\xi_{1.75}$
    and $E_\mathrm{expl}$.
    \label{fig:e_vs_xi}}
\end{figure}

\subsection{Explosion and Remnant Properties -- Standard Case}
\label{sec:expl_prop}

\begin{figure*}
  \includegraphics[width=0.48\linewidth]{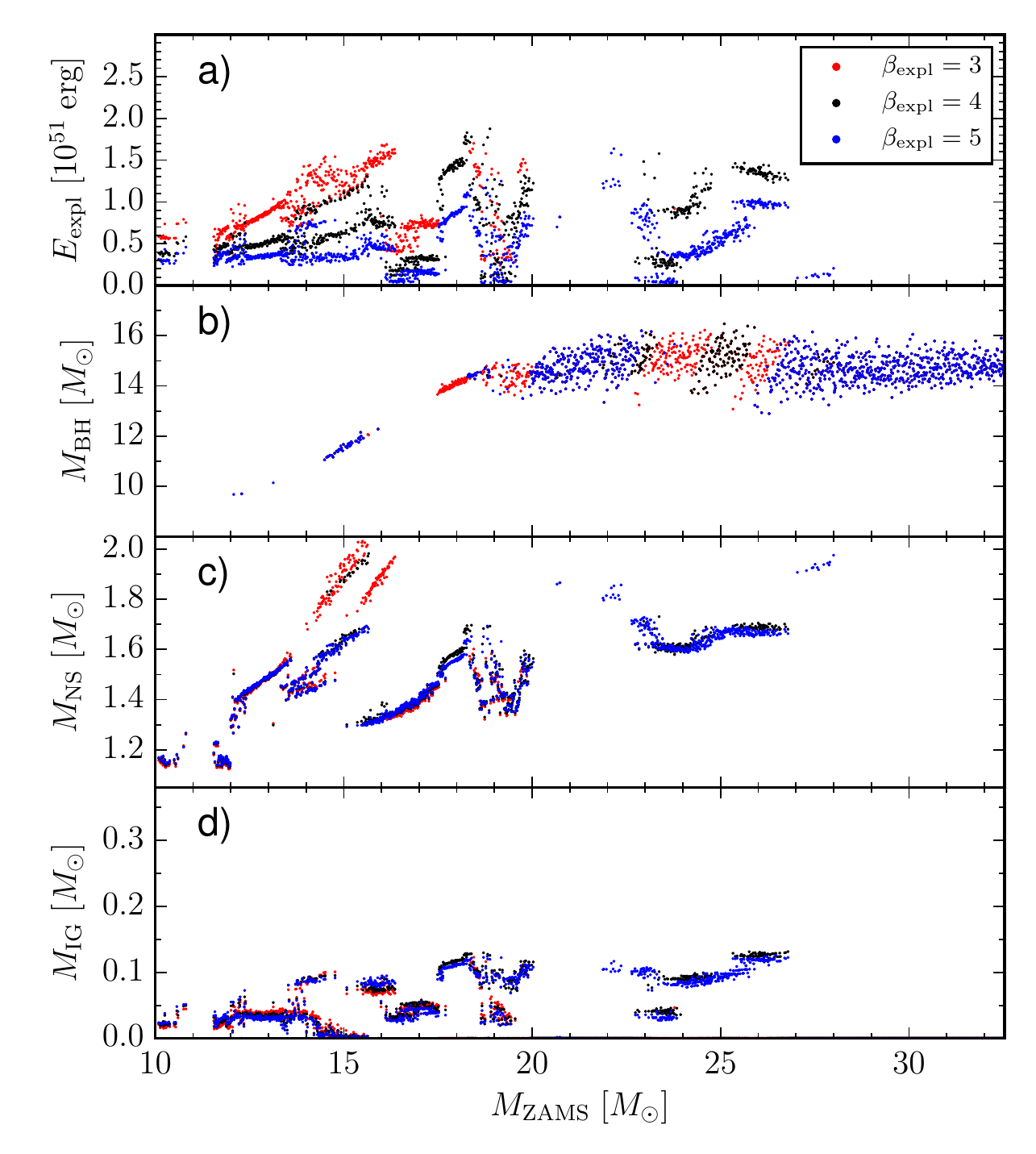}
  \includegraphics[width=0.48\linewidth]{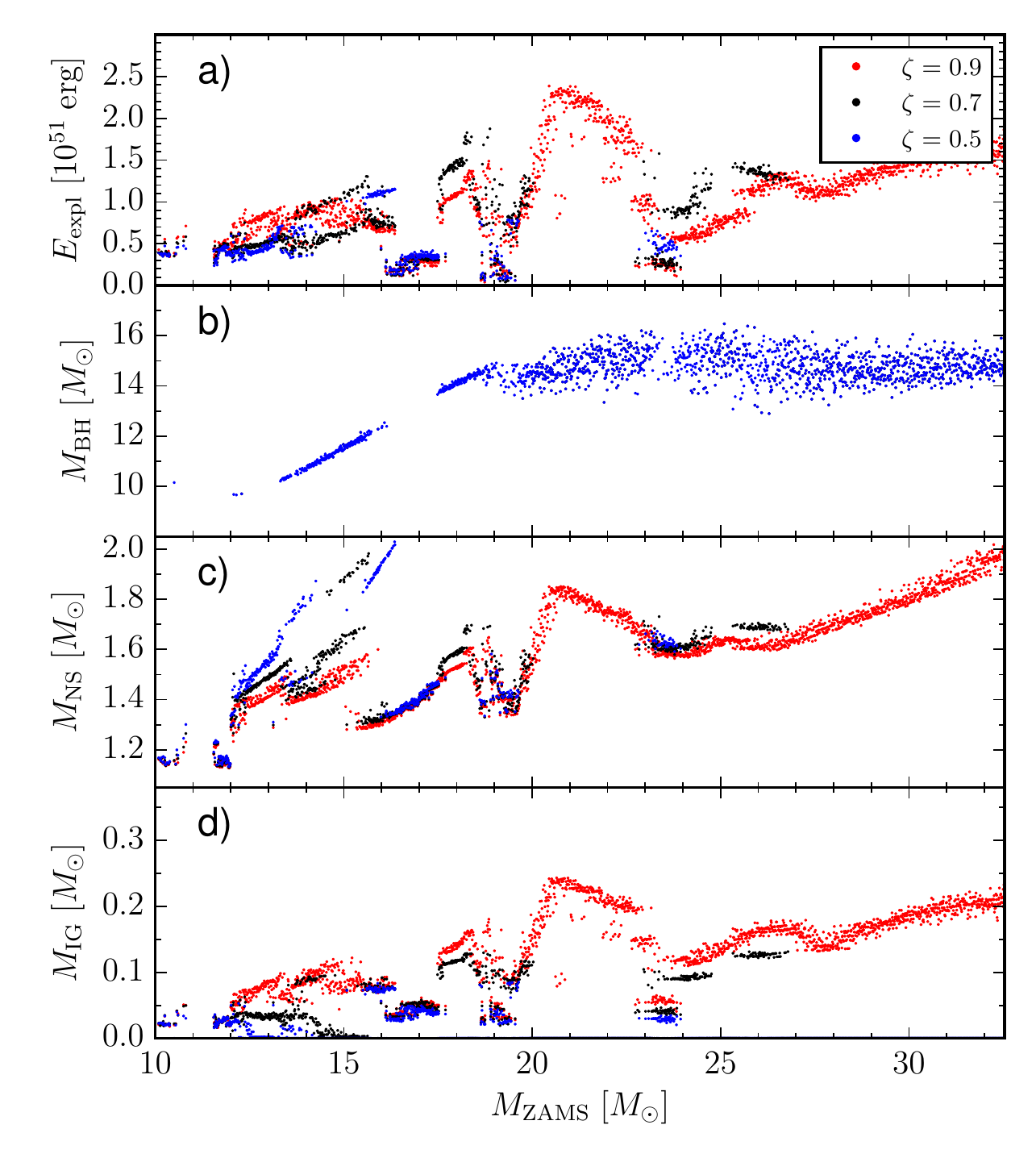} \\
  \includegraphics[width=0.48\linewidth]{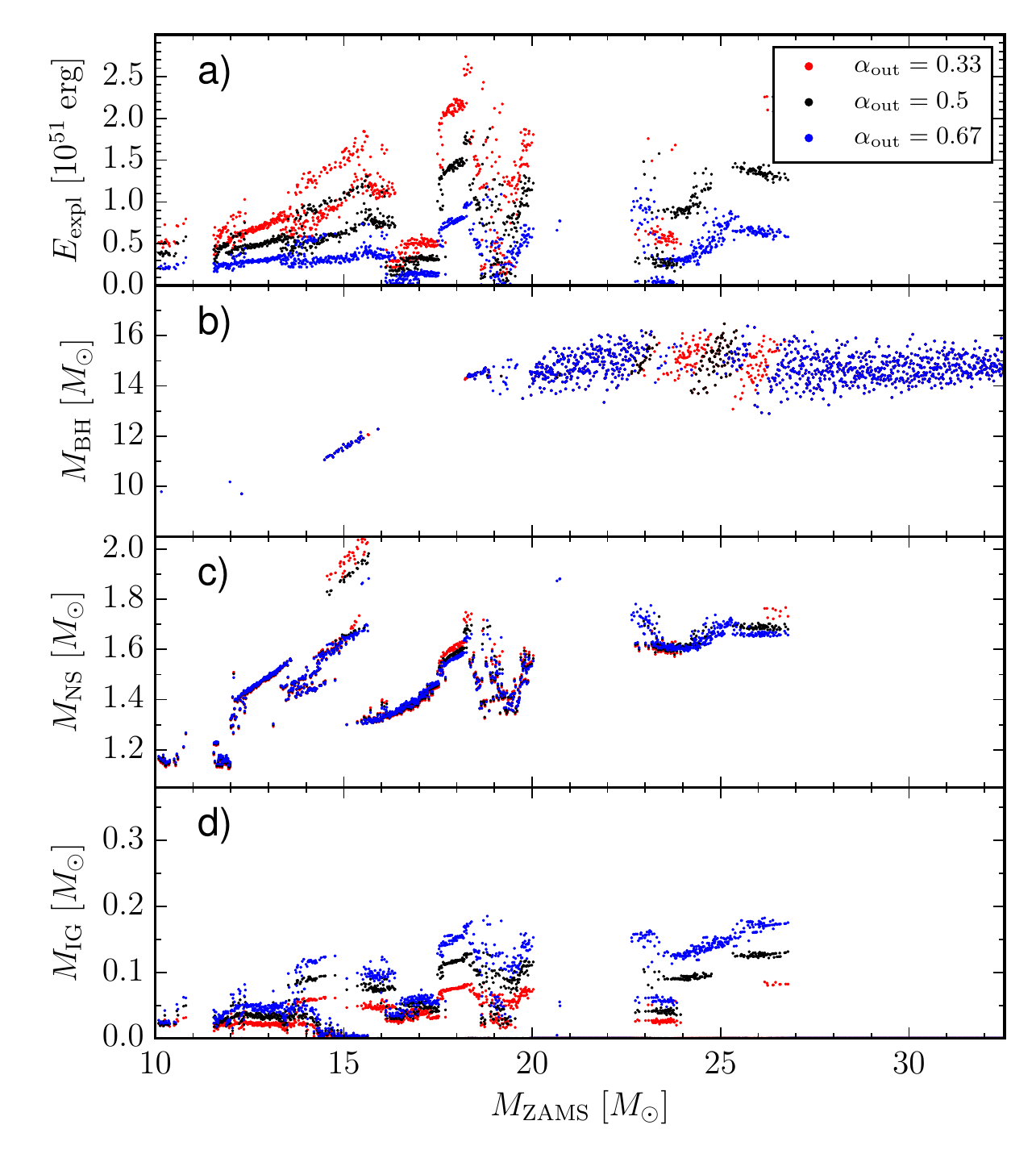}
  \includegraphics[width=0.48\linewidth]{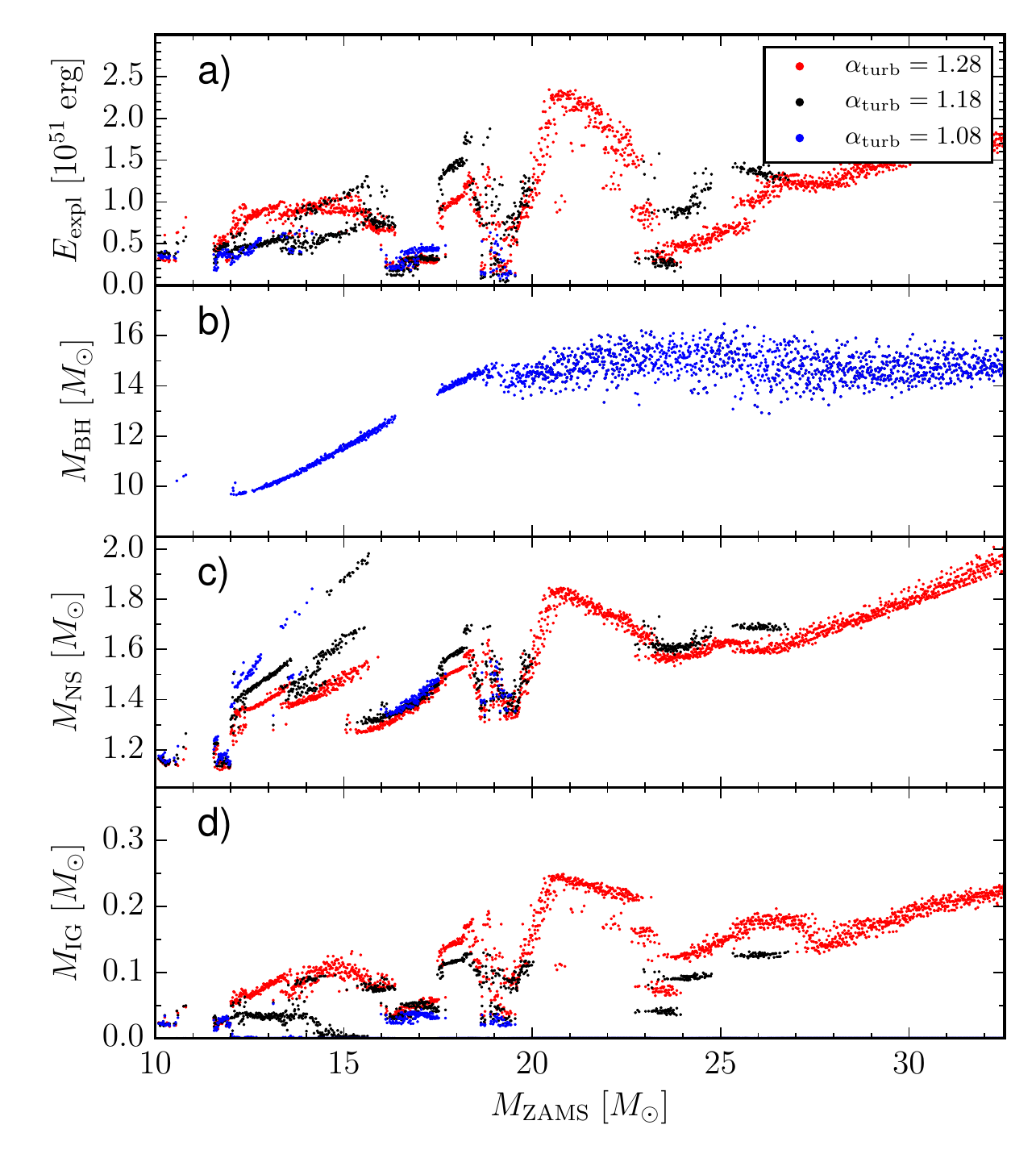}
  \caption{Dependence of the landscape of explosion energies $E_\mathrm{expl}$
    (sub-panel a),
    gravitational remnant masses for black holes
      ($M_\mathrm{BH}$, sub-panel b) and
neutron stars ($M_\mathrm{NS}$, sub-panel c), and iron-group masses (sub-panel d)
  $M_\mathrm{IG}$ on the shock compression ratio
  $\beta_\mathrm{expl}$ (top left), the efficiency factor $\zeta$ for the accretion luminosity (top right),
  the outflow surface fraction $\alpha_\mathrm{out}$ (bottom left), and the factor $\alpha_\mathrm{turb}$ for
  additional shock expansion due to higher turbulent pressure (bottom right).
    \label{fig:sensitivity}}
\end{figure*}

\begin{figure}
  \includegraphics[width=\linewidth]{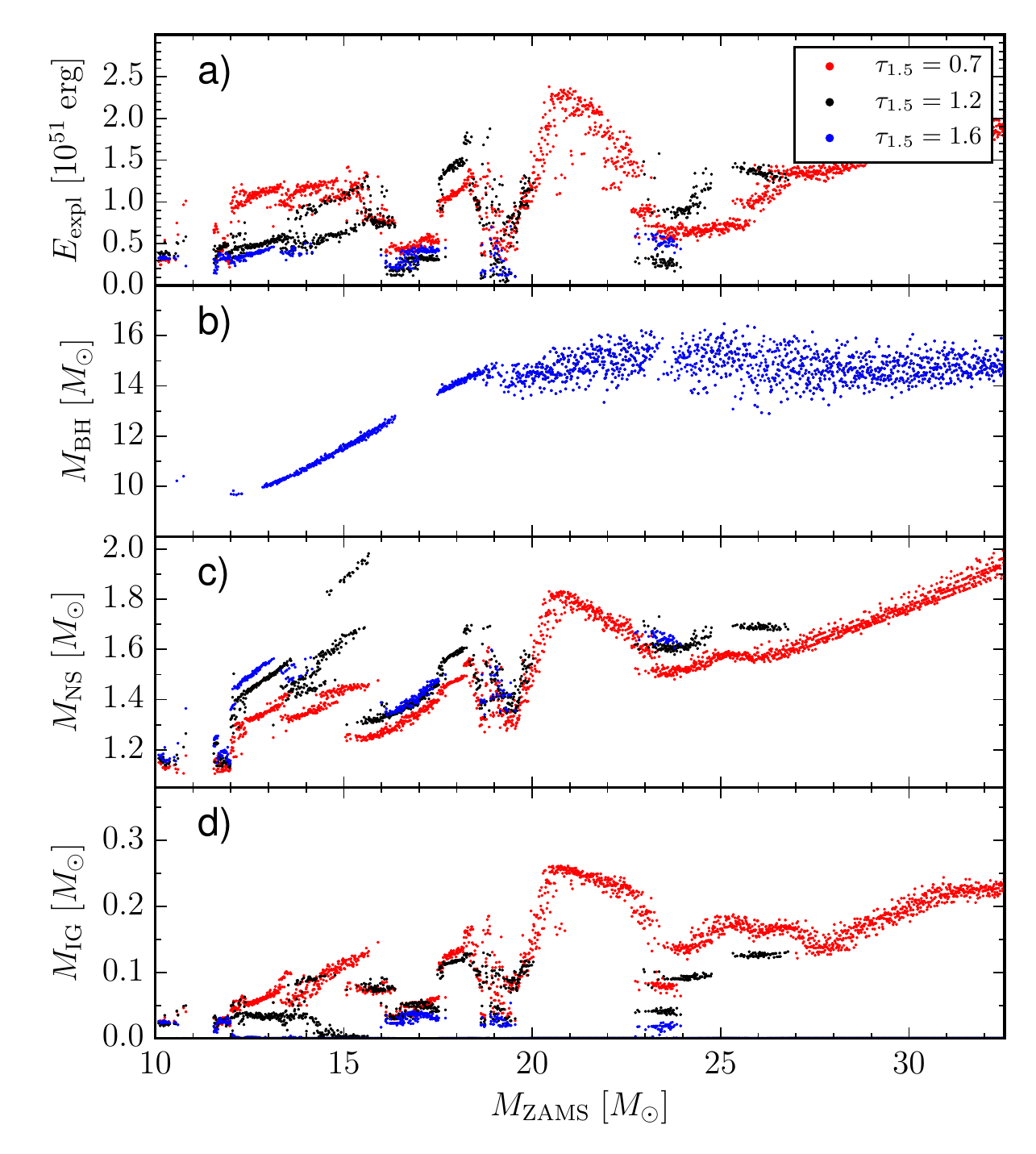}
  \caption{
Dependence of the landscape of explosion energies $E_\mathrm{expl}$ (panel a),
    gravitational remnant masses for black holes
      ($M_\mathrm{BH}$, sub-panel b) and
neutron stars ($M_\mathrm{NS}$, sub-panel c), and iron-group masses
  $M_\mathrm{IG}$ (panel d) on  the cooling time-scale $\tau_\mathrm{1.5}$
for a $1.5\,M_\odot$ neutron star.
    \label{fig:sensitivity2}}
\end{figure}

% In th followin ``while'' could make sense but I don;t think it would
% have the intended meaning ;-)
While our model thus agrees well with the literature when it
comes to predicting the explodability of supernova progenitors, we see
pronounced differences to \citet{ugliano_12} and \citet{pejcha_15a} in
the landscape of explosion and remnant properties.

\subsubsection{Remnant Mass Distribution}
One of the conspicuous features of our model is the prediction of a
multi-modal distribution of neutron star masses
(Figure~\ref{fig:hist}) with peaks around $1.15\,M_\odot$, and
$1.45\,M_\odot$, and possibly a third one at $1.9\,M_\odot$ which is
qualitatively similar to \citet{zhang_08} and case a) of
  \citet{pejcha_15a}, but more conspicuous
than in the work of \citet{ugliano_12}.  The emergence of prominent
peaks at low neutron star masses may simply be due to the better
sampling of small ZAMS masses in our larger set of progenitors, and
the susceptibility of explosions from these progenitors to fallback
due to their extended hydrogen envelope \citep{ugliano_12} could
eventually change the peak structure somewhat.  The location of the
peaks is somewhat different from the bimodal distribution of neutron
stars inferred by \citet{schwab_10} with peaks at $1.25\,M_\odot$ and
$1.35\,M_\odot$, and is pushing the limits of the observed neutron star
mass distribution at low masses $\mathord{\lesssim} 1.2\,M_\odot$,
where we find a far more prominent peak than measured neutron star
masses \citep{lattimer_12} would suggest.  Possible reasons and
remedies for this discrepancy will be discussed later.

Nonetheless, it is noteworthy that a bi- or multi-modal neutron star
mass distribution can naturally be obtained without invoking a
separate stellar evolution channel such as electron capture
supernovae, which have been proposed as the origin of neutron stars
around $1.25\,M_\odot$ \citep{schwab_10}. Structural variations
towards the low-mass end of the iron-core supernova progenitor
population alone might provide an explanation for the observed mass
distribution. At the same time, there is a sufficiently extended tail
of the distribution to produce neutron stars with birth masses
$>1.7\,M_\odot$ mostly from stars between $15\,M_\odot$ and
$20\,M_\odot$.  Such high birth masses are required to account for
cases like the Demorest pulsar, whose birth mass must have been at
least $1.7\,M_\odot$ \citep{tauris_11},

We note, however, that the location of the peaks is somewhat different
to those postulated by \citet{schwab_10} whose inferred distribution
of spin-corrected masses (from 14 well-measured cases) peaks at
$1.25\,M_\odot$ and $1.35\,M_\odot$ with an additional outlier at
$1.6\,M_\odot$.  There is a number of possible reasons for such a
discrepancy; it could point to an overestimation of the Fe and Si core
size in stellar evolution models or a bias in the measured masses due
to binary evolution effects. It could also imply that shock revival
needs to be triggered earlier, i.e., already in the Si shell. Since
the neutron excess in the Si shell dramatically affects the yields
during explosive burning, this is only a viable scenario for a subset
of core-collapse supernovae with supersolar Ni/Fe ratios in the ejecta
\citep{jerkstrand_15} and could not provide a general path towards
smaller neutron star masses due to nucleosynthesis constraints on the
neutron excess in ejecta processed by explosive burning
\citep{woosley_73}.

\subsubsection{Systematics of Explosion Energies and Nickel Masses}
While previous approaches to predict the landscape of supernova
explosion energies using parameterised models have all obtained (by
construction) a similar range for $E_\mathrm{expl}$, some of them are
diametrically opposed concerning the dependence of $E_\mathrm{expl}$
on the ZAMS mass. \citet{ugliano_12} and \citet{pejcha_15a} have
obtained powerful explosions for low-mass progenitors, and in the case
of \citet{pejcha_15a}, there is even an extreme negative correlation
between ZAMS mass and explosion energy. This is due to the major role
of the neutrino-driven wind in powering the explosion in these
studies, which hinged on the neutron star cooling model in the case of
\citet{ugliano_12} and an overly optimistic analytic estimate for the
wind power in \citet{pejcha_15a}. High explosion energies for low-mass
progenitors are, however, inconsistent both with simulations that
point to explosion energies of only a few $ 10^{50}\,\mathrm{erg}$ for
low-mass supernovae
\citep{buras_06b,bruenn_15,melson_15a,mueller_15b}.  Although multi-D
simulations are still limited in their ability to scan the parameter
space systematically, they rather point towards a positive correlation
between ejecta mass and explosion energy
\citep{bruenn_15,nakamura_15}, as does the observational evidence
\citep{poznanski_13,chugai_14,pejcha_15b}.

Among the extant parameterised models, such a positive correlation has
been found by \citet{perego_15}, below ZAMS masses of $15\,M_\odot$ by
\citet{ertl_15}, and below $13\,M_\odot$ by \citet{sukhbold_15}.
\citet{perego_15} relied on a rather \emph{ad hoc} prescription for
boosting the neutrino heating in a pre-specified time interval,
however, and only explored a narrow mass window between $18\,M_\odot$
and $21\,M_\odot$ in ZAMS mass. Similarly, \citet{ertl_15}
introduced a modification of their core cooling model to suppress the
core luminosity depending on $\xi_{1.75}$, which is prompted by an
analysis of the shortcomings of the initial cooling model of
\citet{ugliano_12}, but still savours of a somewhat arbitrary
solution, especially since a parameter characterising a single mass
shell in the progenitor is used to control the diffusion luminosity
from the core at all times.  \citet{sukhbold_15} find a
  correlation between $M_\mathrm{ZAMS}$ and $E_\mathrm{expl}$, but
  only in a very restricted mass range.  They provide some physical
  motivation for a slightly different modification of the cooling
  model, but this comes at the expense of using
  \emph{different} contraction laws for the inner boundary of the grid
  and in the cooling model. Moreover, the parameters of the cooling
  law are still chosen \emph{a priori} based on the mass enclosed in
  the the innermost $3000\,\mathrm{km}$ in the progenitor, and an
  additional ``Crab-like'' calibration model at the lower mass end is
  needed, so, in a sense, the expected result is still put in by
  hand.

Figure~\ref{fig:standard} already suggests that our model is well in
line with the observed correlations without the need for excessive
tweaking. At the low-mass end, we obtain explosion energies as low as
$2.6 \times 10^{50}\,\mathrm{erg}$, whereas all of the energetic
explosion with energies $>10^{51}\,\mathrm{erg}$ occur at higher ZAMS
masses, especially in the islands of explodability around
$18\,M_\odot$ and $22\,M_\odot\ldots25\,M_\odot$.  There is,
however, a subset of low-energy explosions at high masses.  These are
cases on the borderline between neutron star and black hole formation,
where the energy input by neutrinos and nucleon recombination barely
outweighs the binding energy of the envelope. We shall critically
examine this sub-population in more detail below.

In Figure~\ref{fig:e_mej}, we compare the distribution
of ejecta masses $M_\mathrm{ej}$ and
explosion energies $E_\mathrm{expl}$ with fitted power
laws for observed core-collapse events derived
by \citet{pejcha_15b} for two different calibrations of
their model for light curves and expansion velocities.
With a calibration based on \citet{litvinova_85},
they find
\begin{equation}
\label{eq:pejcha1}
\log (E_\mathrm{expl}/10^{50}\,\mathrm{erg})
=
2.09 \log (M_\mathrm{ej}/M_\odot) -1.78,
\end{equation}
while calibrating against \citep{popov_93} yields
\begin{equation}
\label{eq:pejcha2}
\log (E_\mathrm{expl}/10^{50}\,\mathrm{erg})
=
1.81 \log (M_\mathrm{ej}/M_\odot) -1.12\;.
\end{equation}

The bulk of the models fit the power law (\ref{eq:pejcha1}) reasonably
well, although our predicted energies are slightly higher.
Equation~(\ref{eq:pejcha1}) suggests very small explosion energies;
even for the maximum ejecta mass theoretically allowed by our
progenitor models, one would obtain energies only up to
$\mathord{\approx} 4 \times 10^{50}\,\mathrm{erg}$. This simply
reflects calibration problems in the observational determination of
supernova explosion properties. Given these discrepancies between
different approaches for determining explosion energies from light
curves,\footnote{Using detailed Monte Carlo radiative transfer models
  \citet{kasen_09}, for example, obtain a range of values that is
  roughly a factor of two higher than the one given by
  \citet{pejcha_15b}.} the slope of the power laws is arguably to be
trusted more than their normalisation.  If we bear this in mind, the
majority of our models are nicely in line with the observed
correlation between $E_\mathrm{expl}$ and $M_\mathrm{ej}$.

The picture is similar for the nickel mass and
its correlation to the explosion energy that has
already been noted by \citet{hamuy_03}.
In Figure~\ref{fig:e_mig}, we plot the distribution
of our explosion models in the $E_\mathrm{expl}$-$M_\mathrm{IG}$ plane
and compare with the empirical fit obtained
by \citet{pejcha_15b} using \citet{popov_93} for calibration,
\begin{equation}
 \label{eq:e_mni}
\log (M_\mathrm{Ni}/M_\odot)
=
1.13 \log (E_\mathrm{expl}/10^{50}\,\mathrm{erg})
-2.45.
\end{equation}
Except for the subset of low-energy explosions from high ZAMS masses
clustering around $17\,M_\odot$, $19\,M_\odot$ and $24\,M_\odot$, our
model typically predicts iron group masses that agree with the fitted
power law within a factor of two.

The low-energy explosions at high masses are still worrisome.  We
surmise that for these cases the predictions of our model
become rather shaky because one expects considerable fallback.  This
would imply that the observed explosion energy (carried by the ejecta
that avoid fallback) could well be higher, while $M_\mathrm{ej}$ and
$M_\mathrm{IG}$ would be reduced, bringing the models back to the main
branch that fits the power-law dependence of $E_\mathrm{expl}$ on
$M_\mathrm{ej}$.  Moreover, Figures~\ref{fig:e_mej} and
\ref{fig:e_mig} do not provide an adequate picture of the expected
population of observed supernovae: Even if we take the prediction of
such low-energy explosions with high ejecta mass seriously, these
events would be rare because of the steep slope of the IMF, and they
would be faint since the plateau luminosity scales as $L_\mathrm{SN}
\propto E_\mathrm{expl}^{5/6} M_\mathrm{ej}^{-1/2}$
\citep{popov_93,kasen_09}.

In stark contrast to \citet{pejcha_15a}, and in qualitative agreement
with \citet{perego_15} and \citet{nakamura_15}, we find a positive
correlation between neutron star mass and explosion energy
(Figure~\ref{fig:e_mgrav}), at least for the vast majority of
progenitors. Again, the high-mass progenitors with low or moderate
explosion energies on the borderline to black hole formation do not
conform to the general trend; they are the origin of the cluster
around $M_\mathrm{NS}=1.5\,M_\odot$ and $E_\mathrm{expl}=5
\times 10^{50}\,\mathrm{erg}$ and lower. There are also outliers at
$M_\mathrm{NS} \gtrsim 1.8\,M_\odot$ and $E_\mathrm{expl}
\approx 10^{51}\,\mathrm{erg}$.  Moreover, the scatter in the relation
between $M_\mathrm{NS}$ and $E_\mathrm{expl}$ is huge. It is even more
pronounced if we plot $E_\mathrm{expl}$ against the compactness
parameter $\xi_{1.75}$ (see Figure~\ref{fig:e_vs_xi}), the parameter
considered as a correlate to the explosion energy by
\citet{perego_15}. We only find a tendency for very energetic
explosions to occur only at high $\xi_{1.75}$, but no tight
correlation.  This is to be expected because the final explosion
energy is essentially a difference of two quantities that can be of
similar magnitude, i.e., the energy release by nucleon recombination
and nuclear burning and the binding energy of the progenitor.
While either of these will correlate with the proto-neutron star
mass, which directly and indirectly (through correlations with the
structure of the O shell) influences the critical radius where
accretion ceases and hence the amount of material accreted during the
explosion phase, the difference between them will strongly correlate
with the proto-neutron star mass, the mass of the Si core, or the
compactness parameter only over limited ranges of ZAMS mass, where the
structure of the progenitor remains quasi-homologous. Resorting to
single parameters like $\xi$, $M_4$, or $\mu_4 M_4$ as predictors for
the explosion energy therefore seems a somewhat more dubious than
using them as predictors for shock revival only.

\begin{figure}
  \includegraphics[width=\linewidth]{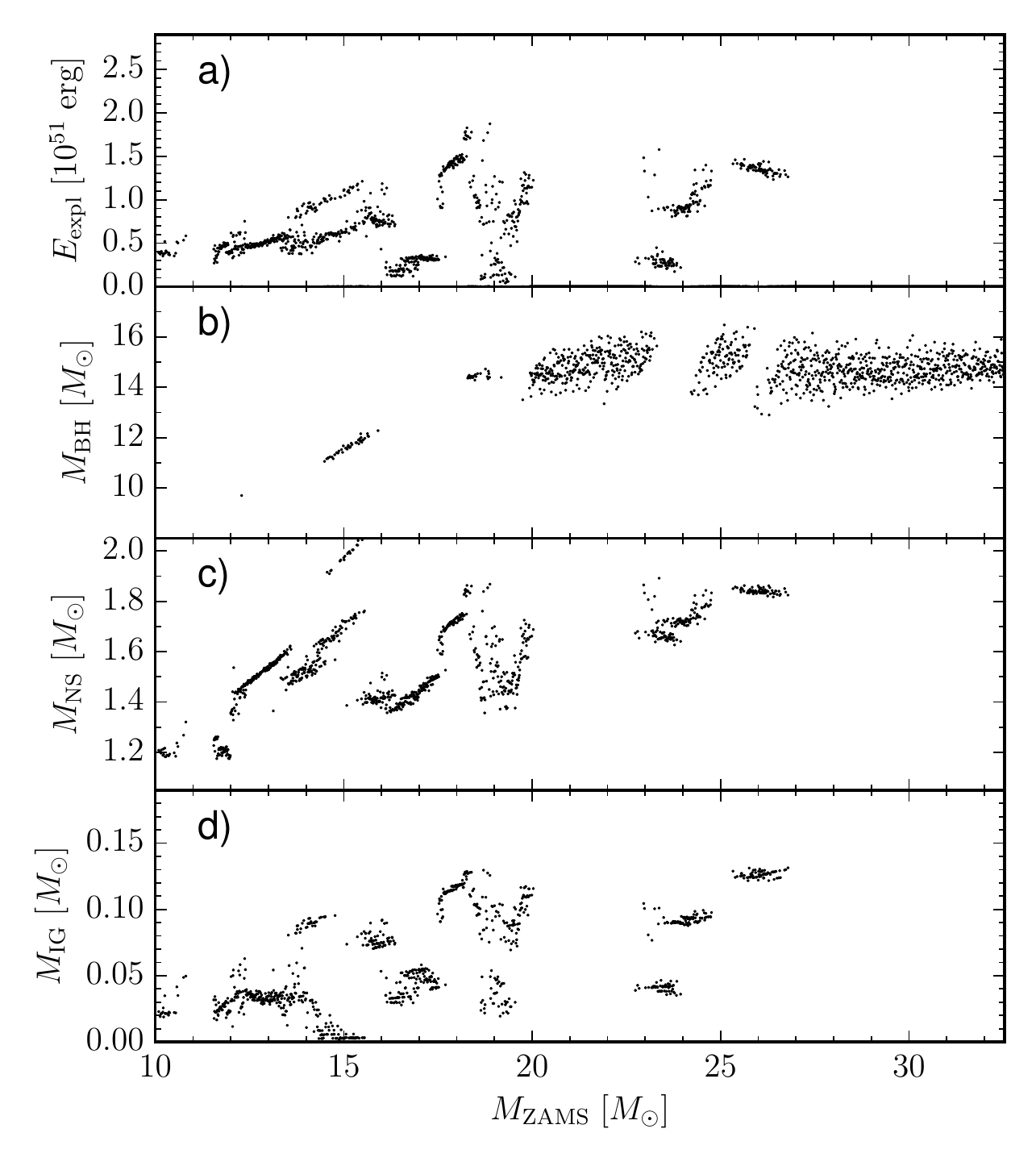}
\caption{Explosion energy $E_\mathrm{expl}$ (panel a),
    gravitational remnant mass for black holes
      ($M_\mathrm{BH}$, sub-panel b) and
neutron stars ($M_\mathrm{NS}$, sub-panel c), and the iron-group mass
  $M_\mathrm{IG}$  (panel d) as a function of
  ZAMS mass for the standard set of parameters,
  but assuming that mass accreted onto the neutron star is not partially
  re-ejected (Equation~\ref{eq:dm_dm_alternative}). Note
that there is a gap in our set of progenitors around $11\,M_\odot$; missing
data points in this region are \emph{not} indicative of black hole formation.
\label{fig:all_accreted}}
\end{figure}

\begin{figure}
  \includegraphics[width=\linewidth]{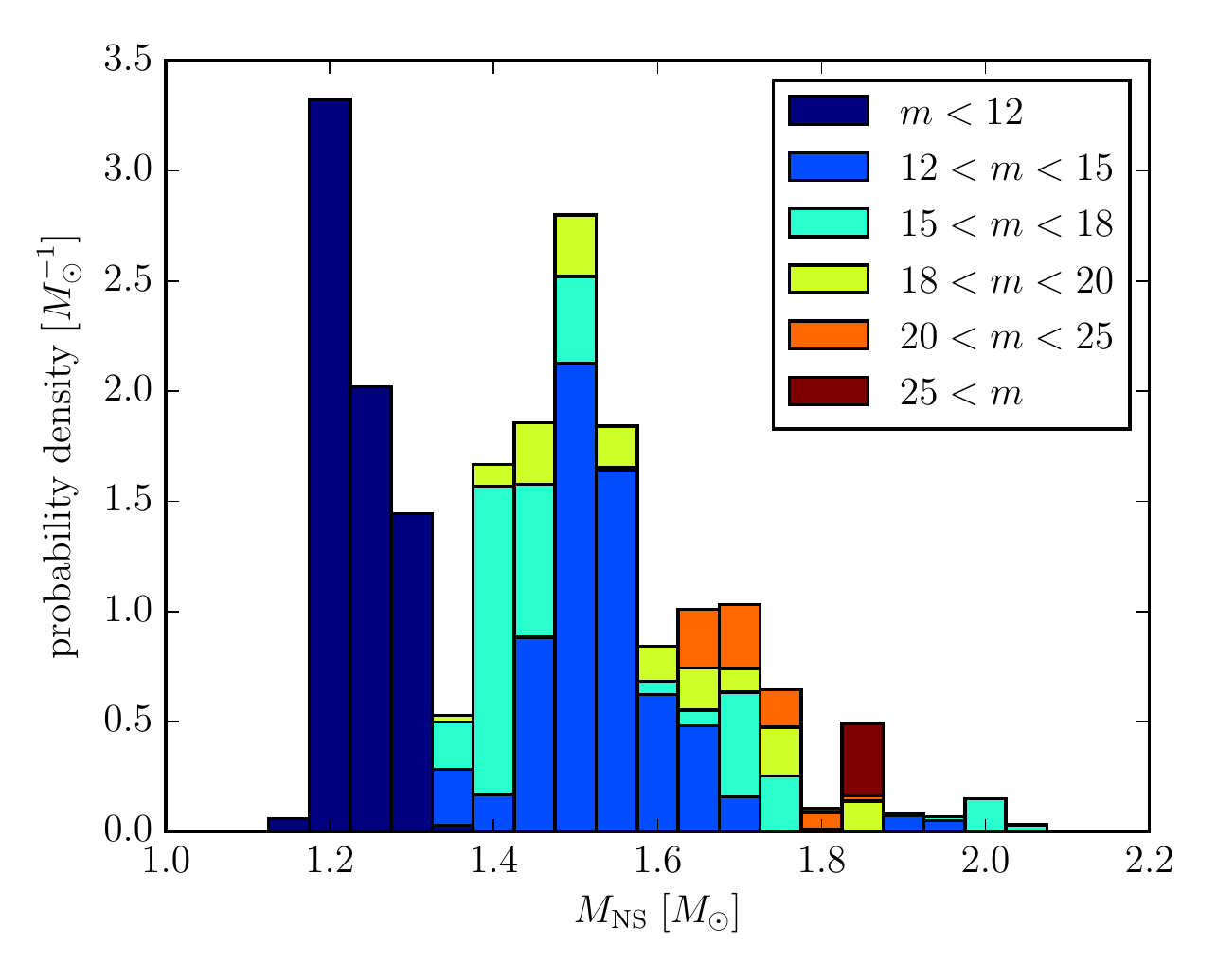}
  \caption{Histogram of the distribution of gravitational neutron star
    masses for for the standard set of parameters, but assuming that
    mass accreted onto the neutron star is not re-ejected
    (Equation~\ref{eq:dm_dm_alternative}) . The stacked bars in
    different colours give the contribution of progenitors from
    different ranges of the ZAMS mass $m$ (measured in solar masses)
    to the average probability density in a given bin.
    \label{fig:hist_acc}}
\end{figure}

\begin{figure}
  \includegraphics[width=\linewidth]{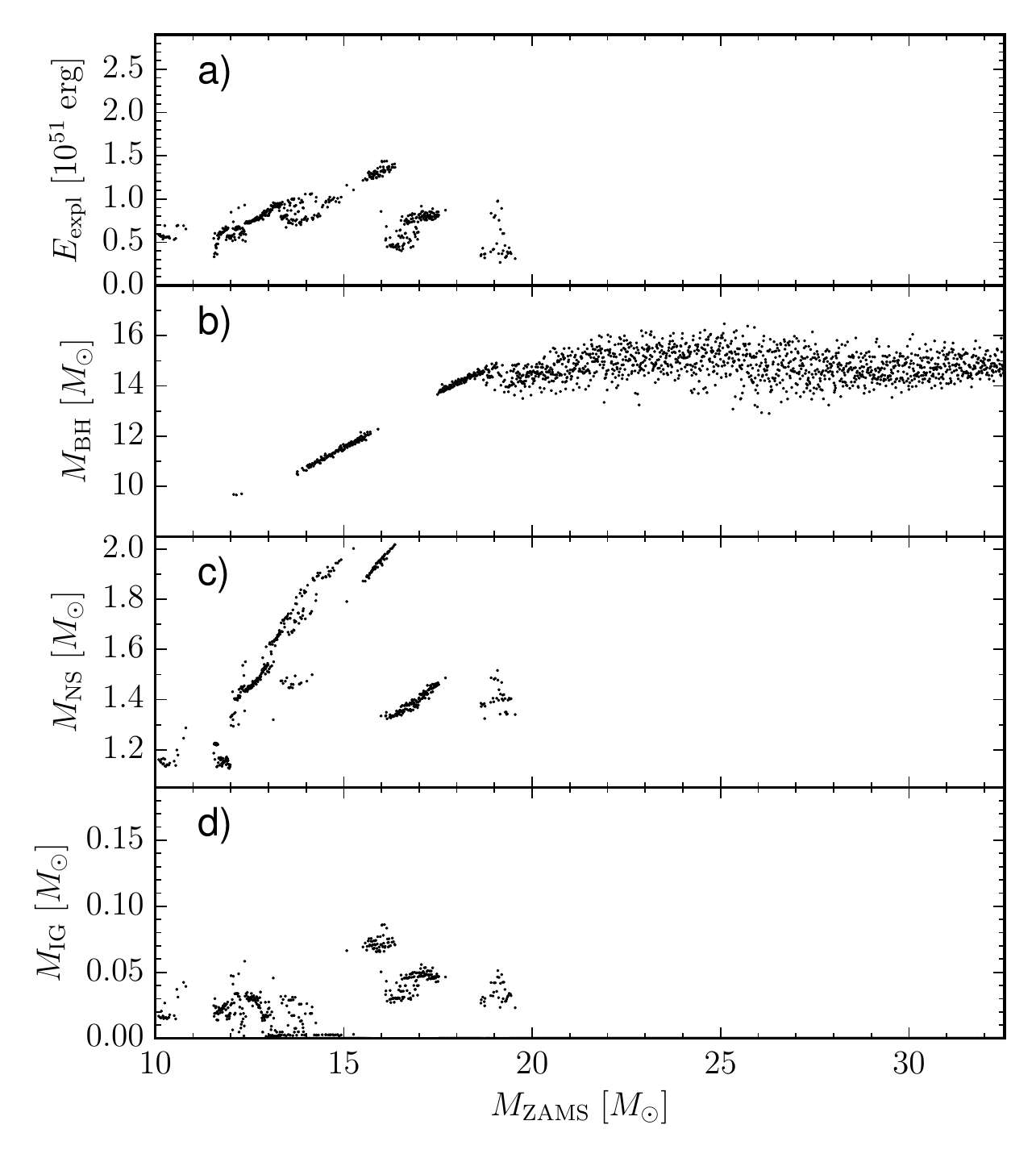}
\caption{Explosion energy $E_\mathrm{expl}$ (panel a),
    gravitational remnant mass for black holes
      ($M_\mathrm{BH}$, sub-panel b) and
neutron stars ($M_\mathrm{NS}$, sub-panel c), and the iron-group mass
  $M_\mathrm{IG}$  (panel d) as a function of
  ZAMS mass for $\alpha_\mathrm{turb}=1.15$
  and $\beta_\mathrm{expl}=3$.
  For this choice of parameters, most progenitors above $\gtrsim 18\,M_\odot$
  form black holes.
\label{fig:many_black_holes}}
\end{figure}

\subsection{Sensitivity to Model Parameters}
\label{sec:variations}
The qualitative agreement of our model with some of the observational
constraints (leaving aside the subset of low-energy explosions
from high-mass stars) is encouraging, but does it actually mean
that the physics of the neutrino-driven explosion mechanism accounts
for the observational trends, or is this just a ``lucky shot''?
What physical ingredients in the model need to be changed
to iron out the remaining tensions with the observational evidence?

This is a critical question for any parameterised approach
to the progenitor-explosion connection, also for calibrated
ones that reproduce the explosion properties of one or a few cases
by construction \citep{ugliano_12,ertl_15,sukhbold_15}. So far, only \citet{pejcha_15a}
have attempted to assess the robustness of their predictions
in a systematic way. Our model has a considerable advantage in that
it allows us to test the impact of variations in \emph{physical} parameters that
characterise physical process in the supernova
core (such as the efficiency of the conversion of accretion
energy into luminosity) rather than abstract exponents in a power law
for the critical luminosity as in \citet{pejcha_15a}.

To assess the sensitivity of our results
to the model parameters, we primarily consider variations of single parameters
around their standard values. The resulting distribution
of the explosion parameters for variations
in $\beta_\mathrm{expl}$, $\zeta$, $\alpha_\mathrm{out}$,
$\alpha_\mathrm{turb}$, and $\tau_\mathrm{1.5}$ are shown  in
Figures~\ref{fig:sensitivity} and \ref{fig:sensitivity2}.

Broadly speaking, the model parameters can be divided
into two classes: $\beta_\mathrm{expl}$ and $\alpha_\mathrm{out}$,
primarily influence $E_\mathrm{expl}$, $M_\mathrm{NS}$,
and $M_\mathrm{IG}$ for \emph{exploding} models but
affect the regions of black hole and neutron star formation
only to a minor degree. $\alpha_\mathrm{turb}$,
$\zeta$, and $\tau_\mathrm{1.5}$ have a larger impact
on success and failure.

\subsubsection{Sensitivity to Accretion After Shock Revival}

Increasing $\beta_\mathrm{expl}$ or decreasing $\alpha_\mathrm{out}$ results
in higher  explosion energies because this allows a larger
amount of mass to be accreted onto the neutron star
and drive an outflow in the process. The overall dependence
of $E_\mathrm{expl}$ on ZAMS mass stays rather similar;
in most  regions the effect is tantamount to a mere rescaling
of the explosion energy. Too much additional accretion onto
the neutron star leads to black hole formation, however, so
that the island of explodability around $23\,M_\odot$
disappears for $\beta_\mathrm{expl}=3$, for example. It is noteworthy
that the distribution of neutron star masses is only
considerably affected for extreme choices of $\beta_\mathrm{expl}$.
This is due to our assumption that a
fraction
$\eta_\mathrm{acc}/|e_\mathrm{g}|$ of the accreted
material is re-channelled into
an outflow (Equation~\ref{eq:dm_dm}), which allows
the cycle of accretion, neutrino heating, and mass
ejection to run without a strong growth
of the neutron star mass if this fraction is high.
Considering the overall uncertainties in the
model, $\eta_\mathrm{acc}/|e_\mathrm{g}|$ may well
be overestimated, which would imply a stronger sensitivity
of $M_\mathrm{NS}$ to $\beta_\mathrm{expl}$ and $\alpha_\mathrm{turb}$
and could shift its distribution to higher masses.
Even if we neglect re-ejection completely
in the mass budget and replace Equation~(\ref{eq:dm_dm})
with
\begin{equation}
  \label{eq:dm_dm_alternative}
  \frac{dM_\mathrm{by}}{dM_\mathrm{sh}}
  =1-\alpha_\mathrm{out}\;,
\end{equation}
however, this does not affect the landscape of explosion properties
qualitatively (Figure~\ref{fig:all_accreted}).  Essentially, the
effect amounts to an upward shift of the peaks of the distribution by
$0.05\,M_\odot$ to $1.2\,M_\odot$ and $1.5\,M_\odot$
(Figure~\ref{fig:hist_acc}). This would bring the low-mass peak more
in line with observations, but increase the tension between the
predictions and the observed neutron star mass distribution
\citep{lattimer_12,schwab_10} at the high-mass end.

The amount of iron group elements $M_\mathrm{IG}$ produced by
explosive burning is little affected by increasing
$\beta_\mathrm{expl}$, on the other hand, because longer accretion
does not increase the shock velocity and post-shock temperature at
early times to allow for explosive burning to the iron group in a more
extended layer. $M_\mathrm{IG}$ also (understandably) decreases for
lower $\alpha_\mathrm{out}$, as a smaller fraction of the burnt
material is swept along with the ejecta. This implies that one can
only trust and expect agreement to empirical correlations between
$E_\mathrm{expl}$ and $M_\mathrm{IG}$ like Equation~(\ref{eq:e_mni})
as far as the power-law index is concerned since the distribution of
these two quantities can easily be rescaled in different directions
within our model.

The \emph{shape} of the distribution of explosion energies
and nickel masses as a function of ZAMS mass emerges as a relatively
robust feature, however. This is an encouraging result
and suggests that the neutrino-driven mechanism can provide
a viable explanation for the observed correlations
between $E_\mathrm{expl}$, $M_\mathrm{ej}$ and $M_\mathrm{Ni}$.

\subsubsection{Sensitivity of Shock Revival to Model Parameters}
$\zeta$, $\tau_{1.5}$, and $\alpha_\mathrm{turb}$ also affect the
heating conditions prior to shock revival, and can change the regions
of neutron star and black hole formation considerably.  The relatively
weak tendency towards black hole formation around $15\,M_\odot$
compared to \citet{ugliano_12} in the standard case is therefore not
indicative of a fundamental disagreement. It merely reflects the
strong sensitivity of shock revival or failure to the assumed physics,
which is perfectly in line with the mixed record of multi-D
simulations, and which also surfaced, albeit to a smaller degree, in
the exploration of different calibration models in \citet{ertl_15} and
\citet{sukhbold_15}.  Given this sensitivity, current parameterised
models can arguably be trusted only to the extent that they predict a
\emph{tendency} towards black hole and neutron star formation for
certain intervals in ZAMS mass, but their extent should be taken as
rather uncertain.  Our result suggests that theoretical models are in
principle compatible with observational evidence that no massive stars
above $\mathord{\approx} 18\,M_\odot$ explode as Type~IIP supernovae
\citep{smartt_15} if we disregard other constraints on explosion
energies and nickel masses for the time being.

Generally (but not invariably), choices for $\zeta$, $\tau_{1.5}$, and
$\alpha_\mathrm{turb}$ that give a large fraction of explosions also
results in higher explosion energies and iron group masses, and
smaller neutron star masses overall. This implies that one needs to
adjust $\beta_\mathrm{expl}$ or $\alpha_\mathrm{out}$ if the overall fraction
of successful explosions goes down in order to obtain reasonable
explosion energies. This is possible for plausible
combinations of parameters, e.g., for
$\alpha_\mathrm{turb}=1.15$ and $\beta_\mathrm{expl}=3$,
which gives a low upper mass limit for successful
explosions in line with \citet{smartt_15}  as shown by
Figure~\ref{fig:many_black_holes}, but unavoidably results
in a more prominent high-mass tail in the distribution of neutron
star masses, which is somewhat at odds with the inferred
birth mass distribution \citep{schwab_10} in binary systems.

Figures~\ref{fig:sensitivity} and \ref{fig:sensitivity2} also
illustrate that better heating conditions during the accretion phase
due to higher $\alpha_\mathrm{turb}$ and $\zeta$ or smaller
$\tau_{1.5}$ can easily result in a landscape of explosion energies
that appears unrealistic not only because of a complete lack of black
hole formation cases, but because of a high incidence of high
nickel/iron group masses among the entire population of core-collapse
supernovae.  Moreover, increasing the heating conditions by shortening
the cooling time-scale $\tau_{1.5}$ tends to destroy the correlation
between ejecta mass and explosion energy for progenitors below
$20\,M_\odot$.  This could be a problem for scenarios for more
efficient shock revival that rely on a faster release of the thermal
energy of the proto-neutron star core, such as active-sterile neutrino
conversion and re-conversion \citep{hidaka_07}.

\subsubsection{Implicitly Fixed Parameters}

The reader should bear in mind that our model still contains a few
  parameters other than $\beta_\mathrm{expl}$, $\zeta$,
  $\alpha_\mathrm{out}$, $\alpha_\mathrm{turb}$, and
  $\tau_\mathrm{1.5}$, which we have implicitly considered as fixed
  because they are arguably not as uncertain as the other ones, or
  because changing them would largely amount to a rescaling of some
  other parameter. For these reasons, variations in these parameters
  do not warrant an extended discussion, and we do not provide plots to
  illustrate them. We nevertheless briefly comment in a non-exhaustive
  fashion on the resulting effects for a few selected
  parameters. Shortening the infall time by using a different
  coefficient in Equation~(\ref{eq:tff}) generally delays shock
  revival and leads to more prevalent black hole formation. Increasing
  the fraction of electron neutrinos and antineutrinos in the
  diffusive flux in Equation~(\ref{eq:ldiff}) tends to increase the
  explosion fraction and make the explosions more energetic across the
  whole mass range. Increasing the effect of gravitational redshift on
  the neutrino luminosity (\ref{eq:ltot}) by computing it for a
  smaller radius $r_\mathrm{PNS} < 5/7 r_\mathrm{g}$ obviously
  decreases the explosion fraction, with a more pronounced effect for
  massive progenitors with large iron and silicon cores.  Increasing
  the coefficient in Equation~(\ref{eq:vshock}) for the shock velocity
  is largely tantamount to increasing $\beta$ and shortening the phase
  of accretion after shock revival (which generally decreases the
  final explosion energies).  Increasing $\epsilon_\mathrm{rec}$ in
  Equation~(\ref{eq:dot_eexpl}), i.e., the contribution of
  neutrino-heated ejecta to the explosion energy per unit mass, leads
  to an earlier termination of accretion and lower neutron star
  masses; the higher asymptotic energy of the ejecta largely balances
  the shorter duration of neutrino-driven mass ejection, so that
  explosion energies are not too strongly affected, especially for low
  progenitor masses. Iron group masses are obviously directly affected
  by the threshold temperature for silicon burning; but other than
  that the threshold temperatures for the different explosive burning
  processes have little effect on the other explosion parameters as
  long as explosion energy is primarily provided by neutrino heating.

  Obviously, some of the independent parameter variations explored
  in this section would already result in a distribution of explosion
  energies and remnant properties that is in conflict with one or more
  observational constraints (range of observed explosion energies,
  explosion properties of SN~1987A, etc.).
  This could suggest that the
  allowed variations in the landscape of explosion properties are
  actually much smaller than this section might suggest, and that even
  their \emph{absolute values} -- and not only general trends and
  correlations -- can be predicted with good accuracy with the help of
  one or two calibration cases. Without a more complete exploration of
  the high-dimensional parameter space this verdict ought to be left
  to the future.

\section{Summary and Conclusions}
\label{sec:summary}
In this paper, we set out to develop a theoretical approach to predict
remnant and explosion properties of neutrino-driven core-collapse
supernovae without an elaborate machinery of parameterised 1D
neutrino-hydrodynamics simulations, let alone multi-D models.  To this
end, we constructed a simple model based on analytic approximations
for the pre-explosion phase up to shock revival and simple ODEs in the
explosion phase.  For the first time, we attempt to take into account
that continued accretion after shock revival plays a major role in
powering the explosion.  While we need to introduce a number of
  parameters, all of these have a physical motivation and
  significance, and multi-D simulations of supernova explosions can be
  used to calibrate them.

Our examination of the model predictions for an unprecedented number
of 2120 progenitor models and their sensitivity to the model
parameters are encouraging: Using plausible choices for physical
parameters based on recent multi-D simulations, we can obtain a
similar landscape of neutron star and black hole formation regions as
time-dependent 1D models with neutrino transport
\citep{ugliano_12,ertl_15,perego_15} or the hybrid approach of
\citet{pejcha_15a} based partly on simulations and partly on analytic
theory: For our standard set of parameters, there are some instances
of black hole formation already at low ZAMS mass ($\approx
15\,M_\odot$), and some islands of explodability at high ZAMS
mass. Good agreement with extant phenomenological explosion criteria
like the progenitor compactness \citep{oconnor_11} and the Ertl
criterion \citep{ertl_15} both validate our model and indicate that
the inclusion of accretion after shock revival does not fundamentally
affect predictions of the explodability of supernova progenitors.

By including a simple estimate for the duration of accretion after
shock revival and the concomitant neutrino heating, we are able to
reproduce observed correlations between the explosion energy and the
ejecta mass \citep{poznanski_13,chugai_14,pejcha_15b} and between the
explosion energy and the nickel mass \citep{hamuy_03,pejcha_15b}
naturally for the bulk of our progenitors. In agreement with the 2D
study of \citet{nakamura_15} and the parameterised 1D study of
\citet{perego_15} (which was restricted to progenitors between
$18\,M_\odot$ and $21\,M_\odot$, however), we also find a loose
correlation between neutron star mass and explosion energy, implying
that the most energetic neutrino-driven explosions with
$E_\mathrm{expl} \approx 2 \times 10^{51}\,\mathrm{erg}$ leave behind
neutron stars with masses $\gtrsim 1.7\,M_\odot$.  For our standard
case, we obtain a multi-modal neutron star mass distribution with
peaks around $1.15\,M_\odot$, $1.45\,M_\odot$ and possibly
$1.9\,M_\odot$. The low-mass peak emerges naturally from stars with
ZAMS masses between $10\,M_\odot$ and $12\,M_\odot$.

An exploration of the sensitivity of our model predictions to
individual parameters revealed that there is considerable leeway in
parameterised approaches like ours to shift explosion energies and
nickel masses up or down \emph{globally} for the entire range of
progenitor masses, but aside from that the functional dependence of
$E_\mathrm{expl}$ and the produced amount of iron group elements on
ZAMS mass appears rather robust. Similarly, plausible variations in
the parameters can easily shift the peaks in the neutron star mass
distribution by $\mathord{\gtrsim} 0.05\,M_\odot$.  Moreover, the
overall fraction of neutron star and black hole formation can change
considerably for reasonable parameter variations, indicating that
empirical parameters for explodability, while useful as a rough
metric, cannot provide a sharp dividing line between exploding and
non-exploding models at the present state of supernova theory.  In
line with \citet{clausen_15}, we also find that the boundaries between
regions of black hole and neutron star formation are fuzzy and both
channels may be similarly prevalent in certain intervals of ZAMS mass
so that a probabilistic description may be more adequate.

All this bodes well for one of the primary purposes of our model.  On
the level of accuracy and reliability that current parameterised 1D
simulations have reached, it appears possible to estimate explosion
properties of massive stars simply based on their structure
\emph{without} the relatively elaborate numerical machinery that
approaches like \citet{ugliano_12,ertl_15,perego_15} and
\citet{sukhbold_15} rely on.  Our model, perhaps with some calibration
against a specific reference case like SN~1987A as in
\citet{ugliano_12} or against other equally important constraints
  on the entire population of explosions, could provide input for
systematic studies of supernova nucleosynthesis or for a quick
exploration of the effect of new or uncertain physics in stellar
evolution on supernovae from massive stars. Even without
  such a calibration, which always faces the dilemma
  of singling out the ``best'' and most important observational
  constraints, it is already
  useful for identifying \emph{trends and tendencies} in the
  explosion properties and checking their robustness against
  parameter variations.
It is remarkable and
informative that a few relatively simple and physically motivated
equations can capture the gist of more complicated simulations to a large
degree. Obviously, this does not render parameterised 1D and 2D
simulations obsolete, however. These are still superior in that they
can treat many aspects of the supernova problem, among them the cooling
and the contraction of the neutron star, explosive burning in the shock,
and fallback (more) self-consistently, and can therefore potentially
provide much firmer quantitative predictions for the observables
discussed here: The range of parameter variations explored here may be
allowed mathematically, but may not be realisable any more in a
more rigorous approach.

Moreover, there are still some caveats and critical issues
that will need to be re-examined in the future.  Aside from some of
the unavoidable oversimplifications and inconsistencies that come
with analytic models, our approach still has two major shortcomings
in terms of missing physics. Moreover, there are some tensions
between observational constraints and the predictions not only of
our model but also other parameteric studies in the literature.

The most obvious shortcoming that could account for some of these
tensions is our all-our-nothing treatment of fallback due to the
deceleration of material by the reverse shock \citep{chevalier_89}.
The recent results of \citet{sukhbold_15}, who found only few cases
with more than $0.01\,M_\odot$ of fallback, suggest that there is
little fallback for low-mass progenitors and for high-mass progenitors
with rather high explosion energies.  Thus, the distribution of
neutron star masses (and even of nickel masses) may not be severely
affected by fallback. We speculate, however, that considerable
fallback could occur for our sub-population of low-energy explosions
from high-mass stars, where the ``initial'' explosion energy and the
binding energy of the envelope almost cancel.  A more consistent
treatment of fallback and the energy transfer from the inner ejecta to
the envelope could help to eliminate this peculiar subset of
explosions, or bring the inordinately high nickel and ejecta masses
down to values that are in line with the observed systematics of
core-collapse supernova explosions.

  Moreover, the nickel (iron group) masses predicted by our very crude
  treatment of explosive burning should be taken with caution even
  though we can obtain a plausible range of values for most
  progenitors for appropriately chosen model parameters.

In the long run, more systematic multi-D studies of supernova
explosions are needed to determine whether multi-D effects can
be subsumed into a simple modification of the heating conditions in 1D
prior to shock revival and a crude budget of mass inflow and outflow
after shock revival. Many light-bulb based models of convection and
the standing accretion shock instability (SASI) in supernova cores \citep{murphy_08b,hanke_12}, as well as the recent
first-principle models of \citet{summa_16}
point in this direction, and there is also
some theoretical justification for this \citep{mueller_15a}. On the
other hand, \citet{cardall_15} recently argued that the threshold for
explosions is smeared out considerably in the SASI-dominated regime
and subject to stochastic variations. Even if a simple rescaling
of the 1D heating conditions were adequate in the SASI-dominated regime, the
reduction in the required heating in 3D might be larger
than for convection-dominated models \citep{fernandez_15}. For the explosion phase, the validity of simple
phenomenological models is even less well explored so far.  Finally,
the (potentially crucial) role of multi-D seed asphericities from
convective burning in shock revival
\citep{couch_13,mueller_15a,couch_15b} is not accounted for in our
model at all.

In fact, additional physics like large seed perturbations for the
hydrodynamic instabilities may be required to resolve the tensions
between model predictions and observations: It appears rather
difficult for parameterised models to produce a prominent second peak
of the neutron star mass distribution at $1.35\,M_\odot$ as suggested
by observations \citep{schwab_10}, while at the same time covering a
range of supernova energies up to $2 \times 10^{51}\,\mathrm{erg}$.
Could the (mild) discrepancy between predictions and observations be
accounted for by selection effects in binaries, or by uncertainties in
the core structure of massive stars, or does shock revival need to be
initiated already in the Si shell for some models
\citep{couch_15b,jerkstrand_15}? Alternatively, fallback in explosions
of low-mass stars might merge the neutron star distribution into a
single peak at the desired value of $1.35\,M_\odot$, and electron
capture supernovae, which are not included here, could provide a
separate peak at lower mass in line with the original idea of
\citet{schwab_10}.

Similarly a cut-off for neutron star formation around $18\,M_\odot$
seems difficult to accommodate without either accepting small
explosion energies or shifting the distribution of neutron stars up to
higher gravitational masses. Could convective seed perturbations help
to explode some progenitors energetically while not affecting most
progenitors above $18\,M_\odot$? Or could uncertainties in the mass
loss, or binary effects change the fate of progenitors above
$18\,M_\odot$?

At present, any attempt to provide a coherent solution for these
problems must remain highly speculative. Nonetheless, our improved
understanding of the neutrino-driven mechanism has, despite some
setbacks in simulations, clearly reached a stage where
it can help to explain the systematics of the observed explosion
and remnant properties.

\section*{Acknowledgements}
% we thank our colleagues first ...
We acknowledge fruitful discussions with H.-T.~Janka, N.~Langer,
O.~Pejcha, P.~Podsiadlowski, T.~Tauris, and S.~Woosley.  This work was
supported in part by an ARC DECRA Fellowship DE150101145 (BM), and by
an ARC Future Fellowship FT120100363 (AH).  This material is based
upon work supported by the National Science Foundation under Grant
No. PHY-1430152 (JINA Center for the Evolution of the Elements).

\bibliography{paper}

\appendix
\section{Dependence of Explosion Parameters on Helium and C/O Core Mass}
\label{sec:appendix}
The primary determinant for the pre-collapse luminosity of Type~II
supernova progenitors is the helium core mass.  The helium core mass
rather than the ZAMS mass is therefore the more appropriate parameter
for interpreting observations of supernova progenitors based on their
position in the Hertzsprung-Russell (HR) diagram
\citep{smartt_09a,smartt_09b,smartt_15} and inferring mass limits for
successful explosions from them.  Different helium core masses may,
e.g., result from a different treatment of mixing in stellar
interiors, which may partly explain the variations of the inferred
maximum ZAMS mass for Type~II supernova progenitors by about
$2\,M_\odot$ \citep{smartt_15}. Obviously, the ZAMS mass is also not a
suitable parameter for incorporating the predictions of parameterised
supernova explosion models into binary evolution and population
synthesis calculations because of the possibility of mass transfer;
again the mass of the helium core or the carbon/oxygen (C/O) core is a
more appropriate parameter \citep{hurley_00,belczynski_08}.

\begin{figure*}
  \includegraphics[width=0.48\linewidth]{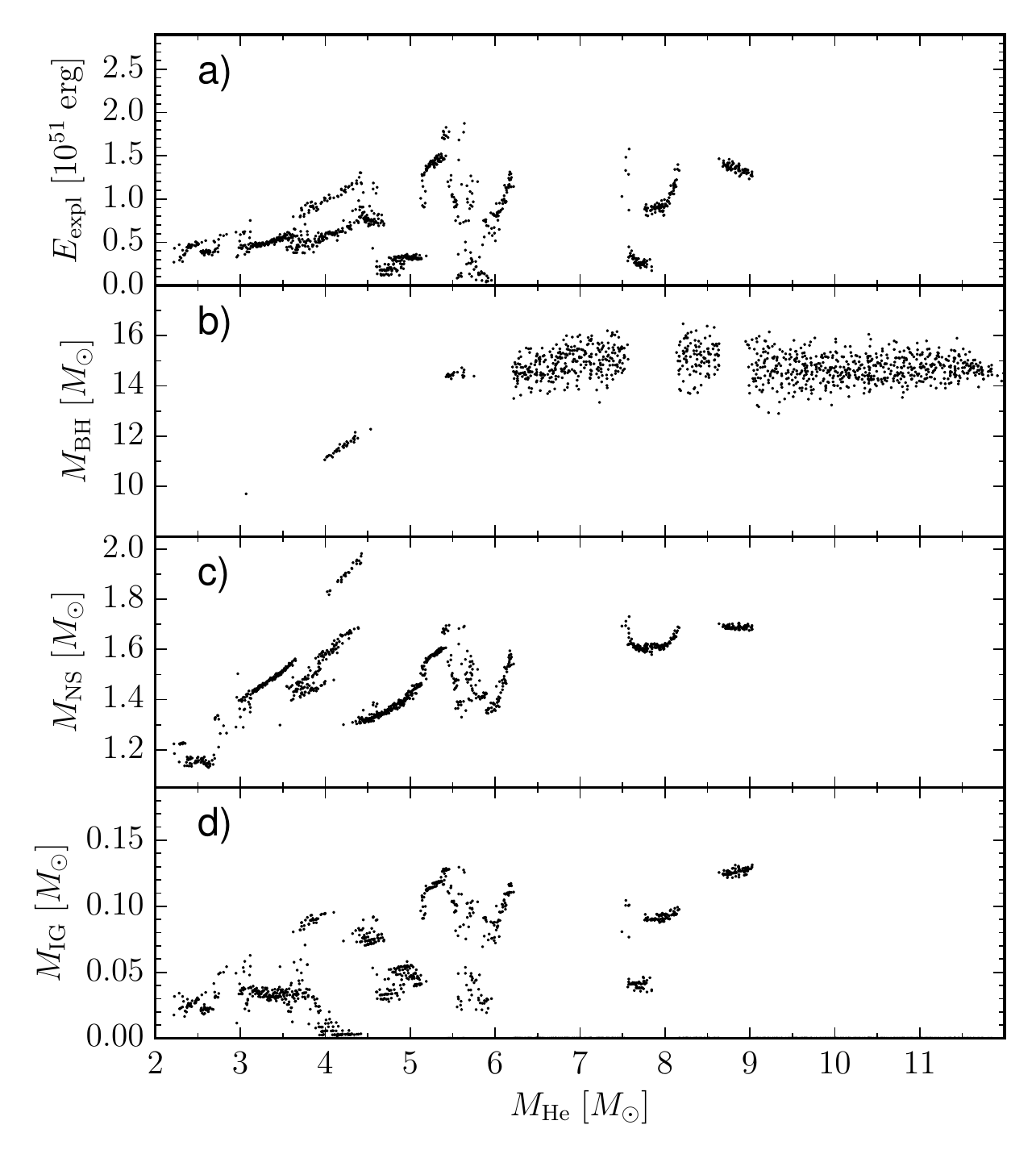}
  \includegraphics[width=0.48\linewidth]{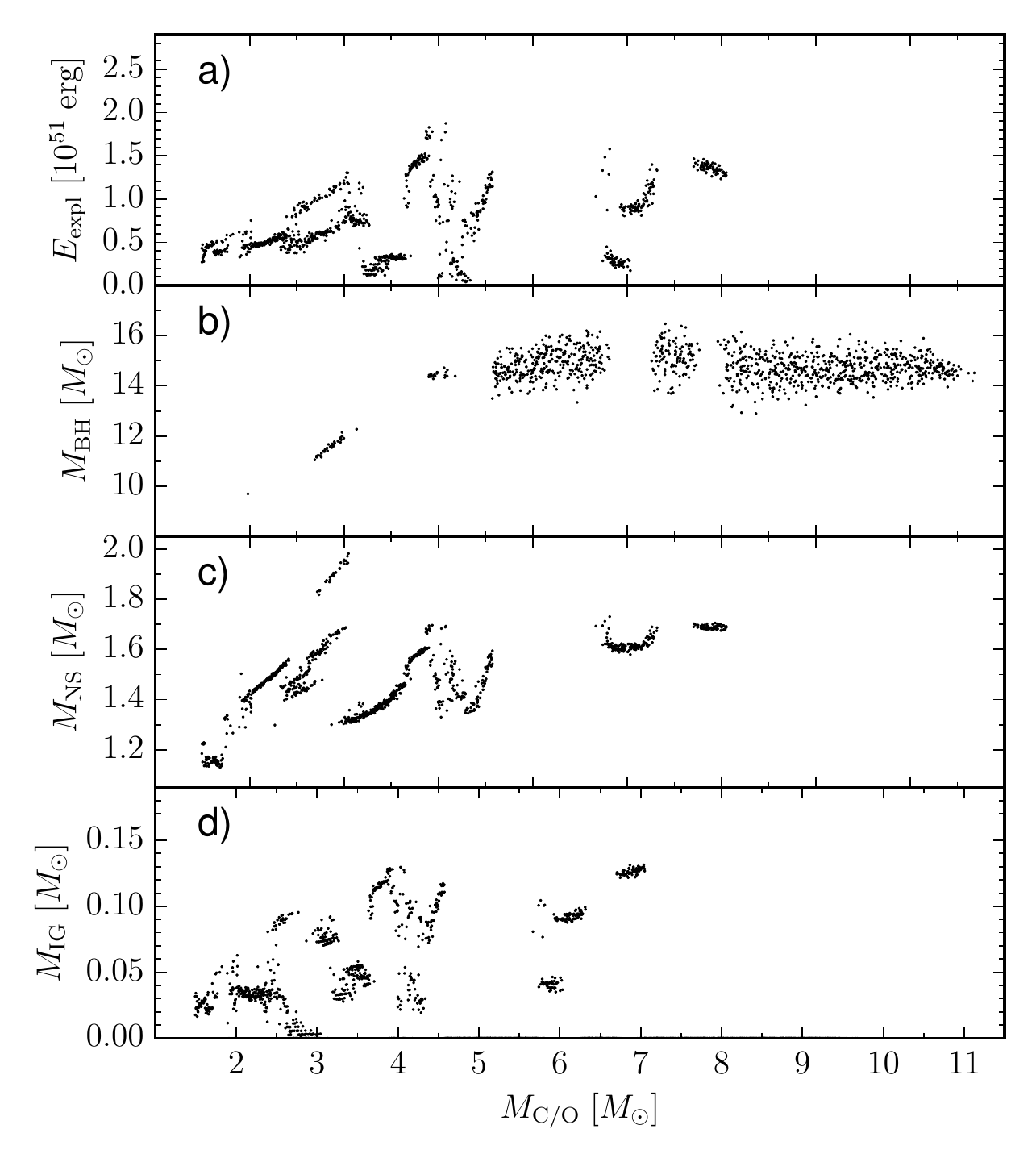}
  \caption{
    % THIS IS NOW MORE CONSISTENT
    Explosion energy ($E_\mathrm{expl}$, Panel a),
    gravitational remnant mass for black holes ($M_\mathrm{BH}$,
      Panel b) and neutron stars ($M_\mathrm{NS}$, sub-panel c), and
    the iron-group mass ($M_\mathrm{IG}$, Panels d) as a function of
    the helium core mass ($M_\mathrm{He}$ left) and the C/O core mass
    ($M_\mathrm{C/O}$, right) at collapse for the standard case.  Note
    that there is a gap in our set of progenitors around $11\,M_\odot$;
    missing data points in this region are \emph{not} indicative of
    black hole formation.
\label{fig:standard_he_co}}
\end{figure*}

\begin{figure*}
  \includegraphics[width=0.48\linewidth]{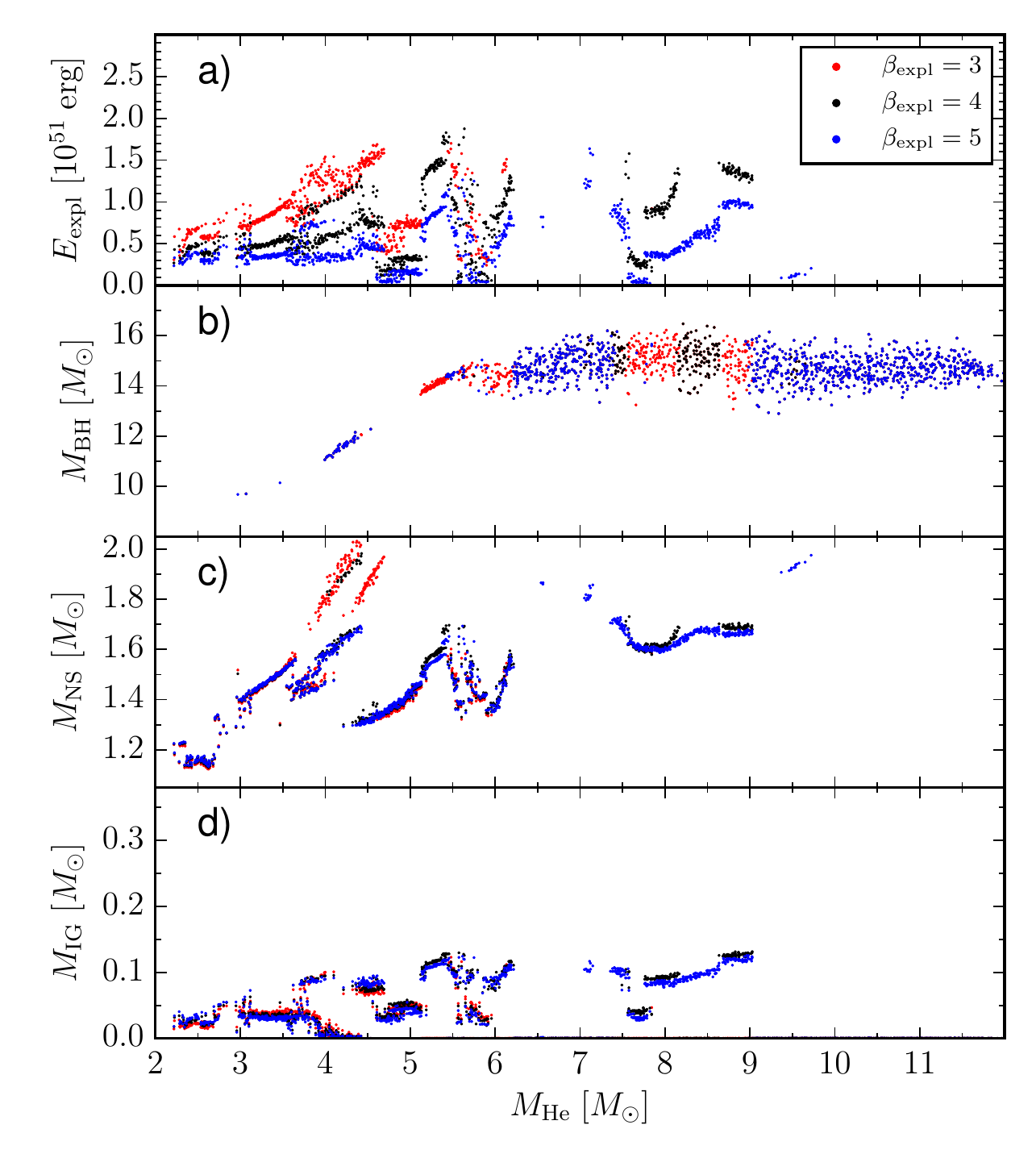}
  \includegraphics[width=0.48\linewidth]{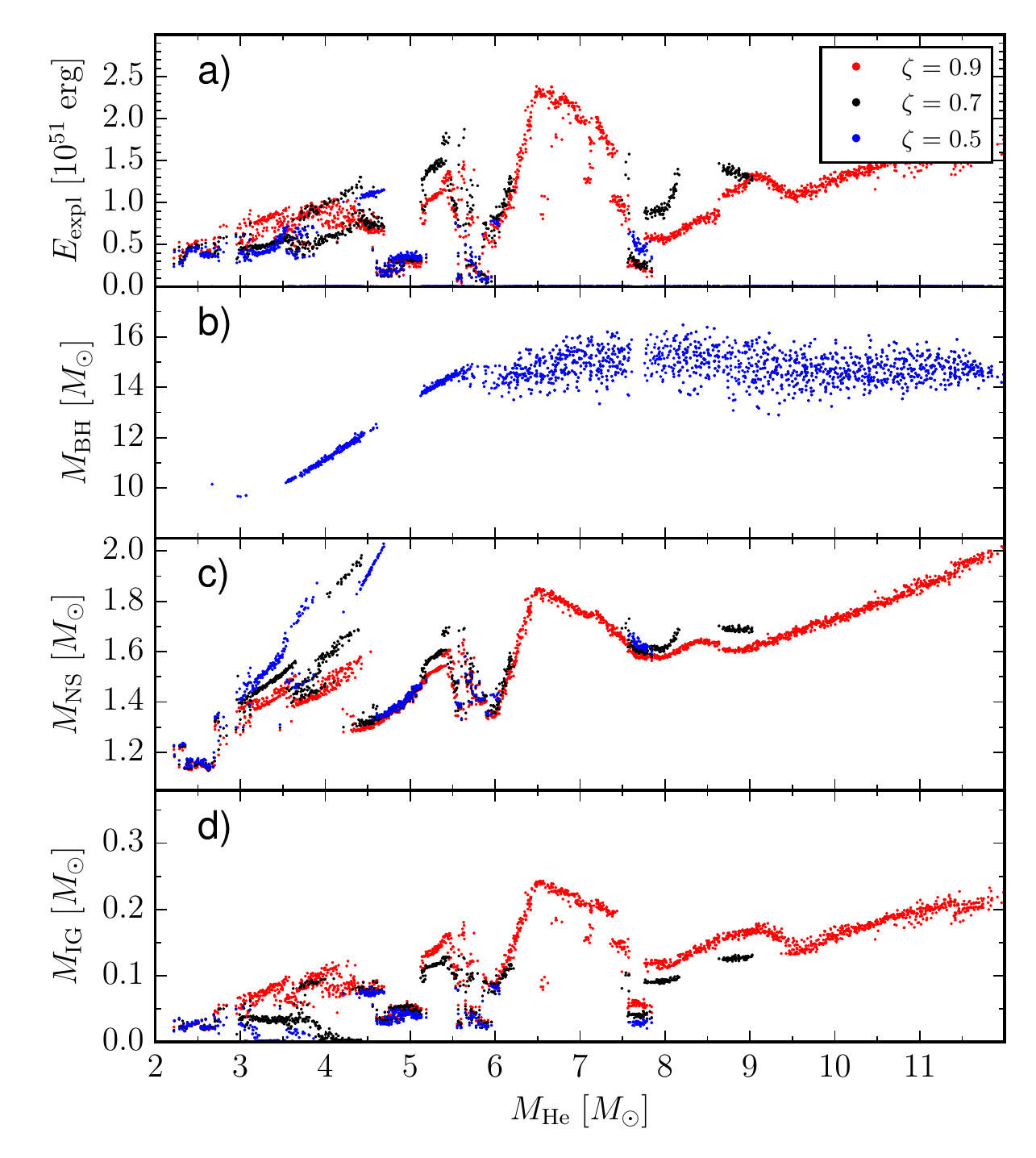} \\
  \includegraphics[width=0.48\linewidth]{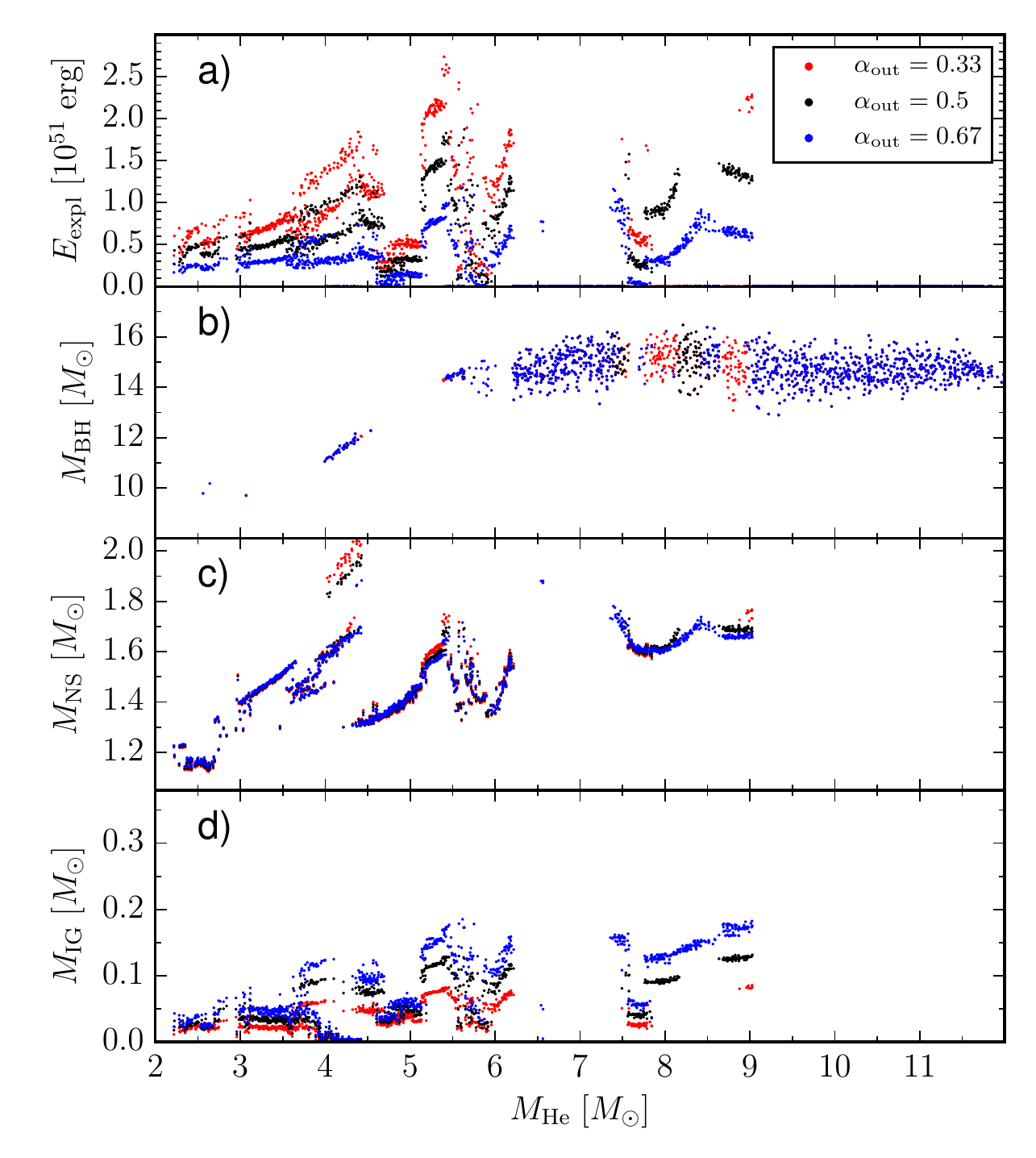}
  \includegraphics[width=0.48\linewidth]{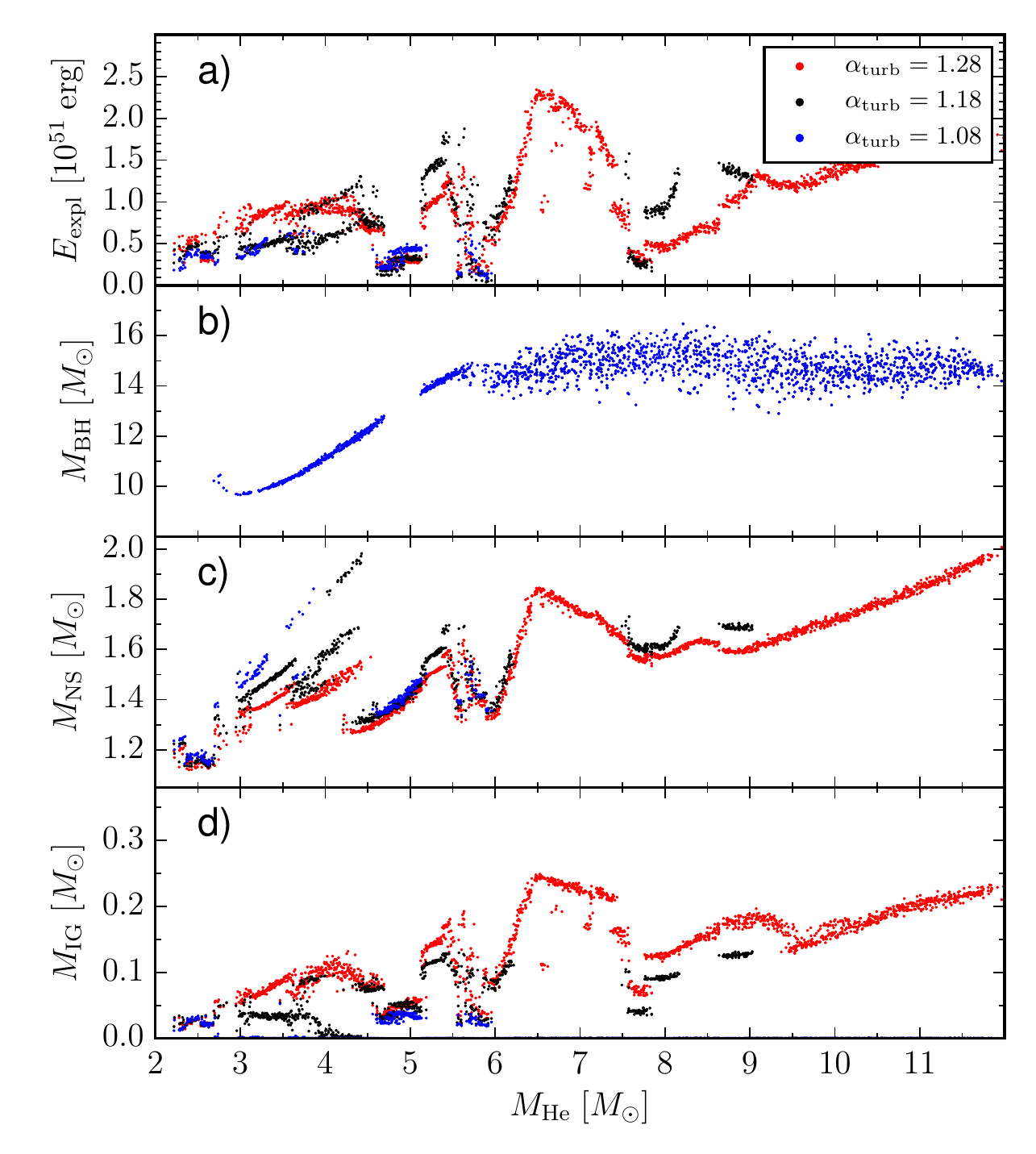}
  \caption{
    % THIS IS NOW MORE CONSISTENT
    Dependence of the landscape of explosion energies
    ($E_\mathrm{expl}$, Panels a), gravitational remnant masses for
      black holes ($M_\mathrm{BH}$, Panels b) and neutron stars
      ($M_\mathrm{NS}$, Panels c), and iron-group masses
    ($M_\mathrm{IG}$, Panels d) on the shock compression ratio
    ($\beta_\mathrm{expl}$, top left), the efficiency factor for the
    accretion luminosity ($\zeta$, top right), the outflow surface
    fraction ($\alpha_\mathrm{out}$, bottom left), and the factor for
    additional shock expansion due to higher turbulent pressure
    ($\alpha_\mathrm{turb}$, bottom right).  Different from
    Figure~\ref{fig:sensitivity}, the explosion parameters are given
    as a function of helium core mass at collapse, $M_\mathrm{He}$,
    instead of ZAMS mass.
    \label{fig:sensitivity_he}}
\end{figure*}

\begin{figure}
  \includegraphics[width=\linewidth]{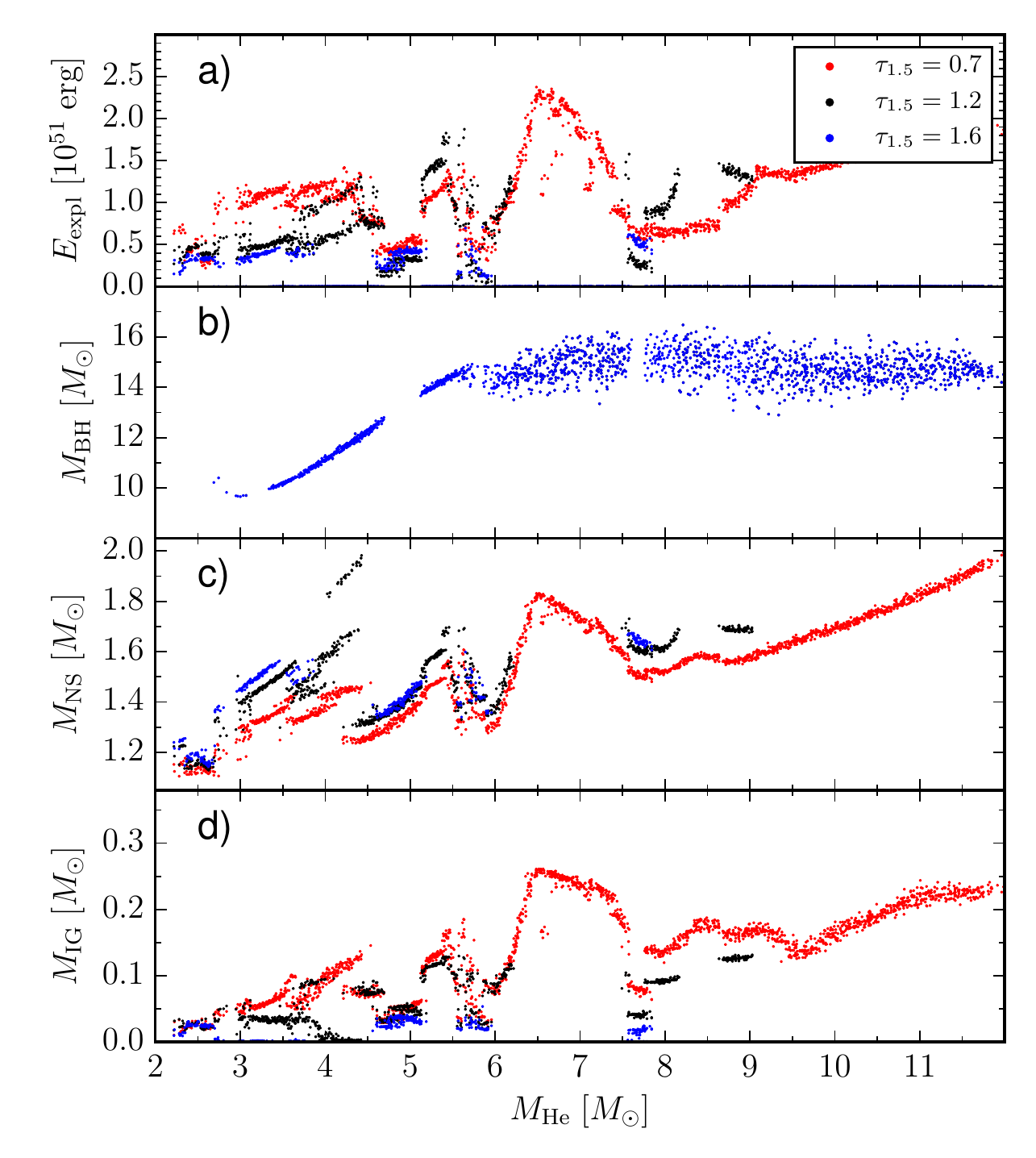}
  \caption{
    % THIS IS NOW MORE CONSISTENT
    Dependence of the landscape of explosion energies
    ($E_\mathrm{expl}$,  Panel a), gravitational remnant masses for
      black holes ($M_\mathrm{BH}$, Panel b) and neutron stars
      ($M_\mathrm{NS}$, Panel c), and iron-group masses
    ($M_\mathrm{IG}$, Panel d) on the cooling time-scale
    $\tau_\mathrm{1.5}$ for a $1.5\,M_\odot$ neutron star.  Different
    from Figure~\ref{fig:sensitivity2}, the explosion parameters are
    given as a function of helium core mass at
    collapse, $M_\mathrm{He}$, instead of ZAMS mass.
    \label{fig:sensitivity2_he}}
\end{figure}

For these reasons, we also provide versions of our key plots showing the
dependence of the explosion properties on the helium and C/O core mass at
collapse instead of the ZAMS mass.  Explosion energies, remnant
masses, and iron group masses for the standard scenario are shown in
Figure~\ref{fig:standard_he_co} (corresponding to
Figure~\ref{fig:standard}) both for the helium and the C/O core mass.
The dependence of the landscape of
explosion properties on the model parameters is illustrated in
Figure~\ref{fig:sensitivity_he} (for $\beta_\mathrm{expl}$, $\zeta$,
$\alpha_\mathrm{out}$, and $\alpha_\mathrm{turb}$) and
Figure~\ref{fig:sensitivity2_he} (for $\tau_\mathrm{1.5}$)
using only the helium core mass as abscissa (since
the picture is very similar for the C/O core mass).

\begin{figure}
  \includegraphics[width=\linewidth]{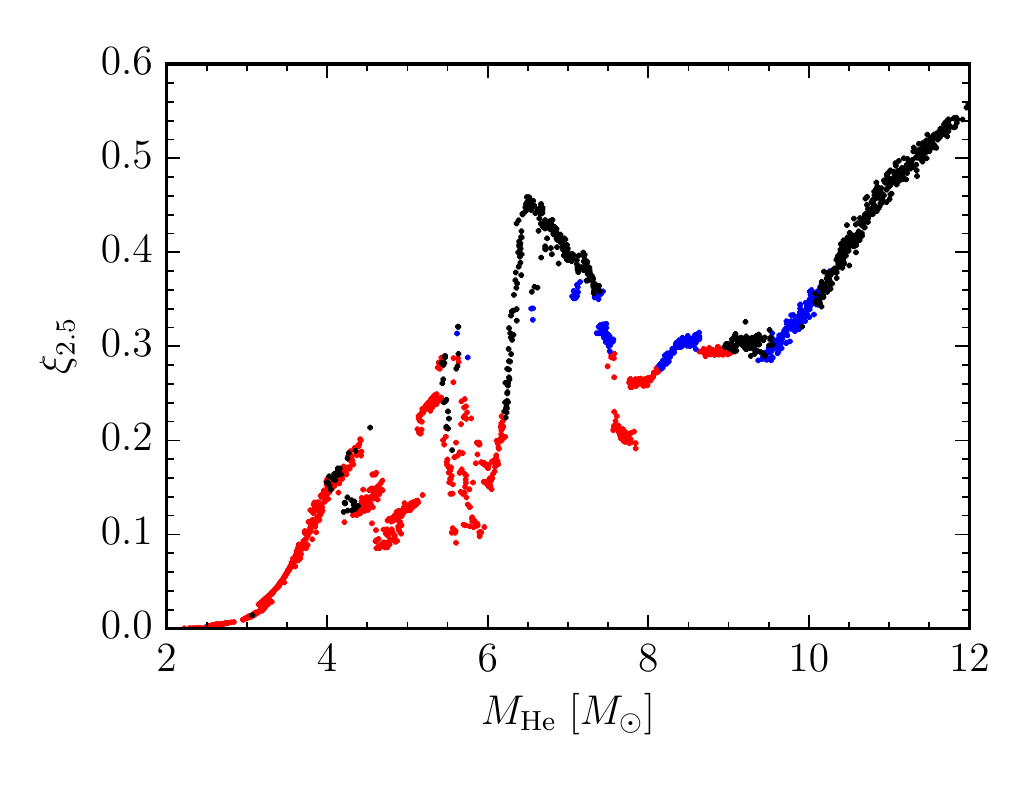}
  \caption{ Compactness parameters, $\xi_{2.5}$, for exploding (red)
    and non-exploding (black) models as a function of helium core mass
    $M_\mathrm{He}$ at collapse instead of ZAMS mass as in
    Figure~\ref{fig:xi}.  Blue dots denote models where shock revival
    is initiated, but the explosion eventually fails because the
    diagnostic energy becomes negative as the shock propagates out or
    the neutron star mass exceeds the maximum neutron star mass due to
    ongoing accretion in the explosion phase.
    \label{fig:xi_he}}
\end{figure}

In the mass range considered here ($10\,M_\odot\ldots 32\,M_\odot$),
the dependence of the helium core mass (as well as the C/O core mass)
on ZAMS mass is largely monotonic (\citealp{sukhbold_14}).  Plotting
the landscape of explosion properties as a function of helium or C/O
core mass instead of ZAMS mass therefore does not lead to major
qualitative changes.  The scatter, however, is somewhat reduced in
certain regions, e.g., in the islands of explodability at ZAMS masses
at $M=23\,M_\odot\ldots27\,M_\odot$ (corresponding to
$M_\mathrm{He}=7.5\,M_\odot\ldots9\,M_\odot$) in the standard case.
This reduction of the scatter is even more evident in the structural
features that determine the explosion properties, e.g., in the plot of
the compactness parameter $\xi_{2.5}$ versus $M_\mathrm{He}$ in
Figure~\ref{fig:xi_he}.  Considering that the uncertainties inherent
in our phenomenological supernova model dwarf the relatively small
scatter induced by the use of $M$ instead of $M_\mathrm{He}$ as
abscissa coordinate, however, the reduction of the scatter may be
noteworthy for future studies, but has no implications for the
conclusions of our present study.

\label{lastpage}

\end{document}